\documentclass[showpacs,preprintnumbers,amsmath,amssymb,floatfix]{revtex4}
\usepackage[german,english]{babel}
\usepackage{graphicx}
\usepackage{amsmath}
\usepackage{amsfonts}
\usepackage{amssymb}
\usepackage{epsfig}
\usepackage[hang,nooneline]{subfigure}
\newcommand{\insertplot}[5]{\begin{figure}
 \hfill\hbox to 0.05in{\vbox to #5in{\vfill
 \inputplot{#1}{#4}{#5}}\hfill}
 \hfill\vspace{-.1in}
 \caption{#2}\label{#3}
 \end{figure}}
\newcommand{\inputplot}[3]{
 \special{ps: plotfile #1}
}
 
\newcounter{fig}

\begin{document}

\title{\bf Symmetric and Asymmetric Wormholes Immersed In Rotating Matter}

\author{{\bf Christian Hoffmann$^{1,2}$}}
\email[{\it Email:}]{christian.hoffmann@uni-oldenburg.de}
\author{{\bf Theodora Ioannidou$^3$}}
\email[{\it Email:}]{ti3@auth.gr}
\author{{\bf Sarah Kahlen$^1$}}
\email[{\it Email:}]{sarah.kahlen@uni-oldenburg.de}
\author{{\bf Burkhard Kleihaus$^1$}}
\email[{\it Email:}]{b.kleihaus@uni-oldenburg.de}
\author{{\bf Jutta Kunz$^1$}}
\email[{\it Email:}]{jutta.kunz@uni-oldenburg.de}
\affiliation{
$^1$Institut f\"ur Physik, Universit\"at Oldenburg, Postfach 2503,
D-26111 Oldenburg, Germany\\
$^2$Department of Mathematics and Statistics, University of Massachusetts, Amherst,
Massachusetts, 01003-4525, USA\\
$^3$Faculty of Civil Engineering,  School of Engineering,  
Aristotle University of Thessaloniki, 54249, Thessaloniki, Greece
}

\date{\today}
\pacs{04.20.Jb, 04.40.-b}

\begin{abstract}
We consider four-dimensional wormholes immersed in bosonic matter.
While their existence is based on the presence of a phantom field,
many of their interesting physical properties are bestowed upon them
by an ordinary complex scalar field, which carries only a mass term,
but no self-interactions. For instance, the rotation of
the scalar field induces a rotation of the throat as well. Moreover,
the bosonic matter need not be symmetrically distributed in both
asymptotically flat regions, leading to symmetric and asymmetric
rotating wormhole spacetimes. The presence of the rotating
matter also allows for wormholes with a double throat.
\end{abstract}

\maketitle

\section{Introduction}

In General Relativity the non-trivial topology of wormholes
can be achieved by allowing for the presence of exotic matter 
\cite{Ellis:1973yv,Bronnikov:1973fh,Kodama:1978dw,Ellis:1979bh,Morris:1988cz,Morris:1988tu,Lobo:2005us,Lobo:2017oab}.
In the simplest case, a massless phantom (scalar) field can be chosen,
whose Lagrangian carries the opposite sign as compared to the one of
an ordinary scalar field. The resulting Ellis wormholes are static
spherically symmetric solutions, connecting two asymptotically flat
regions of space-time.

Ellis wormholes may also be coupled to matter fields.
For instance, 
they may be filled by nuclear matter 
\cite{Dzhunushaliev:2011xx,Dzhunushaliev:2012ke,Dzhunushaliev:2013lna,Dzhunushaliev:2014mza,Aringazin:2014rva,Dzhunushaliev:2016ylj},
they may be threaded by chiral (Skyrmionic) matter
\cite{Charalampidis:2013ixa}, 
or they may be immersed in bosonic matter
consisting of an ordinary complex boson field
with self-interaction
\cite{Dzhunushaliev:2014bya,Hoffmann:2017jfs},
or with only a mass term present \cite{Hoffmann:2017vkf}.


As in the case of non-topological solitons and boson stars 
\cite{Jetzer:1991jr,Lee:1991ax,Schunck:2003kk,Liebling:2012fv}
the boson field then has a harmonic time-dependence, 
allowing for localized matter fields surrounding a
static throat. Since the time-dependence cancels in the stress-energy
tensor, the resulting metric is static.
Clearly, there are symmetric wormholes, 
where the matter is distributed
symmetrically on either side of the throat.
Thus these symmetric  wormholes possess reflection symmetry.
Due to the non-trivial topology, however, also
wormhole solutions appear, where the matter is distributed unevenly
with respect to the spacetime regions on either side of the throat.
These asymmetric wormholes always appear in pairs,
where the two solutions of a pair are again related via 
reflection symmetry.

Ellis wormholes may also rotate 
\cite{Kashargin:2007mm,Kashargin:2008pk,Kleihaus:2014dla,Chew:2016epf,Kleihaus:2017kai}.
A rotation of the throat can be induced by an appropriate choice of the
boundary conditions, allowing for asymptotic flatness in the two
asymptotic regions, which, however, are rotating with respect to 
one another.
On the other hand, a rotation of the throat can also be induced
by immersing the throat into rotating matter
\cite{Hoffmann:2017vkf}.
In that case, the boundary conditions can be chosen symmetrically in
the two asymptotic regions, and thus
these need not rotate with respect to one another.

These wormholes immersed in rotating matter possess interesting properties \cite{Hoffmann:2017vkf}.
Depending on their physical parameters, their geometry can change
from exhibiting a single throat to developing an equator 
surrounded by a throat on either side, i.e., 
these wormholes then feature a double throat geometry.
These wormholes also possess ergoregions and an interesting
lightring structure.
Here we show that, 
like their non-rotating counterparts,
they also come in two versions, namely symmetric and asymmetric wormholes,
where the asymmetric wormholes again always come in pairs related to each
other via reflection symmetry.

The study of rotating wormholes is rather attractive from an
astrophysical point of view, since rotation is ubiquitous in the
Universe, and first astrophysical searches for wormholes
have already been carried out
\cite{Abe:2010ap,Toki:2011zu,Takahashi:2013jqa}.
In this respect 
studies of the properties of wormholes as gravitational lenses
are highly relevant
\cite{Cramer:1994qj,Safonova:2001vz,Perlick:2003vg,Nandi:2006ds,Nakajima:2012pu,Tsukamoto:2012xs,Kuhfittig:2013hva,Tsukamoto:2016zdu},
and so are studies of their shadows
\cite{Bambi:2013nla,Nedkova:2013msa}
together with the signatures of accretion disks
surrounding them \cite{Zhou:2016koy,Lamy:2018zvj}.

Here we investigate the basic physical properties of wormholes immersed 
in rotating bosonic matter, which does not possess any self-interaction,
focusing on the new aspect of the presence of symmetric and asymmetric solutions. 
The set of field equations is symmetric with respect
to reflection of the radial coordinate at the center,
and the boson field and the metric possess the same
boundary conditions in both asymptotically flat regions.
The solutions, however, may be either symmetric or asymmetric
with respect to such a reflection, and asymmetric solutions
exist at least within a certain domain
of the parameter space. 

We remark that the solutions investigated here are quite different from those (non-rotating)
solutions studied before, which are based on a complex boson field
with a sextic self-interaction \cite{Hoffmann:2017jfs}.
There, the mechanism of spontaneous symmetry breaking arises,
which is known in diverse contexts in physics, 
ranging from ferromagnetic materials to
the generation of mass of the elementary particles via the
Higgs mechanism (see e.g., \cite{Strocchi:2008gsa}).
It applies to systems, where the ground state of a system does not
display the full symmetry of the underlying set of equations.

The paper is organized as follows.
In section II we provide the theoretical setting for the wormhole solutions.
We present the action, the Ans\"atze, the equations of motion,
the boundary conditions, as well as the expressions for the
global charges, the geometrical properties, and the lightrings.
In section III we first demonstrate that for the present case
of a boson field with a mass term only, there are no non-trivial
solutions in the probe limit. Subsequently we address the
solutions of the fully gravitating system, starting with the
non-rotating case. For the rotating case we then consider
solutions with the lowest values of the rotational quantum number,
entering the Ansatz for the boson field analogously to the
well-known case of rotating boson stars 
\cite{Schunck:1996,Schunck:1996he,Ryan:1996nk,Yoshida:1997qf,Schunck:1999pm,Kleihaus:2005me,Kleihaus:2007vk}.
In particular,
we analyze their global charges,
discuss their bifurcations,
illustrate their geometrical properties, 
and address their lightring structure.
We end with our conclusions in section IV.

\section{Theoretical setting}

In the following we first present the action employed for obtaining 
wormholes in the presence of rotating bosonic matter.
We then discuss the Ans\"atze for the metric and the scalar fields,
exhibit the resulting set of field equations,
and present an adequate set of symmetric boundary conditions.
Subsequently we discuss the global charges of the solutions.
We present the formulae for the analysis of their geometrical properties, 
including their throat(s) and equator,
and we provide the basis for the analysis of their lightrings.

\subsection{Action}

We start from an action in four spacetime dimensions,
where General Relativity is minimally coupled to 
a complex scalar field $\Phi$ and a real phantom field $\Psi$.
The action $S$ 
\begin{equation}
S=\int \left[ \frac{1}{2 \kappa}{\cal R} + 
{\cal L}_{\rm ph} +{\cal  L}_{\rm bs} \right] \sqrt{-g}\  d^4x  
 \label{action}
\end{equation}
then contains besides the Einstein-Hilbert action
with curvature scalar $\cal R$, coupling constant $\kappa=8\pi G$
and metric determinant $g$,
the respective matter contributions,
the Lagrangian ${\cal L}_{\rm bs}$ of the complex scalar field $\Phi$
\begin{equation}
{\cal L}_{\rm bs} = 
-\frac{1}{2} g^{\mu\nu}\left( \partial_\mu\Phi^* \partial_\nu\Phi
                            + \partial_\nu\Phi^* \partial_\mu\Phi 
 \right) - m_{\rm bs}^2 |\Phi|^2  \ ,
\label{lphi}
\end{equation}
where the asterisk denotes complex conjugation
and $m_{\rm bs}$ denotes the boson mass,
as well as the Lagrangian ${\cal L}_{\rm ph}$ of the phantom field $\Psi$,
\begin{equation}
 {\cal L}_{\rm ph} = \frac{1}{2}\partial_\mu \Psi\partial^\mu \Psi \ ,
\label{lpsi}
\end{equation}
which carries the reverse sign as compared to the kinetic term 
of the complex scalar field $\Phi$
\cite{Ellis:1973yv,Bronnikov:1973fh,Kodama:1978dw,Ellis:1979bh,Morris:1988cz,Morris:1988tu,Lobo:2005us,Lobo:2017oab}.

By varying the action with respect to the metric
we obtain the Einstein equations
\begin{equation}
G_{\mu\nu}= {\cal R}_{\mu\nu}-\frac{1}{2}g_{\mu\nu}{\cal R} =  \kappa T_{\mu\nu}
\label{ee} 
\end{equation}
with stress-energy tensor
\begin{equation}
T_{\mu\nu} = g_{\mu\nu}{{\cal L}}_M
-2 \frac{\partial {{\cal L}}_M}{\partial g^{\mu\nu}} \ ,
\label{tmunu} 
\end{equation}
where we have denoted the sum of the scalar field Lagrangians by 
${\cal L}_{M} = {\cal L}_{\rm ph}+{\cal L}_{\rm bs} $.
By varying with respect to the
scalar fields we obtain the phantom field equation
\begin{equation}
\nabla^\mu \nabla_\mu \Psi =0 \ , 
\label{epsi} 
\end{equation}
and the equation for the complex scalar field
\begin{equation}
\nabla^\mu \nabla_\mu \Phi 
   = 
   m_{\rm bs}^2 \Phi \ .
\label{ephi} 
\end{equation}

For convenience we introduce dimensionless scalar fields 
and a dimensionless coordinate
\begin{equation}
\hat{\Phi} = \sqrt{\kappa}  \Phi \ , \ \ \ 
\hat{\Psi} = \sqrt{\kappa}  \Psi \ , \ \ \ 
\hat{\eta} = \frac{1}{\lambda} \eta \ ,
\label{scaling}
\end{equation}
yielding
\begin{equation}
\hat{G}_{\mu\nu} =  \hat{T}_{\mu\nu} \ ,
\label{ees} 
\end{equation}
where the mass term of the complex boson field 
$\hat{m}_{\rm bs}^2 |\hat{\Phi}|^2$ now contains the
dimensionless mass parameter 
\begin{equation}
\hat{m}_{\rm bs}=\left(\frac{m_{\rm bs}}{m_{\rm P}}\right)\left(\frac{M_0}{m_{\rm P}}\right) \ ,
\label{hatm} 
\end{equation}
with the mass scale $M_0$ being related 
to the length scale by $M_0 = \lambda /G$,
and $m_{\rm P}$ denotes the Planck mass.
Throughout the paper we will choose $\hat{m}_{\rm bs}=\sqrt{1.1}$.
In the following we will omit the hats again for simplicity.

\subsection{Ans\"atze}

The line element for the metric incorporates the non-trivial topology of the wormhole solutions,
\begin{equation}
ds^2 = -e^{f} dt^2 
    +e^{q-f}\left[e^b(d\eta^2 + h d\theta^2)+ h \sin^2\theta
    (d\varphi -\omega dt)^2\right] \ ,
\label{lineel}
\end{equation}
where  $f$, $q$, $b$ and $\omega$ are functions of 
the radial coordinate $\eta$ and the polar angle $\theta$,
and $h = \eta^2 +\eta_0^2$ is an auxiliary function, which contains the
throat  parameter $\eta_0$.
The radial coordinate $\eta$ takes positive and negative 
values, i.e. $-\infty< \eta < \infty$. 
The two limits $\eta\to \pm\infty$
correspond to two distinct asymptotically flat regions,
associated with ${\cal M}_+$ and ${\cal M}_-$, respectively.

For  the complex scalar field $\Phi$ we adopt the same Ansatz
as the one usually employed for rotating $Q$-balls and boson stars
\cite{Schunck:1996,Schunck:1996he,Ryan:1996nk,Yoshida:1997qf,Schunck:1999pm,Kleihaus:2005me,Kleihaus:2007vk},
\begin{equation}
\Phi(t,\eta,\theta, \varphi) 
  =  \phi (\eta,\theta) ~ e^{ i\omega_s t +  i n \varphi} \ ,   \label{ansatzp}
\end{equation}
where $\phi (\eta,\theta)$ is a real function,
$\omega_s$ denotes the real boson frequency, and $n$ is the integer winding number
or rotational quantum number. Finally, the Ansatz for
the phantom field $\Psi$ is chosen to depend only on the coordinates $\eta$ and $\theta$,
\begin{equation}
\Psi(t,\eta,\theta, \varphi) 
=  \psi (\eta,\theta) \ .
\label{ansatzph}
\end{equation}

\subsection{Einstein and Matter Field Equations}

After substituting the above Ans\"atze into the Einstein equations 
$E_\mu^\nu=G_\mu^\nu- T_\mu^\nu=0$ we obtain the following set of coupled field equations
\begin{equation}
f_{,\eta\eta} 
+ \frac{f_{,\theta\theta}}{h} 
+ f_{,\eta} \frac{h q_{,\eta} + 4 \eta}{2 h} 
+f_{,\theta} \frac{2 \cot\theta + q_{,\theta} }{2 h} 
- e^{q-2f} \sin^2\theta \left(h \omega_{,\eta}^2 + \omega_{,\theta}^2\right)
 =  
4 \left[2 (n \omega + \omega_s)^2 - m_{\rm bs}^2  e^{f} \right] e^{b+q-2f} \phi^2 \ ,
\label{eq1}
\end{equation}
\begin{equation}
q_{,\eta\eta} 
+ \frac{q_{,\theta\theta}}{h}
+\frac{q_{,\eta}^2}{2} 
+ \frac{3\eta q_{,\eta}}{h} 
+ \frac{q_{,\theta}^2}{2 h} 
+ \frac{2 q_{,\theta} \cot\theta}{h} 
 = 
8 \left(
  e^{q-2f} h\left[(n \omega + \omega_s)^2 - m_{\rm bs}^2  e^{f} \right]\sin^2\theta 
- n^2 
\right)
 \phi^2\frac{e^{b}}{h \sin^2\theta} \ ,
\label{eq2}
\end{equation}
\begin{eqnarray}
& &
b_{,\eta\eta} 
+ \frac{b_{,\theta\theta}}{h} 
+b_{,\eta} \frac{\eta}{h} 
\nonumber\\
&  &
+ \frac{1}{2}\left(
 f_{,\eta}^2+\frac{f_{,\theta}^2}{h}-q_{,\eta}^2-\frac{q_{,\theta}^2}{h}
 - \frac{4}{h} \left( \cot\theta  q_{,\theta} + \eta q_{,\eta} \right) 
+ 
 \frac{4}{h}\left( 1-\frac{\eta^2}{h}\right)
- 3 e^{q-2f} h \left(\omega_{,\eta}^2 +\frac{\omega_{,\theta}^2}{h}\right)\sin^2\theta 
\right)
\nonumber\\
& = &
2  \left(\psi_{,\eta}^2 +\frac{\psi_{,\theta}^2 }{h}\right)
-4 \left(\phi_{,\eta}^2 + \frac{\phi_{,\theta}^2}{h} \right)
-4 
\left( e^{q-2f} h \left[(n \omega + \omega_s)^2 - m_{\rm bs}^2 e^f\right]\sin^2\theta 
- 3n^2
\right)\phi^2
\frac{e^{b}}{ h \sin^2\theta} \ ,
\label{eq3}
\end{eqnarray}
\begin{equation}
\omega_{,\eta\eta} 
+ \frac{\omega_{,\theta\theta}}{h}
+\frac{\omega_{,\eta}}{2} \left( 3q_{,\eta}- 4 f_{,\eta} + \frac{8\eta}{h} \right)
+\frac{\omega_{,\theta} }{2h}
\left(3 q_{,\theta} - 4 f_{,\theta} +6 \cot\theta \right)
  =  
  8 n (n\omega + \omega_s)\phi^2\frac{e^b}{h\sin^2\theta} \ ,
\label{eq4}
\end{equation}
which result from $E^t_t=0$, $E^\eta_\eta+E^\theta_\theta=0$, $E^\varphi_\varphi=0$
and  $E^t_\varphi=0$.
The matter field equations are obtained from the Euler-Lagrange equations,
\begin{equation}
\partial_\mu\left(\frac{\partial L_{\rm m}}{\partial \phi_{,\mu}}\right)
-\frac{\partial L_{\rm m}}{\partial \phi} = 0 \ , \ \ \
\partial_\mu\left(\frac{\partial L_{\rm m}}{\partial \psi_{,\mu}}\right)
-\frac{\partial L_{\rm m}}{\partial \psi} = 0 \ , 
\label{ELeqs}
\end{equation}
where $L_{\rm m} = \left({\cal L}_{\rm bs}+{\cal L}_{\rm ph}\right)\sqrt{-g}$.
They read
\begin{equation}
  \phi_{,\eta\eta} 
+ \frac{\phi_{,\theta\theta}}{h}
+\frac{\phi_{,\eta}}{2} \left( q_{,\eta} + 4 \frac{\eta}{h}\right)
+ \frac{\phi_{,\theta} }{2h} \left( q_{,\theta} +2 \cot\theta \right)
+\left( 
e^{q-2f}h\left[(n \omega+ \omega_s )^2
- m_{\rm bs}^2  e^{f} \right]\sin^2\theta
- n^2
\right)\phi\frac{e^{b} }{h \sin^2\theta} = 0 \ ,
\label{eq5}
\end{equation}
\begin{equation}
 \psi_{,\eta\eta} 
+ \frac{\psi_{,\theta\theta}}{h}
+\frac{\psi_{,\eta}}{2} \left( q_{,\eta} + 4 \frac{\eta}{h}\right)
+ \frac{\psi_{,\theta}}{2h} \left(q_{,\theta}+2 \cot\theta \right)
  =  0 \ .
\label{eq6}
\end{equation}

This system of equations is symmetric with respect to reflection, $\eta \to - \eta$.
Thus it allows for reflection symmetric
solutions, i.e., solutions whose functions are either symmetric
or antisymmetric under reflection symmetry, $\eta \to - \eta$.
However, as we will see below, it also allows for solutions, which are asymmetric,
when $\eta \to - \eta$. These asymmetric solutions, however, then always come in pairs,
where the two solutions of a pair are related via the transformation $\eta \to - \eta$.

\subsection{Boundary Conditions}

In order to solve the above set of six coupled partial differential equations (PDEs)
of second order, we have to impose boundary conditions for each function
at the boundaries of the domain of integration consisting of the two asymptotic
regions $\eta \to \pm \infty$, the axis of rotation $\theta = 0$, and the 
equatorial plane $\theta = \pi/2$.

Our choice of boundary conditions is guided by considerations of symmetry. In particular,
we here would like to impose symmetric boundary conditions for the metric
and the complex boson field, i.e., boundary conditions which are the same
for $\eta \to \infty$ and $\eta \to -\infty$. Then also symmetric solutions will be found,
as briefly discussed in \cite{Hoffmann:2017jfs}. However, asymmetric solutions arise
as well, and in that case the asymmetry appearing in the
solutions is not enforced via the boundary conditions. 

Let us now detail our choice of boundary conditions.
We demand in both asymptotic regions a Minkowski metric and a vanishing boson field
\begin{equation}
\left. f(\eta,\theta)\right|_{\eta \to \pm \infty} = 
\left. q(\eta,\theta)\right|_{\eta \to \pm \infty} = 
\left. b(\eta,\theta)\right|_{\eta \to \pm \infty} = 
\left. \omega(\eta,\theta)\right|_{\eta \to \pm \infty} = 
\left. \phi(\eta,\theta)\right|_{\eta \to \pm \infty} = 0 \ .
\label{bc_pminfty}
\end{equation}
Along the rotation axis regularity requires
\begin{equation}
\left.\partial_\theta f(\eta,\theta)\right|_{\theta = 0} = 
\left.\partial_\theta q(\eta,\theta)\right|_{\theta = 0} = 
\left.\partial_\theta \omega(\eta,\theta)\right|_{\theta = 0} = 0 \ , \ \ \
\left. b(\eta,\theta)\right|_{\theta = 0} = \left. \phi(\eta,\theta)\right|_{\theta = 0} =0  \ .
\label{bc_axis}
\end{equation}
Imposing reflection symmetry with respect to the equatorial plane yields
\begin{equation}
\left.\partial_\theta f(\eta,\theta)\right|_{\theta = \frac{\pi}{2}} = 
\left.\partial_\theta q(\eta,\theta)\right|_{\theta = \frac{\pi}{2}} = 
\left.\partial_\theta b(\eta,\theta)\right|_{\theta = \frac{\pi}{2}} = 
\left.\partial_\theta \omega(\eta,\theta)\right|_{\theta = \frac{\pi}{2}} =
\left.\partial_\theta \phi(\eta,\theta)\right|_{\theta = \frac{\pi}{2}} = 0 \ .
\label{bc_eqplane}
\end{equation}

For the phantom field the boundary conditions need special care. Analogous to the other functions,
regularity and  reflection symmetry with respect to the equatorial plane require
\begin{equation}
\left.\partial_\theta \psi(\eta,\theta)\right|_{\theta = 0} = 
\left.\partial_\theta \psi(\eta,\theta)\right|_{\theta = \frac{\pi}{2}} = 0 \ .
\label{bcpsi_axiseqplane}
\end{equation}
However, since only derivatives of the phantom field enter the PDEs, we can
make any choice $\psi(\eta,\theta) \to \psi_0$ as $\eta \to \infty$.
The boundary condition at $\eta = -\infty$ is then determined from the 
asymptotic form of the solutions as $\eta \to -\infty$,
\begin{equation}
\left\{2\eta^2\partial_\eta \left(\eta^2\partial_\eta b\right)
+\sin^2\theta \left[\left(\eta^2\partial_\eta f\right)^2-4 \left(\eta^2\partial_\eta \psi\right)^2
                    +4\eta^2\partial_\eta \left(\eta^2\partial_\eta q\right)
		    +4\eta_0^2\right] \right\}_{\eta \to - \infty} = 0 \ .
\label{bspsi_minfty}
\end{equation}

\subsection{Mass, angular momentum and particle number}

With each of the two distinct asymptotically flat regions,
${\cal M}_\pm$,
we can associate a mass, $M_\pm$,
an angular momentum, $J_\pm$, and a particle number, $Q_\pm$, 
where the signs of the global charges refer to the signs of ${\cal M}_\pm$.
The mass and the angular momentum can be obtained from the asymptotic behaviour of 
the metric functions
\begin{equation}
f \longrightarrow \mp\frac{2 M_{\pm}}{\eta} \ , \ \ \ 
\omega \longrightarrow \frac{2 J_{\pm}}{\eta^3}  \ \ \ \ {\rm as} \ \eta \to \pm \infty \ .
\label{MJinfty}  
\end{equation}
%

For symmetric solutions the global charges are the same in both asymptotically flat regions.
Therefore, we will omit the index $\pm$ for symmetric solutions.
Their particle number $Q$ is related to the angular momentum $J$ via the
well-known relation $J=nQ$
\cite{Schunck:1996,Schunck:1996he,Ryan:1996nk,Yoshida:1997qf,Schunck:1999pm,Kleihaus:2005me,Kleihaus:2007vk}
which also holds in this case \cite{Hoffmann:2017vkf}.

For the asymmetric solutions the values of the mass in the two asymptotic regions differ from each other,
and so do the values of the angular momentum.
However, because the two asymmetric solutions of a pair are related via $\eta \to - \eta$,
their masses are related via $M_\pm \to M_\mp$, and likewise their angular momenta via
$J_\pm \to J_\mp$.

For the asymmetric solutions also the extraction of the respective particle numbers $Q_\pm$  is more involved, 
since the choice of the inner boundary  is ambiguous,
when the usual integral over the time-component  
of  the conserved current is performed, as discussed in \cite{Hoffmann:2017jfs}.
Therefore a trick may be applied, introducing a coupling  to a fictitious electromagnetic field,
such that the particle number can be read off asymptotically as the electric charge associated with
the gauge field. This approach is outlined in Appendix A.

\subsection{Geometrical properties}

From the geometrical side it is most interesting to analyze the throat structure
for these wormholes.
To that end we consider the
circumferential radius $R_e(\eta)$ in the equatorial plane,
\begin{equation}
R_e(\eta) = \sqrt{h}\left. e^{\frac{q-f}{2}}\right|_{\theta = \pi/2} \ .
\label{Rade}  
\end{equation}
The minima of the circumferential radius $R_e$ then correspond to throats, 
while the local maxima correspond to equators. 
Since the circumferential radius $R_e(\eta)$ increases without bound in the asymptotic regions,
${\cal M}_\pm$,
any equator must reside between two throats.

For symmetric solutions the center $\eta=0$ will always correspond to either a throat
or an equator. For asymmetric solutions this will no longer be the case.
Here the location of the single throat will shift away from $\eta=0$, and when an equator arises,
this can happen far from $\eta=0$, as well, as we will see below.

\subsection{Ergoregion and lightrings}

Further physical quantities of interest are the location of the ergoregion and the presence of lightrings.
The condition  $g_{tt}>0$ defines the ergoregion, i.e., it represents the region
where the time-time component of the metric is positive. Its boundary is referred to as ergosurface.
Here the condition
\begin{equation}
g_{tt}(\eta,\theta) = 0 \ 
\label{ergobound}  
\end{equation}
holds.

To find the lightrings we consider
the geodesic motion of massless particles in the equatorial plane.
This leads to the equation of motion
\begin{eqnarray}
\dot{\eta}^2 & = &
\frac{L^2}{e^{(b-q)}}\left(\frac{E}{L} -V_+(\eta)\right)
                     \left(\frac{E}{L} -V_-(\eta)\right) \ ,
\label{doteta2} \\
& & {\rm with} \ \ V_\pm(\eta) = \left(\omega\mp\frac{e^{f-q/2}}{\sqrt{h}}\right) \ ,	     
\label{Vpm} 
\end{eqnarray}
where $E$ and $L$ are the particle energy and angular momentum, respectively.
For circular orbits the derivatives $\dot{\eta}$ and $\frac{d\dot{\eta}}{d\eta}$ must vanish.
Consequently, the location of the lightrings $\eta_{\rm lr}$
is determined by the extrema of the potentials $V_\pm(\eta)$,
and by the ratio $\frac{E}{L}$, i.e. $\frac{E}{L}=V_\pm(\eta_{\rm lr})$.

\section{Solutions}

In this section we present and analyze the wormhole solutions immersed in bosonic matter.
First we will argue that in the probe limit no solutions with a non-trivial boson field exist.
Then we will turn to the gravitating solutions, starting with the non-rotating case
and subsequently address the rotating case, which represents the main focus of the present work.
In both cases we will consider the properties of the symmetrical and asymmetrical solutions.

The gravitating wormhole solutions depend on three continuous parameters,
the boson mass $m_{\rm bs}$, the boson frequency $\omega_s$,
and the throat parameter $\eta_0$, and on the integer winding number $n$. 
Note that the set of Einstein equations
and matter field equations is
invariant under the scaling transformation
\begin{equation}
\eta \to \lambda \eta \ , \ \ \ 
\eta_0 \to \lambda \eta_0 \ , \ \ \ 
\omega \to \frac{1}{\lambda} \omega \ , \ \ \ 
\omega_s \to \frac{1}{\lambda} \omega_s \ , \ \ \ 
m_{\rm bs} \to \frac{1}{\lambda} m_{\rm bs} \ .
\end{equation}
Here we choose $m_{\rm bs}=\sqrt{1.1}$ for the boson mass to break the scaling invariance. 
The remaining free parameters are then 
the boson frequency $\omega_s$, the throat parameter $\eta_0$, and the winding number $n$.

\subsection{Probe limit}

In the probe limit we assume a fixed spacetime background. For the given
boundary conditions the background is just the static Ellis wormhole,
since a rotating Ellis wormhole would need asymmetric boundary conditions
\cite{Kleihaus:2014dla,Chew:2016epf}.
In this case
all the metric functions (except for the auxiliary function $h=\eta^2 + \eta_0^2$) are zero. 

Integration of the field equation for the boson field,
Eq.~(\ref{eq5}), over the whole space then leads to 
\begin{equation}
\left. 2 h \phi_{,\eta}\right|_{-\infty}^{\infty} = 
\int{\left[ \left(m_{\rm bs}^2 -\omega_s^2\right)+\frac{n^2}{h\sin^2\theta}\right]h \phi(\eta,\theta) \sin\theta } 
         d\eta  d\theta \ .
\label{prblim}
\end{equation}
Assuming first that $m_{\rm bs}^2>\omega_s^2$ as to allow for exponentially decaying solutions,
the right hand side is positive unless $\phi$ vanishes identically. 
Consequently, the left hand side must be non-vanishing as well, which implies that
the boson field should behave like $B_{\pm}/\eta$, with constants $B_{\pm}$, in the asymptotic regions.
In this case, however, we would find $h\phi \to \eta B_{\pm}$, and the integral in Eq.~(\ref{prblim})
would not exist. 
Consequently, only the trivial solution $\phi=0$ remains in the probe limit.

We note, however, that this result does not hold true, if the bosonic potential includes
adequate self-interaction terms. Then a non-trivial probe limit exists,
as has been demonstrated for the non-rotating case \cite{Hoffmann:2017jfs}.
A non-trivial probe limit exists also in the rotating case in five dimensions,
when such self-interactions are present \cite{Hoffmann}, 
and we expect it in the corresponding four-dimensional rotating case, as well.

This very different behaviour for the case without and with self-interaction is not unexpected.
On the contrary, we know that also for the case of a trivial spacetime background,
such as a simple Minkowski spacetime, there are no localized finite energy solutions of the boson field,
when there is only a mass term for the boson field. The localized $Q$-ball 
(or non-topological soliton) solutions,
in contrast, require a sextic potential for their existence.

In this latter case of a sextic self-interaction of the matter fields,
the replacement of the Minkowski background with a topologically non-trivial
wormhole background has brought forward the interesting phenomenon of
spontaneous symmetry breaking \cite{Hoffmann:2017jfs}.
Here in a certain range of the parameter space, 
the equations allow not only for symmetric solutions, where the boson field
configuration is symmetric under reflection symmetry, but it allows also
for asymmetric solutions, where the latter come in pairs, which are
energetically favoured as compared to the symmetric solutions \cite{Hoffmann:2017jfs}.

When gravity is coupled, all these solutions continue to exist,
and thus the phenomenon of spontaneous symmetry breaking
is retained in the presence of gravity for wormholes immersed
in self-interacting bosonic matter.
Here we will show that the absence of a non-trivial probe limit
for the case of bosons without self-interaction
is associated with the absence of the analogous phenomenon of 
spontaneous symmetry breaking in the gravitating case.
Still there will be symmetric and asymmetric gravitating solutions,
but the asymmetric  solutions are no longer always energetically favoured.

\subsection{Non-rotating solutions}

Let us now turn to the gravitating solutions.
For vanishing rotational quantum number, $n=0$, the Ansatz simplifies considerably
and so does the set of Einstein equations and matter field equations.
Now all functions depend only on the radial coordinate $\eta$, and the 
metric functions $b$ and $\omega$ are trivial, $b=0$ and $\omega=0$. The system of
partial differential equations reduces to a system of
ordinary differential equations.

\begin{figure}[t!]
\begin{center}
\mbox{\hspace{0.2cm}
\subfigure[][]{\hspace{-1.0cm}
\includegraphics[height=.25\textheight, angle =0]{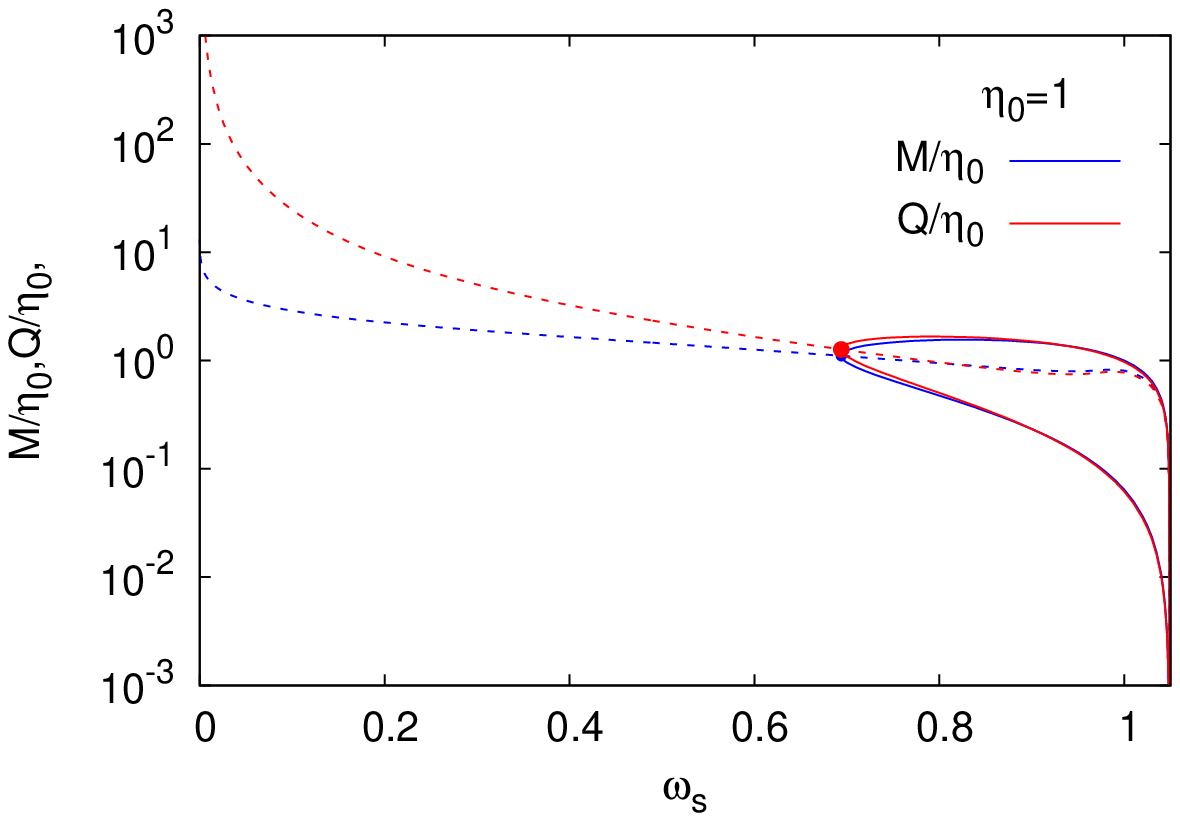}
\label{Fig1a}
}
\subfigure[][]{\hspace{-0.5cm}
\includegraphics[height=.25\textheight, angle =0]{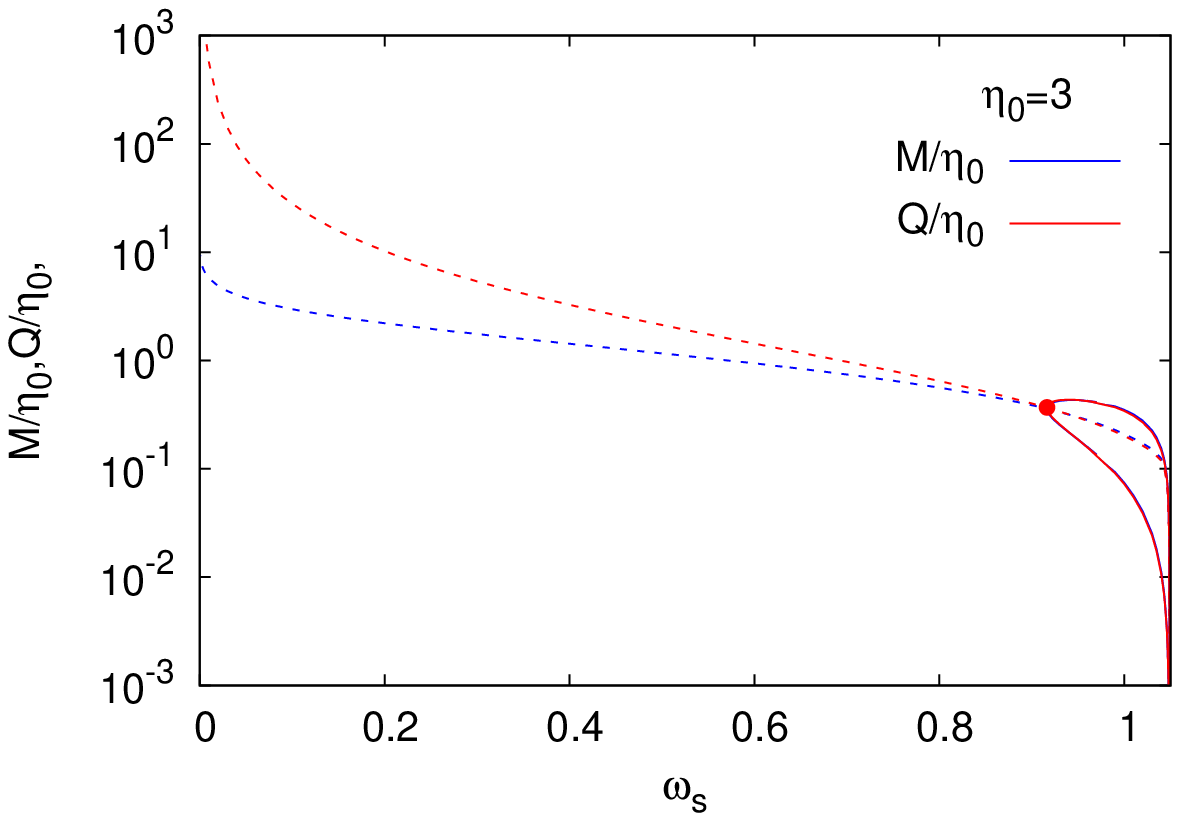}
\label{Fig1b}
}
}
\mbox{\hspace{0.2cm}
\subfigure[][]{\hspace{-1.0cm}
\includegraphics[height=.25\textheight, angle =0]{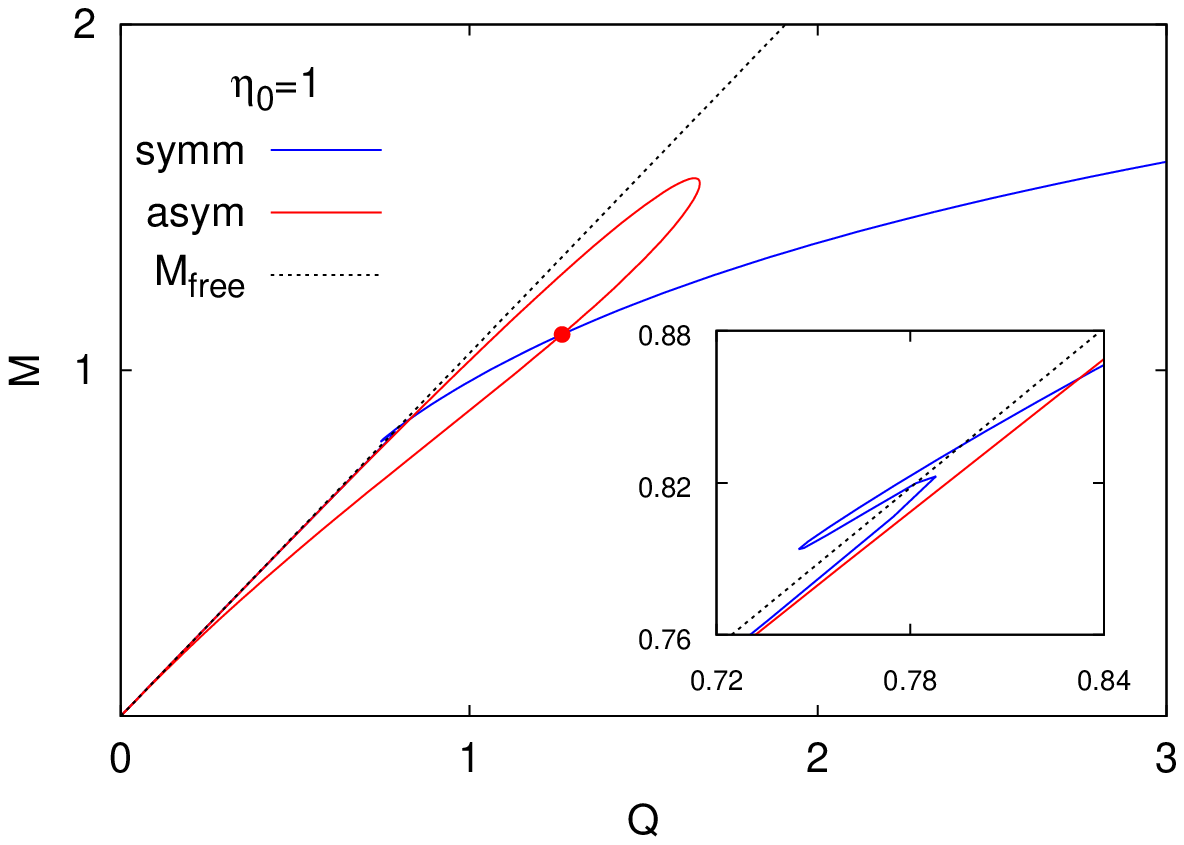}
\label{Fig1c}
}
\subfigure[][]{\hspace{-0.5cm}
\includegraphics[height=.25\textheight, angle =0]{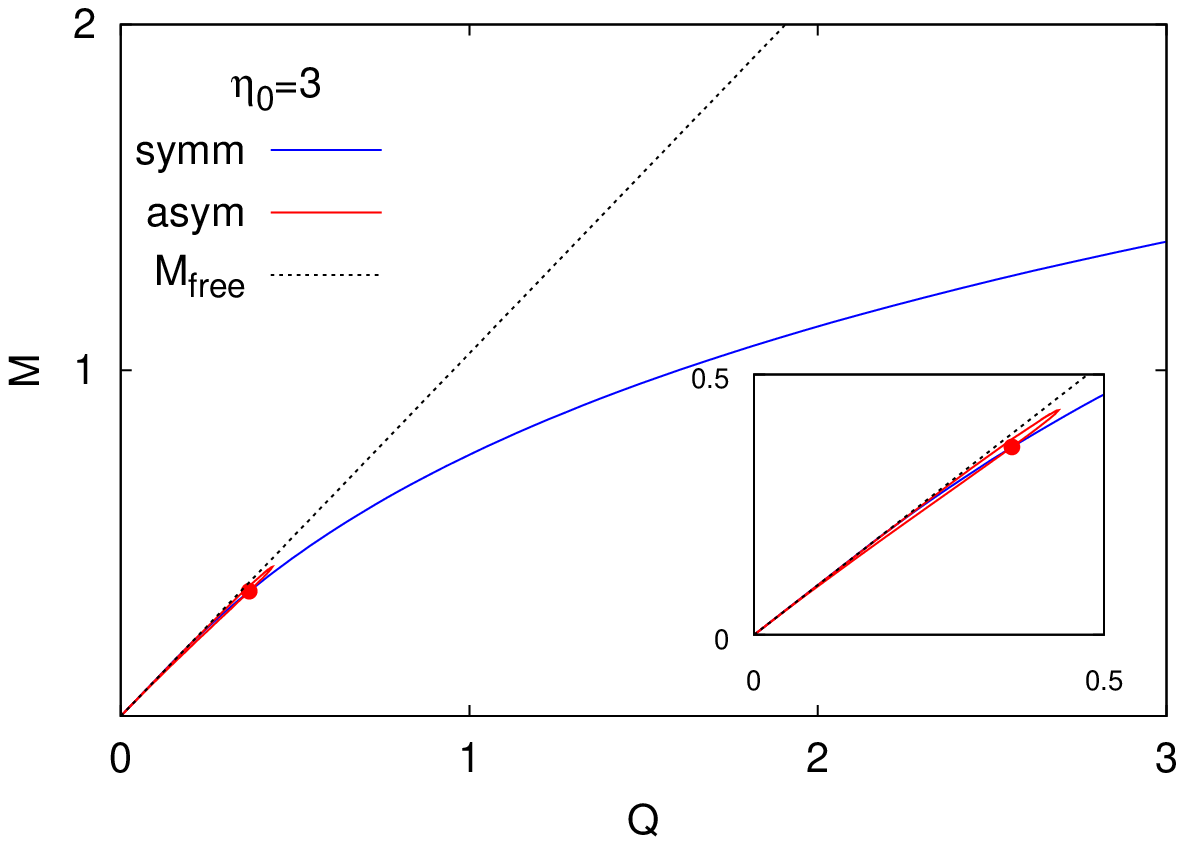}
\label{Fig1d}
}
}
\mbox{\hspace{0.2cm}
\subfigure[][]{\hspace{-1.0cm}
\includegraphics[height=.25\textheight, angle =0]{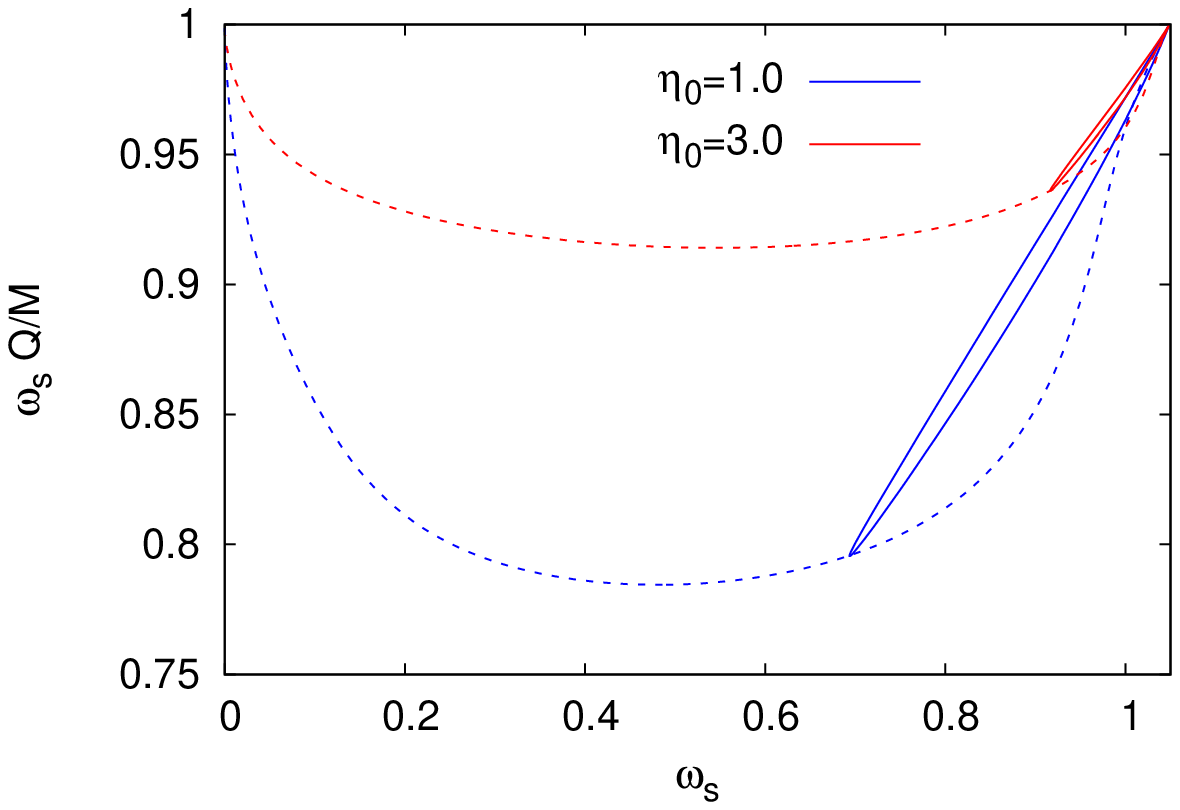}
\label{Fig1e}
}
\subfigure[][]{\hspace{-0.5cm}
\includegraphics[height=.25\textheight, angle =0]{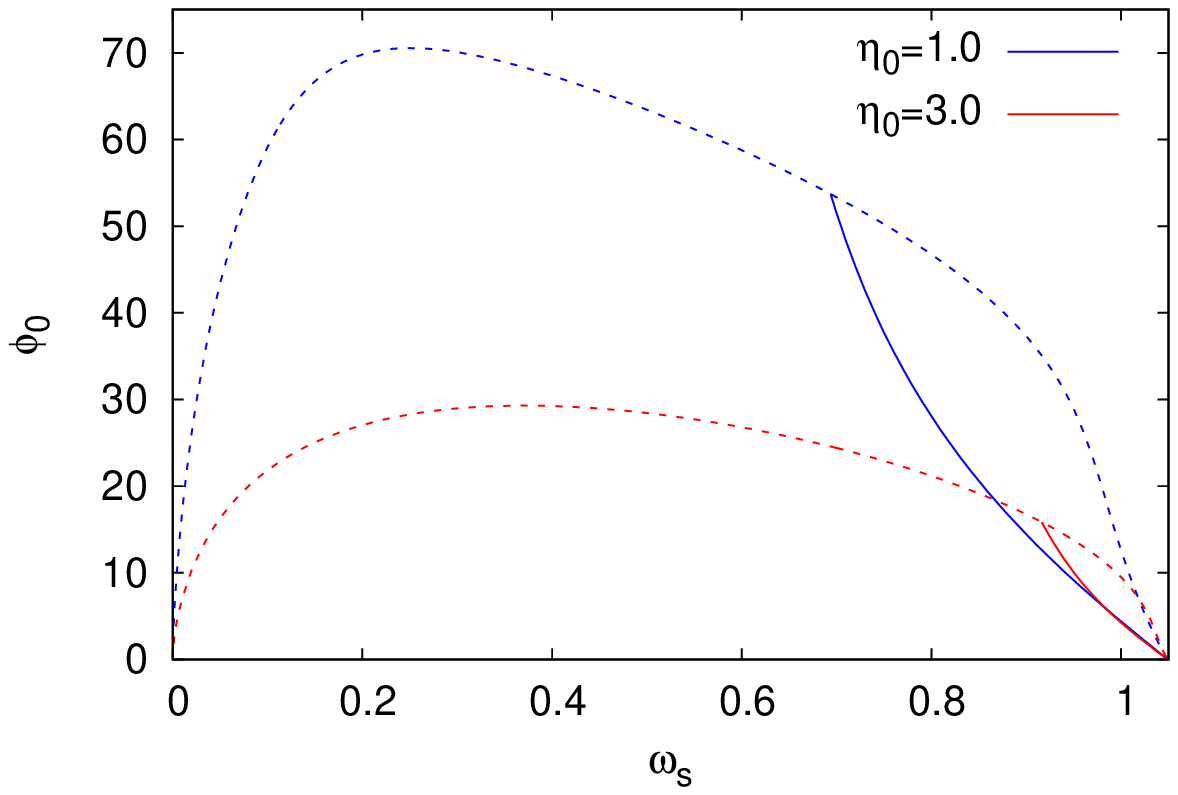}
\label{Fig1f}
}
}
\end{center}
\vspace{-0.5cm}
\caption{
Non-rotating solutions ($n=0$):
(a) the scaled mass $M/\eta_0$ and the scaled particle number $Q/\eta_0$
 versus the boson frequency $\omega_s$ for  throat parameter $\eta_0=1$;
(b) same as (a) for  $\eta_0=3$;
(c) the mass $M$ versus the particle number $Q$ for throat parameter $\eta_0=1$;
(d)  same as (c) for  $\eta_0=3$;
(e) the quantity $\omega_s Q/M$ versus the boson frequency $\omega_s$ for  
throat parameters $\eta_0=1$ and $3$;
(f) the value of the boson field at the center $\phi_0=\phi(0)$ versus the boson frequency $\omega_s$ for  
throat parameters $\eta_0=1$ and $3$.
The dashed and solid curves correspond to the symmetric and asymmetric solutions, respectively.
The dots indicate the bifurcation point of the asymmetric solutions.
\label{Fig1}
}
\end{figure}

We have solved these ordinary differential equations in the full range of the boson frequency
$0< \omega_s < m_{\rm bs}$, selecting for the throat parameter the values $\eta_0=1$ and $3$.
In Fig.\ref{Fig1} we present the main features of these solutions.
Figs.\ref{Fig1a} and \ref{Fig1b} show the mass $M$ and particle number $Q$ versus
the boson frequency $\omega_s$ for 
$\eta_0=1$ and $\eta_0=3$, respectively.
Also shown is the mass of $Q$ free particles, $M_{\rm free} = m_{\rm bs} Q$,
for comparison.

For large values of the boson frequency $\omega_s$ both symmetric and asymmetric solutions
exist. As noted above, the latter always appear in pairs, since for any asymmetric solution
we obtain a second asymmetric solution by the transformation $\eta \to -\eta$.
Then we can consider the three masses shown in the figures
as representing the masses of these three solutions
at either asymptotically flat region. By going  from ${\cal M}_+$ to ${\cal M}_-$
the masses $M_+$ and $M_-$ will only be interchanged.

In the limit $\omega_s \to m_{\rm bs}$, the mass and particle number tend to zero. In fact, the
boson field vanishes in this limit. The limiting solution, then, is the massless Ellis
wormhole.
Whereas the symmetric solutions extend down to arbitrarily small values of $\omega_s$,
the asymmetric solutions merge with the symmetric ones at some critical value $\omega_s^{\rm cr}$,
which depends on $\eta_0$, as seen in the figures.

Figs.\ref{Fig1c} and \ref{Fig1d} show the mass as a function of the particle number for 
the same sets of solutions.
We observe that for $\eta_0=1$ the mass of the symmetric solutions exhibits two spikes 
corresponding to three branches of solutions. On the first branch the mass increases with
the particle number up to a local maximum. A second branch then extends
back to smaller values of the mass and particle number until a local minimum is reached. 
Here a third branch arises along which the mass increases again monotonically with
the particle number. In contrast,
for $\eta_0=3$ the mass of the symmetric solutions increases monotonically with the
particle number, forming only a single branch.
For the pair of asymmetric solutions the mass forms a loop, both for $\eta_0=1$ and $\eta_0=3$.
The bifurcation point at $\omega_s^{\rm cr}$, where the asymmetric branches end, 
is indicated by a dot in the figures.

When considering the phenomenon of spontaneous symmetry breaking, observed
before in the presence of self-interactions of the boson field \cite{Hoffmann:2017jfs},
we realize the absence of this phenomenon in the present sets of solutions.
While there are symmetric and asymmetric solutions in a certain interval
of the parameter space here, as well, we cannot interpret these as arising 
on energetic grounds.
In the case of self-interaction, for a given particle number
both asymmetric solutions possess a smaller mass than
the symmetric solution. Thus the asymmetric solutions are energetically favoured over
the symmetric solutions in the interval where they exist.
(Note, that only in the small interval close to $\omega_{\rm max}$
one cannot discern any difference between the three masses.)

In contrast, without the self-interaction present, only the mass of one of the sets of asymmetric 
solutions is always smaller than the mass of the symmetric solutions for a given
particle number. The mass of the other set of asymmetric solutions, however, is larger 
than the mass of the symmetric solutions for a given particle number
in a large range of the interval. In particular, at the bifurcation point the mass of
one set of asymmetric solutions approaches the mass of the symmetric solution
from below, while the mass of the other set of asymmetric solutions approaches from above.

The energetics of the boson field does not seem to be sufficiently relevant
in the present case to give rise to the phenomenon of spontaneous symmetry breaking.
Here both symmetric and asymmetric solutions seem to exist, 
at least in a part of the parameter space,
simply because the field equations
and the boundary conditions allow for them to exist, when the spacetime
topology is non-trivial.

Inspecting  Figs.\ref{Fig1a} and \ref{Fig1b} again, we note that
in the limit $\omega_s \to 0$ the mass and particle number seem to diverge for the symmetric solutions.
(The asymmetric ones have ceased to exist below $\omega_s^{\rm cr}$.)
Remarkably, however, the quantity $\omega_s Q/M$ 
tends to one in this limit, corresponding to $M = \omega_s Q + \cdots$, as seen 
in Fig.\ref{Fig1e}. On the other hand
the boson field vanishes as $\omega_s \to 0$, as seen in Fig.\ref{Fig1e}, 
where its central maximal value $\phi_0$ is exhibited versus the boson frequency,
which seems hard to reconcile with a diverging particle number and mass. 
To gain some understanding of this limit
we will address the behaviour of the solutions in this limit in Appendix B.

\subsection{Rotating solutions}

We now turn to the rotating solutions.
What makes these solutions so special is that the rotation is not imposed via the boundary
conditions, as done in the case of the rotating Ellis wormholes 
\cite{Kleihaus:2014dla,Chew:2016epf}.
Instead the rotation of the wormholes is generated by the rotation of the bosonic matter
into which the wormholes are immersed.
This allows for wormhole configurations, where both asymptotically
flat regions are completely symmetric \cite{Hoffmann:2017vkf}.
However, besides these symmetric solutions the non-trivial topology allows for asymmetric
solutions, as well, as we show below.

We have constructed numerically solutions for the rotational quantum numbers $n=1$ and $n=2$, 
and for the values of the throat parameter $\eta_0=1$ and $\eta_0=3$,
covering the interval of the boson frequency
$0.3 \leq \omega_s \leq  1.03$. 
For values of $\omega_s$ outside this interval
the numerical errors have increased too much, making the solutions
no longer fully reliable.

We have solved the system of coupled partial differential equations
with the help of the routine FIDISOL/CADSOL \cite{schonauer:1989},
a finite difference solver based on a Newton-Raphson scheme.
To obtain a finite coordinate patch, we have made a coordinate
transformation to a compactified coordinate
$x=\arctan(\eta/\eta_0)$. We have then chosen a
non-equidistant grid, employing typically $100 \times 40$ grid points
in radial direction $x$ and angular direction $\theta$, respectively.

In the following we first consider the global charges of the resulting
wormhole solutions. Then we address their ergoregions, their geometry,
and their lightrings.
We always consider symmetric and asymmetric solutions.

\subsubsection{Global charges}

\begin{figure}[t!]
\begin{center}
\mbox{\hspace{0.2cm}
\subfigure[][]{\hspace{-1.0cm}
\includegraphics[height=.25\textheight, angle =0]{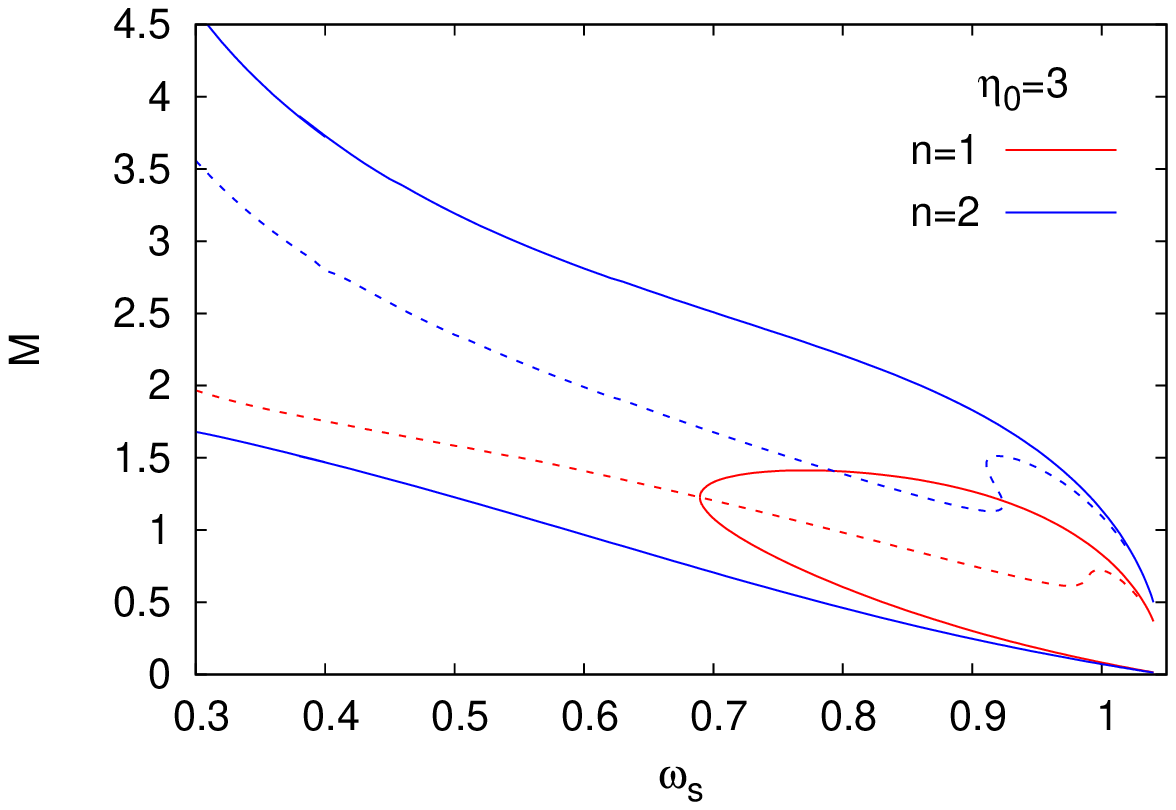}
\label{Fig2a}
}
\subfigure[][]{\hspace{-0.5cm}
\includegraphics[height=.25\textheight, angle =0]{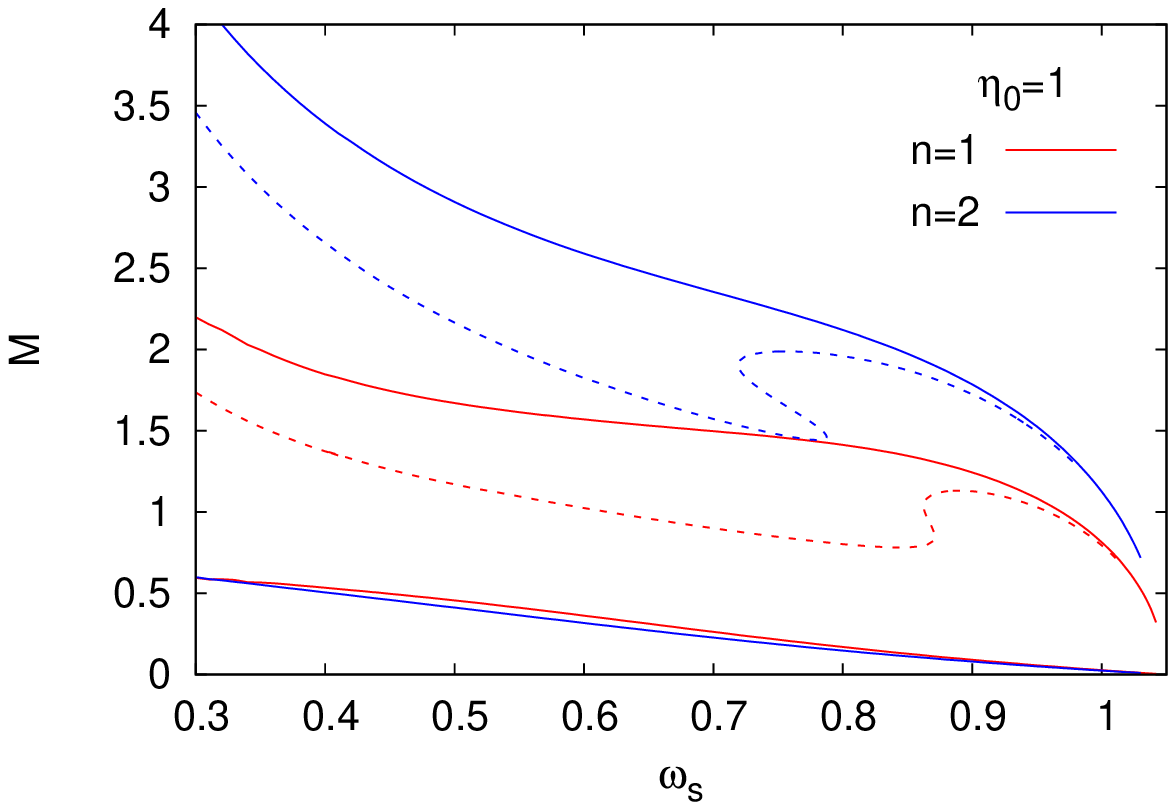}
\label{Fig2b}
}
}
\mbox{\hspace{0.2cm}
\subfigure[][]{\hspace{-1.0cm}
\includegraphics[height=.25\textheight, angle =0]{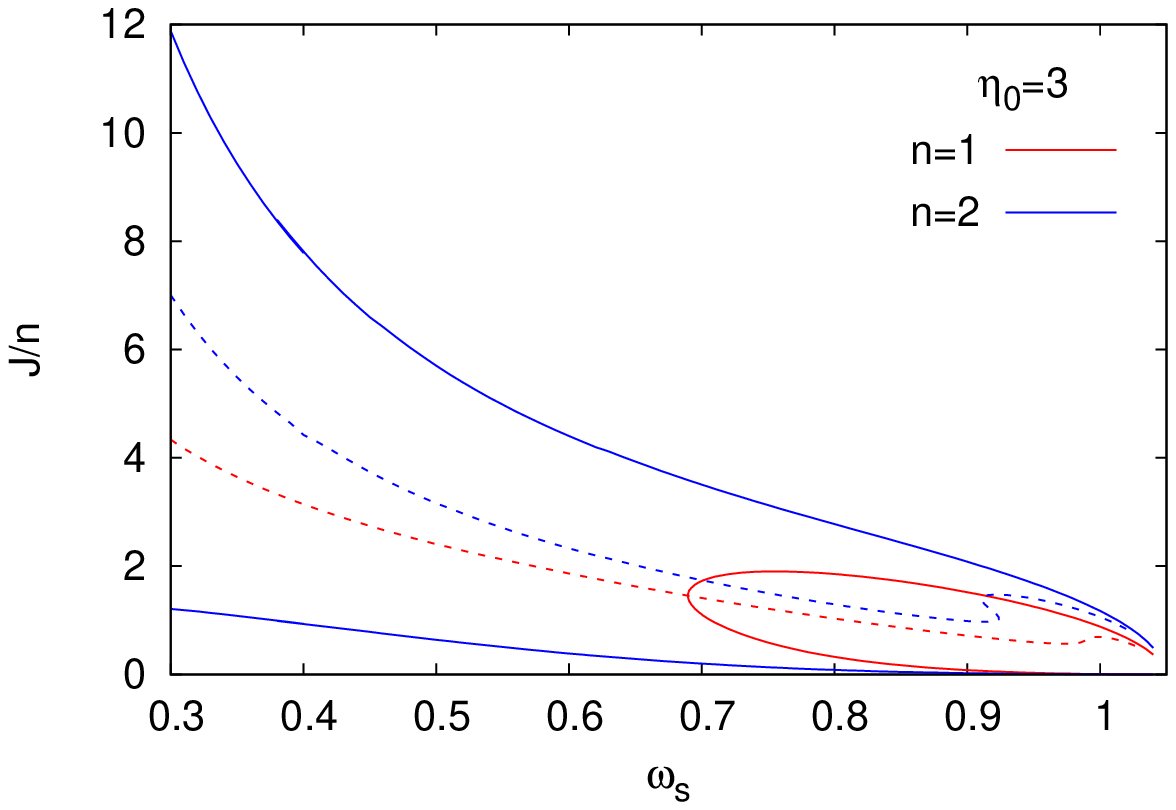}
\label{Fig2c}
}
\subfigure[][]{\hspace{-0.5cm}
\includegraphics[height=.25\textheight, angle =0]{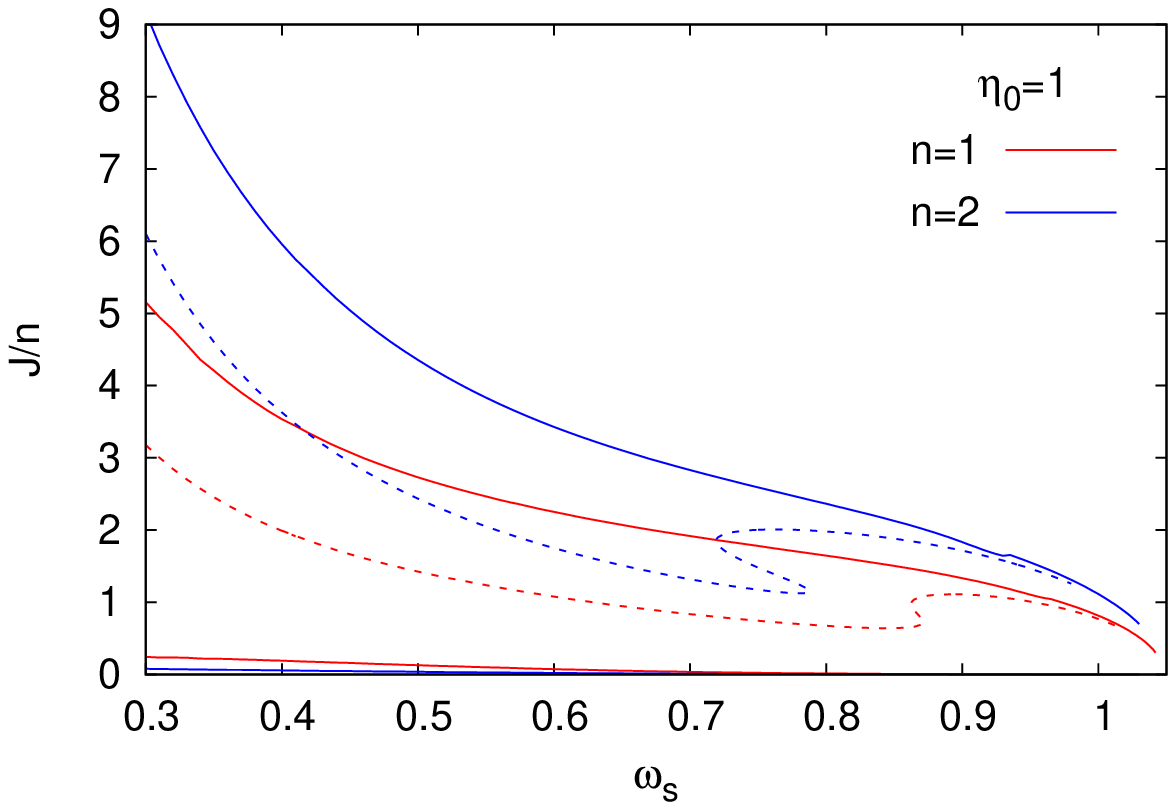}
\label{Fig2d}
}
}
\mbox{\hspace{0.2cm}
\subfigure[][]{\hspace{-1.0cm}
\includegraphics[height=.25\textheight, angle =0]{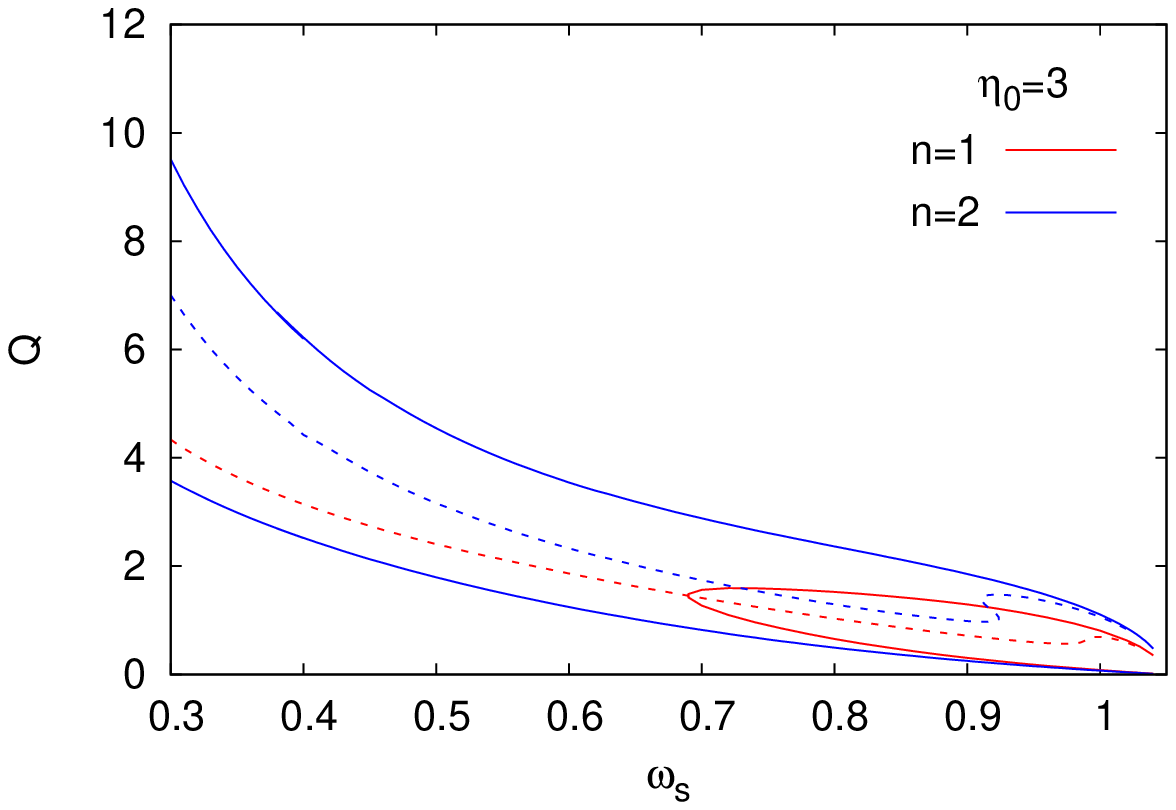}
\label{Fig2e}
}
\subfigure[][]{\hspace{-0.5cm}
\includegraphics[height=.25\textheight, angle =0]{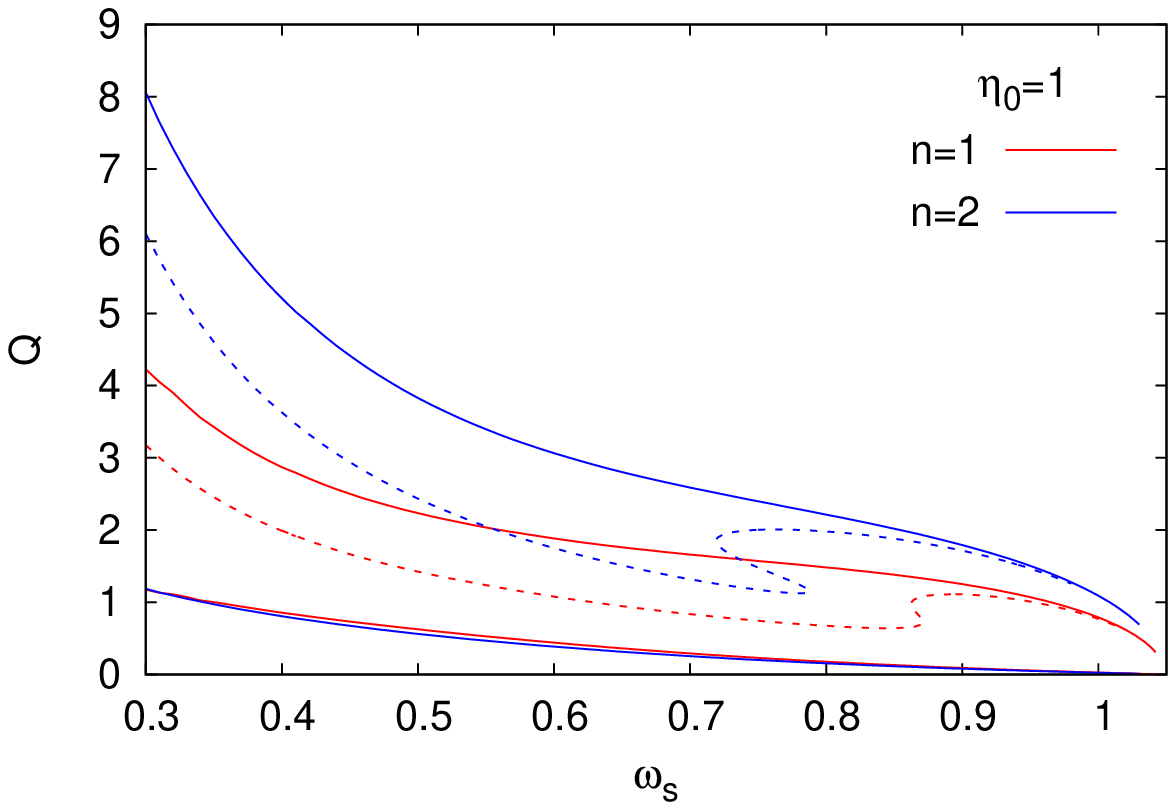}
\label{Fig2f}
}
}

\end{center}
\vspace{-0.5cm}
\caption{Global charges of rotating solutions:
(a) mass $M$ versus boson frequency $\omega_s$ for rotational quantum number $n=1$ and $n=2$ 
for the symmetric (dashed) and 
asymmetric (solid) solutions for throat parameter $\eta_0=3$;
(b) same as (a) for  $\eta_0=1$;
(c) scaled angular momentum $J/n$ versus the boson frequency $\omega_s$ 
for rotational quantum number $n=1$ and $n=2$ 
for the symmetric (dashed) and 
asymmetric (solid) solutions for throat parameter $\eta_0=3$;
(d) same as (c) for  $\eta_0=1$.
(e) particle number $Q$ versus the boson frequency $\omega_s$ 
for rotational quantum number $n=1$ and $n=2$ 
for the symmetric (dashed) and 
asymmetric (solid) solutions for throat parameter $\eta_0=3$;
(f) same as (e) for  $\eta_0=1$.
\label{Fig2}
}
\end{figure}

Let us start our discussion  of the wormhole solutions immersed in rotating bosonic matter
by addressing their global charges, their mass, angular momentum
and particle number.
In Fig.\ref{Fig2} we show the mass and the angular momentum versus the boson frequency $\omega_s$.
From Fig.\ref{Fig2a} we observe that for the smallest (non-trivial) rotational quantum number
$n=1$ and throat parameter $\eta_0=3$ the situation  is analogous to the non-rotating case.
The asymmetric solutions exist only down to a critical value of the boson frequency
$\omega_s^{\rm cr}$, where they bifurcate with the symmetric solutions.
The latter again seem to exist in  the full interval of the boson frequency.

However, for rotational quantum number $n=2$
and throat parameter $\eta_0=3$, we do not observe such a bifurcation
of the asymmetric and symmetric solutions any longer.
Instead all three sets of solutions, the symmetric one and the pair of asymmetric ones,
appear to persist in the full interval of the boson frequency, 
as seen in  the figure.
The same holds true for throat parameter $\eta_0=1$ and both rotational
quantum numbers, $n=1$ and $n=2$, which is illustrated in Fig.\ref{Fig2b}.

Fig.\ref{Fig2c} and \ref{Fig2d} show the angular momentum $J$, scaled by the
respective rotational quantum number $n$,  for the same sets of solutions.
We note, that the scaled angular momentum $J/n$ exhibits a very similar behaviour
as the mass of the solutions. 

Fig.\ref{Fig2e} and \ref{Fig2f} show the particle number $Q$ computed as outlined
in Appendix A for the same sets of solutions.
Also the  particle number  exhibits a very similar behaviour
as the mass and the scaled angular momentum of the solutions. 

We note, that for the symmetric solutions in Fig.\ref{Fig2} additional structures are present,
which can be traced back to the presence of a spiral in boson star solutions 
\cite{Jetzer:1991jr,Lee:1991ax,Schunck:2003kk,Liebling:2012fv}.
This spiral, however, unwinds when a negative energy density is present
\cite{Dzhunushaliev:2014bya,Hoffmann:2017jfs,Hartmann:2013tca}.
In particular, we observe that backbending of the curves occurs (except for $n=1$ and $\eta_0=3$),
i.e. mass and angular momentum are then no longer uniquely characterized by 
the boson frequency $\omega_s$. 
This is in contrast to the asymmetric solutions, where the mass, angular momentum and particle number
are unique functions of the boson frequency. In fact, the mass, angular momentum and particle number of all asymmetric solutions
decrease monotonically with increasing boson frequency $\omega_s$, 
except for $M_+$, $J_+$, and $Q_+$ for $\eta_0=3$ and $n=1$.
In the latter case  $M_+$, $J_+$, and $Q_+$ possess (local) maxima, but these do not occur at the same value of 
the boson frequency $\omega_s$.

\begin{figure}[t!]
\begin{center}
\mbox{\hspace{0.2cm}
\subfigure[][]{\hspace{-1.0cm}
\includegraphics[height=.25\textheight, angle =0]{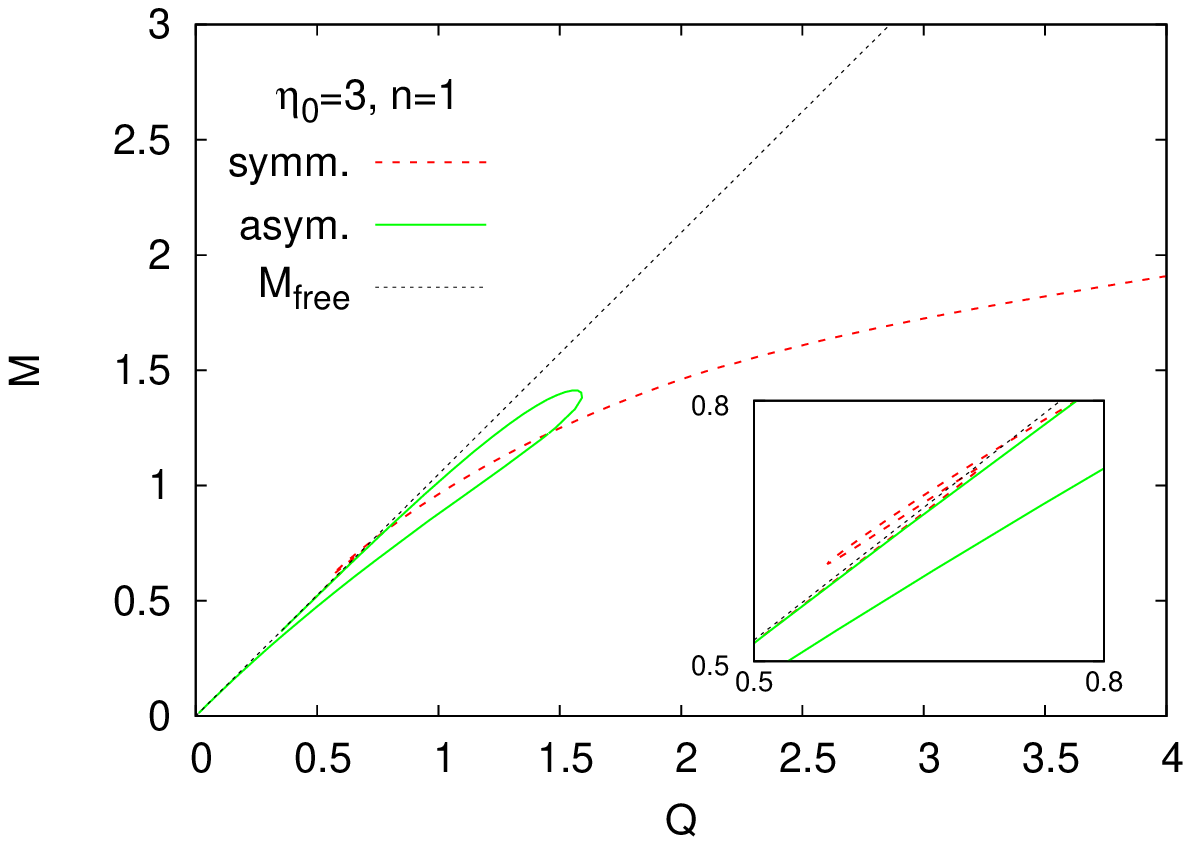}
\label{Fig3a}
}
\subfigure[][]{\hspace{-0.5cm}
\includegraphics[height=.25\textheight, angle =0]{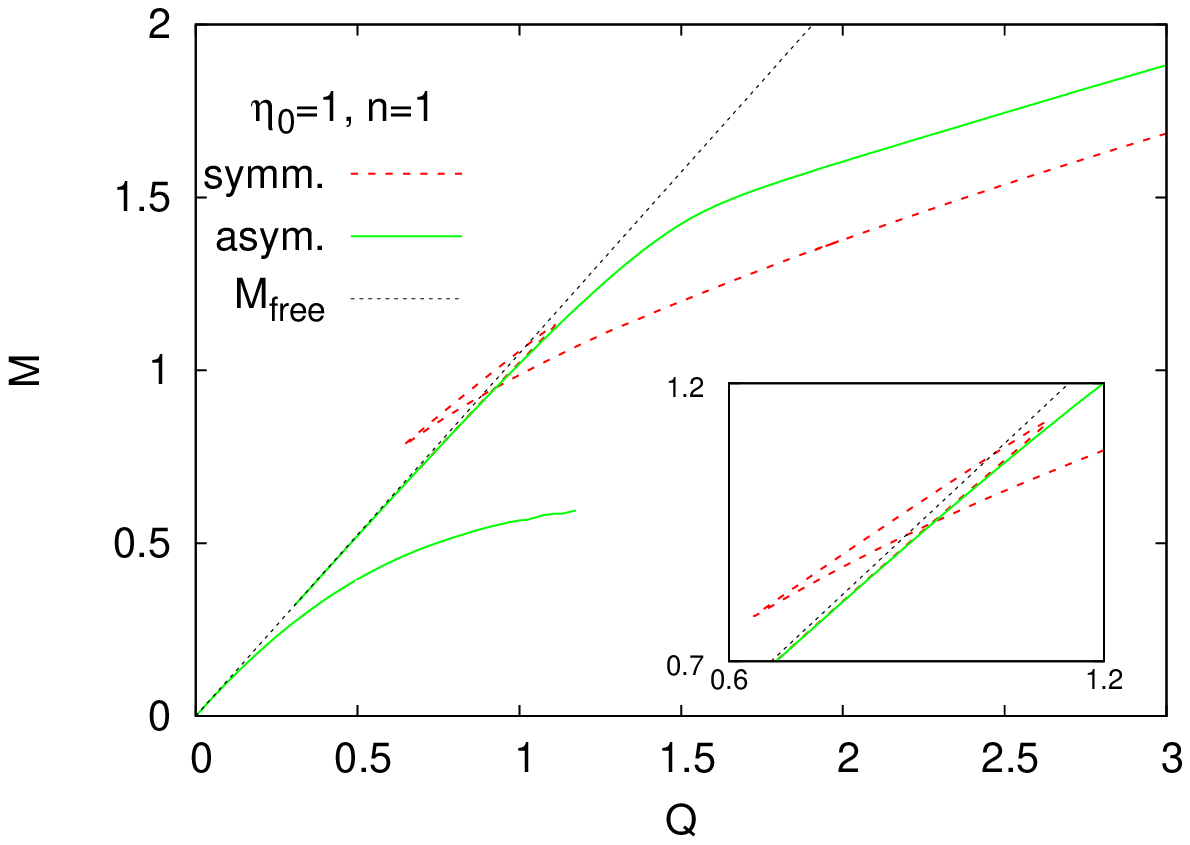}
\label{Fig3b}
}
}
\mbox{\hspace{0.2cm}
\subfigure[][]{\hspace{-1.0cm}
\includegraphics[height=.25\textheight, angle =0]{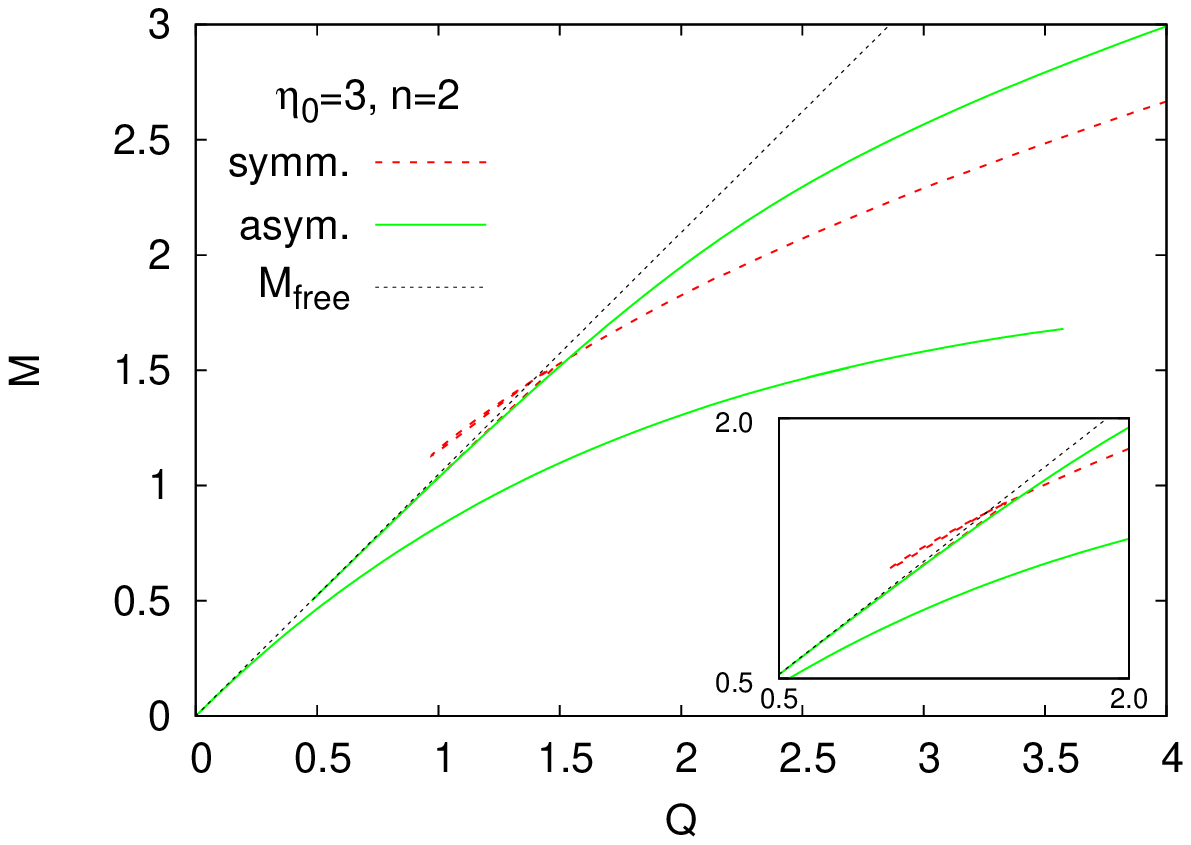}
\label{Fig3c}
}
\subfigure[][]{\hspace{-0.5cm}
\includegraphics[height=.25\textheight, angle =0]{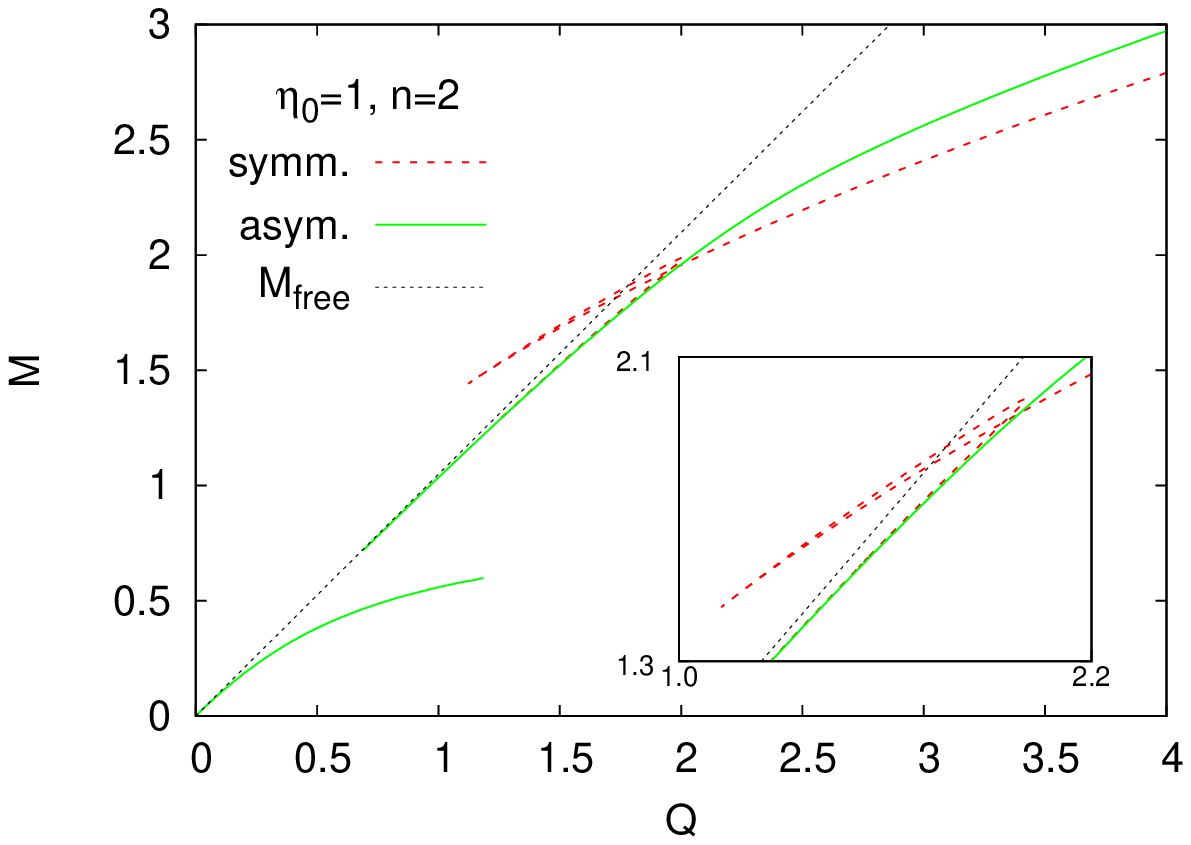}
\label{Fig3d}
}
}
\end{center}
\vspace{-0.5cm}
\caption{Global charges of rotating solutions:
(a) Mass $M$ versus particle number  $Q$ for rotational quantum number $n=1$ 
for the symmetric (dashed) and 
asymmetric (solid) solutions with throat parameter $\eta_0=3$;
(b) same as (a) for  $\eta_0=1$;
(c) same as (a) for $n=2$;
(d) same as (c) for  $\eta_0=1$.
Also shown is the mass of $Q$ free bosons (thin dotted).
\label{Fig3}
}
\end{figure}

Let us now consider the dependence of the mass on the particle number for these sets of solutions.
In Fig.\ref{Fig3} we exhibit the mass $M$ versus the particle number $Q$
for rotational quantum number $n=1$ and throat parameter $\eta_0=3$ (Fig.\ref{Fig3a}) 
respectively $\eta_0=1$ (Fig.\ref{Fig3b}),
as well as for $n=2$, $\eta_0=3$ (Fig.\ref{Fig3c}) respectively $\eta_0=1$ (Fig.\ref{Fig3d}).
Also shown is the mass of $Q$ free particles, $M_{\rm free} = m_{\rm bs} Q$,
for comparison.

The case of rotational quantum number $n=1$ and throat parameter $\eta_0=3$ (Fig.\ref{Fig3a}) 
is analogous to the one of the non-rotational case discussed above. Here 
the set of symmetric solutions
forms three branches. 
On each branch the mass changes monotonically with the particle number.
A first branch extends from the vacuum up to some (local) maximum value of the particle number
$Q$, where it merges with a second branch, which bends back to some (local)
minimum value of $Q$.
At this point a third branch arises and extends up to arbitrary large values of $Q$. 
On the interval where all three branches co-exist the mass is smallest on the first branch and 
largest on the third branch. 
The line representing the mass of free particles intersects the second and the third branch.
Thus $M < M_{\rm free}$ on the first branch and on the third branch for large enough $Q$,
whereas  $M > M_{\rm free}$ on most of the second branch and a small part of the third branch.

Let us now turn to the asymmetric solutions. Both $M_+(Q_+)$ and $M_-(Q_-)$ emerge from the
vacuum, with $M_+ \geq M_-$. They merge and bifurcate with the symmetric solutions on the third
branch of the symmetric solutions.
We note that  $M_+(Q_+)$ and $M_-(Q_-)$ are smaller than  $M_{\rm free}$.
Comparing with the mass of the symmetric solutions for some fixed particle number, we note that
$M_-$ is always the smallest, all the way up to the bifurcation point.
$M_+$ is smaller than the mass $M$ of the symmetric solutions on the first and 
on the second branch, although it is
remarkably close to the mass $M$ on the first branch.
Finally, $M_+$ intersects the mass $M$ on the third branch, and exceeds $M$ from the
intersection point until the bifurcation point is met.

For small particle number there exist three solutions, a symmetric solution and a pair of asymmetric
solutions. In the interval where the second branch of symmetric solutions is present,
there are five solutions, three symmetric ones and again a pair of asymmetric
solutions. For larger values of the particle number, up to the bifurcation point again three solutions exist.
Between the bifurcation point and the maximal value of $Q_+$ also three solutions
exist. Here, in addition to the symmetric one there are two asymmetric solutions
corresponding to two branches of $M_+(Q_+)$. For even larger values of $Q$ only symmetric solutions
are present.

Let us now turn to
the case $\eta_0=1$ and $n=1$ presented in Fig.\ref{Fig3b}.
Again the symmetric solutions form three branches. 
But now the mass is largest on the second branch. 
The first and the third branch then intersect at a certain value $Q_{\rm int}$. For smaller values of
$Q$ the mass is lowest on the first branch, for larger $Q$ it is lowest on the third branch.
As in the case of Fig.\ref{Fig3a} for $\eta_0=3$ 
the line showing the mass of $Q$ free particles intersects the second and the
third branch.
Since the asymmetric solutions do not bifurcate with the symmetric ones for $\eta_0=1$ and $n=1$,
$M_+$ and $M_-$ extend up to arbitrarily large values of $Q$. Both increase monotonically with
increasing particle number, and $M_+ \geq M_-$. 
$M_+$ is close to the mass $M$ of the symmetric solutions
on the first branch, and exceeds the mass $M$ on the third branch for $Q > Q_{\rm int}$.
Similarly to the previous case, for small and large values of $Q$ three solutions are present,
while five solutions exist for an intermediate interval.

Now we turn to winding number $n=2$. The case 
for throat parameter $\eta_0=3$ is shown in  Fig.\ref{Fig3c}.
Starting with the symmetric solutions we observe again three branches. In this case,
however, the second branch almost coincides with (part of) the third branch, and can hardly 
be distinguished in the figure. 
The asymmetric solutions also behave similarly to the case $\eta_0=1$ and $n=1$. Again $M_+$ is close 
to the mass of the symmetric solutions on the first branch, but exceeds the mass on the third branch for
sufficiently large values of the particle number.
The case $\eta_0=3$ and $n=2$ (Fig.\ref{Fig3d}) is qualitatively similar, as well.
 
We remark, that the phenomenon of spontaneous symmetry breaking remains absent
also in the case of rotation. This is not surprising, since the phenomenon is based
on a sufficiently strong self-interaction of the boson field, as we have discussed above.
If we were to allow for such self-interactions, we would expect this phenomenon to
make its reappearance in the presence of rotation.

\subsubsection{Ergoregions}

\begin{figure}[t!]
\begin{center}
\mbox{\hspace{0.2cm}
\subfigure[][]{\hspace{-1.0cm}
\includegraphics[height=.25\textheight, angle =0]{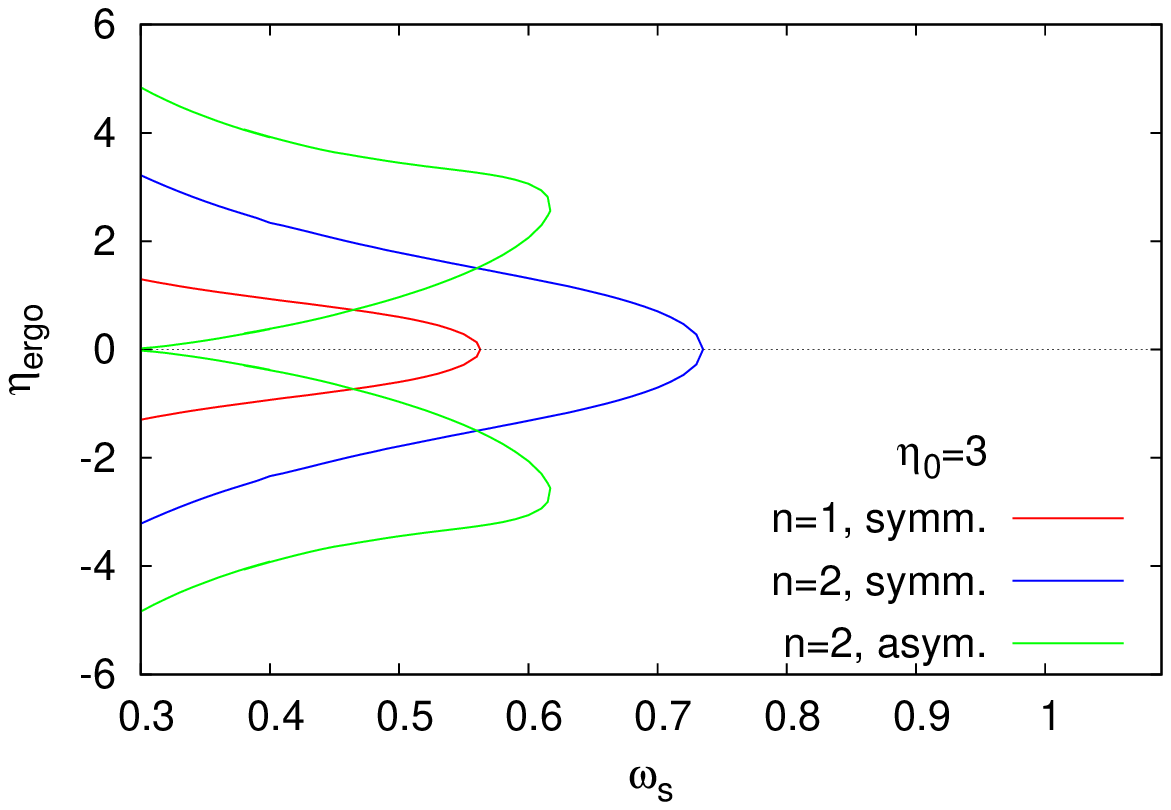}
\label{Fig4a}
}
\subfigure[][]{\hspace{-0.5cm}
\includegraphics[height=.25\textheight, angle =0]{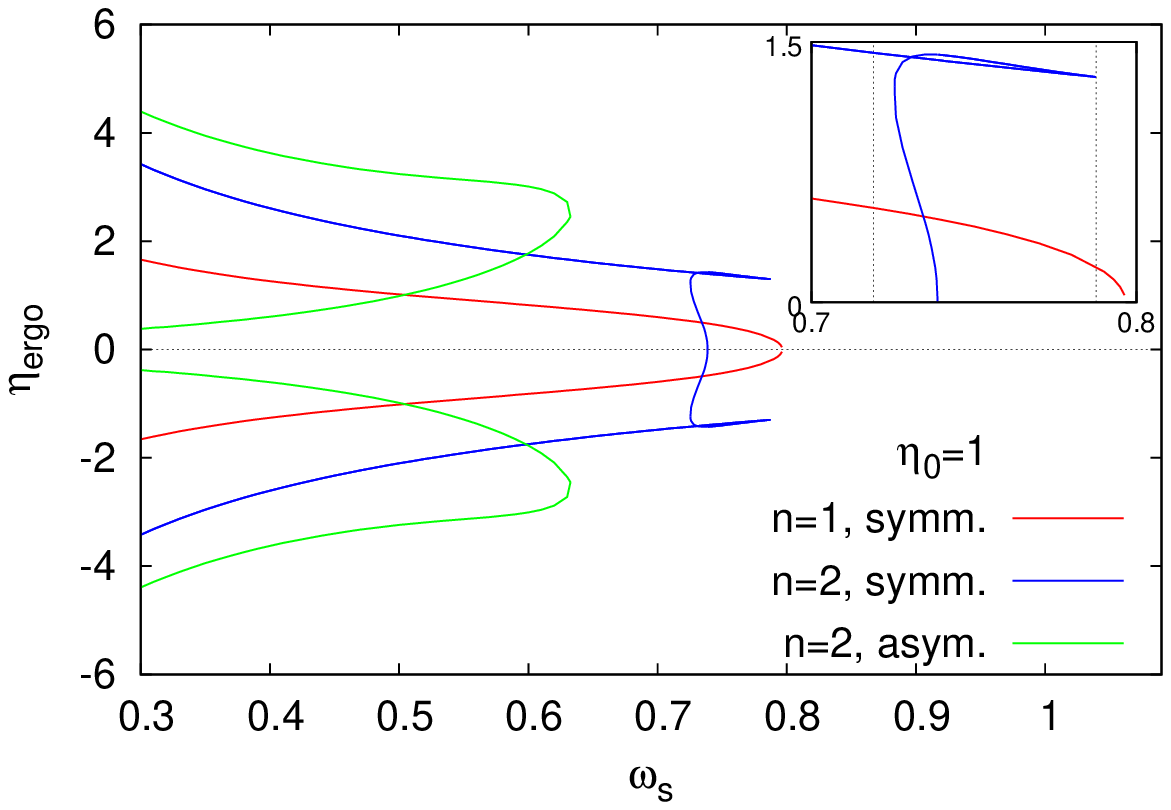}
\label{Fig4b}
}
}
\mbox{\hspace{0.2cm}
\subfigure[][]{\hspace{-1.0cm}
\includegraphics[height=.25\textheight, angle =0]{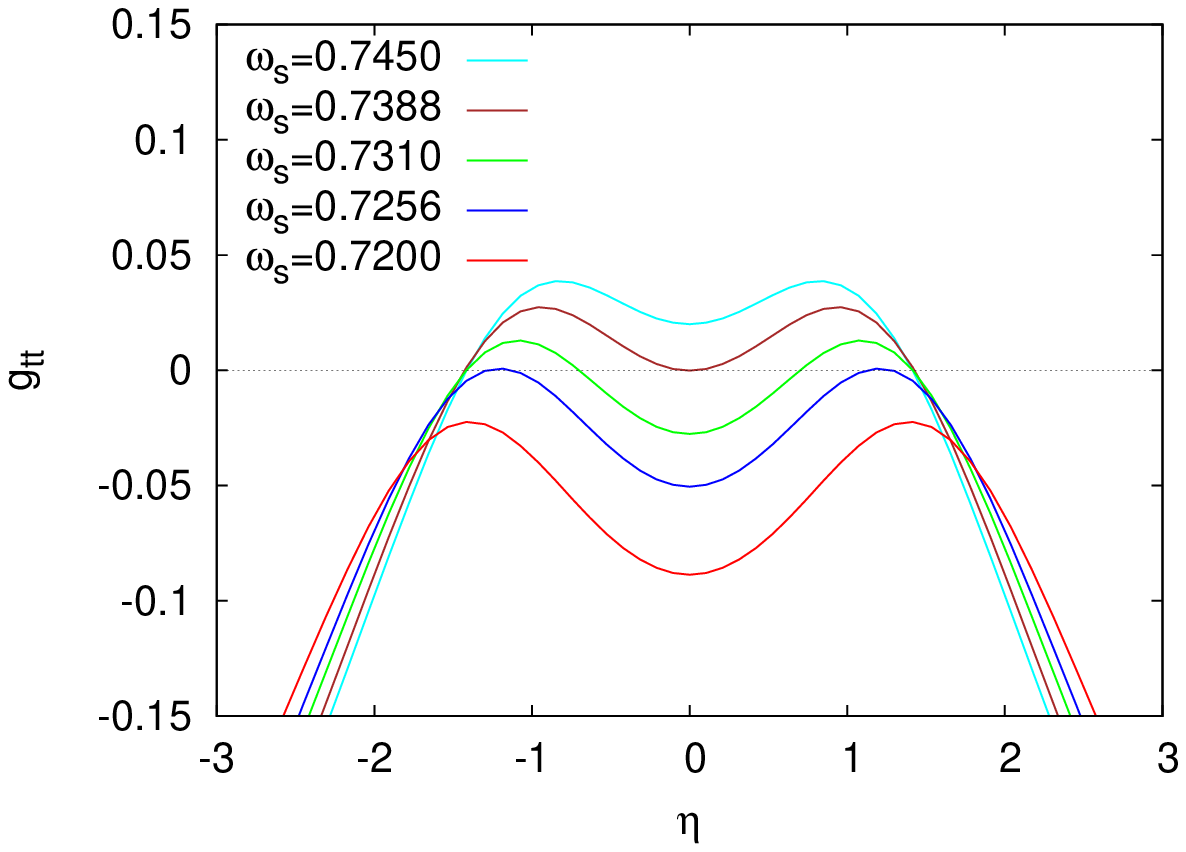}
\label{Fig4c}
}
}
\end{center}
\vspace{-0.5cm}
\caption{Ergoregion(s) of rotating solutions:
(a) the coordinate $\eta_{\rm ergo}$ of the ergosurface(s) in the equatorial
plane versus the boson frequency $\omega_s$ for 
rotational quantum number $n=1$ and $n=2$ 
for the symmetric and asymmetric solutions 
for throat parameter $\eta_0=3$;
(b) same as (a) for $\eta_0=1$;
(c) the metric component $g_{tt}$ in the equatorial plane versus the radial coordinate $\eta$
for throat parameter $\eta_0=1$, rotational quantum number $n=2$ and several values of 
the boson frequency $\omega_s$.
\label{Fig4}
}
\end{figure}

We now turn to the ergoregions of the rotating solutions. 
In Fig.\ref{Fig4} we exhibit the coordinate $\eta_{\rm ergo}$ of the ergosurface(s), 
in the equatorial plane versus the boson frequency $\omega_s$
for the same sets of solutions as discussed above.
We immediately observe that ergoregions exist only if 
the boson frequency $\omega_s$
is smaller than some maximal value,
which depends on the throat parameter $\eta_0$, 
on the rotational quantum number $n$, and on whether 
the solutions are symmetric or asymmetric. 

Let us now consider the ergoregions in more detail,
starting with the case of throat parameter $\eta_0=3$,
shown in Fig.\ref{Fig4a}. We observe that
the ergoregions decrease in size with increasing 
boson frequency $\omega_s$, until they
degenerate into a circle, when the 
respective maximal value of $\omega_s$ is reached. 
For the symmetric solutions, this circle then always resides
at the throat, i.e., $\eta_{\rm ergo}=0$.
For the asymmetric solutions this circle is considerably shifted.
But of course there is such a circle for either of the two
asymmetric solutions, located symmetrically with respect to each other.
We note that the asymmetric solutions for $n=1$, which only exist if
$\omega_s$ exceeds a critical value, do not possess an ergoregion.

We observe the analogous behaviour for the case of
throat parameter $\eta_0=1$, shown in Fig.\ref{Fig4b}, 
except for the symmetric solutions with rotational quantum number $n=2$. 
This latter case is considerably more involved, because the region
where the backbending of the solutions occurs along with its
multiple branch structure, resides in this case at sufficiently small values
of the boson frequency to fall into the range of frequencies,
where ergoregions are present. 

Let us therefore look into more detail for this case.
Again, at first the ergoregion decreases in size with
increasing boson frequency $\omega_s$,
until the third branch ends, when the second branch begins.
On the second branch $\eta_{\rm ergo}$ increases with decreasing $\omega_s$,
reaches a maximum, and then decreases rapidly. In fact,
the ergoregion disappears before the second branch reaches the third branch.

As the boson frequency decreases,
in the small interval $0.7256 \leq \omega_s \leq 0.7388$ the 
surface of the ergoregion consists of two disconnected parts, 
each one located on one side of the throat,
symmetrically with respect to each other.
This is illustrated in Fig.\ref{Fig4c},
where the metric component $g_{tt}$ in the equatorial plane 
is exhibited versus the radial coordinate $\eta$.
At $\omega_s=0.7388$, an inner boundary ring appears at the throat,
whereas at $\omega_s = 0.7256$ the ergoregion degenerates into two rings,
located symmetrically with respect to each other. Beyond this latter frequency
there are no ergoregions found any more along the second branch,
and neither along the first branch.

\subsubsection{Geometrical  properties}

\begin{figure}[t!]
\begin{center}
\mbox{\hspace{0.2cm}
\subfigure[][]{\hspace{-1.0cm}
\includegraphics[height=.25\textheight, angle =0]{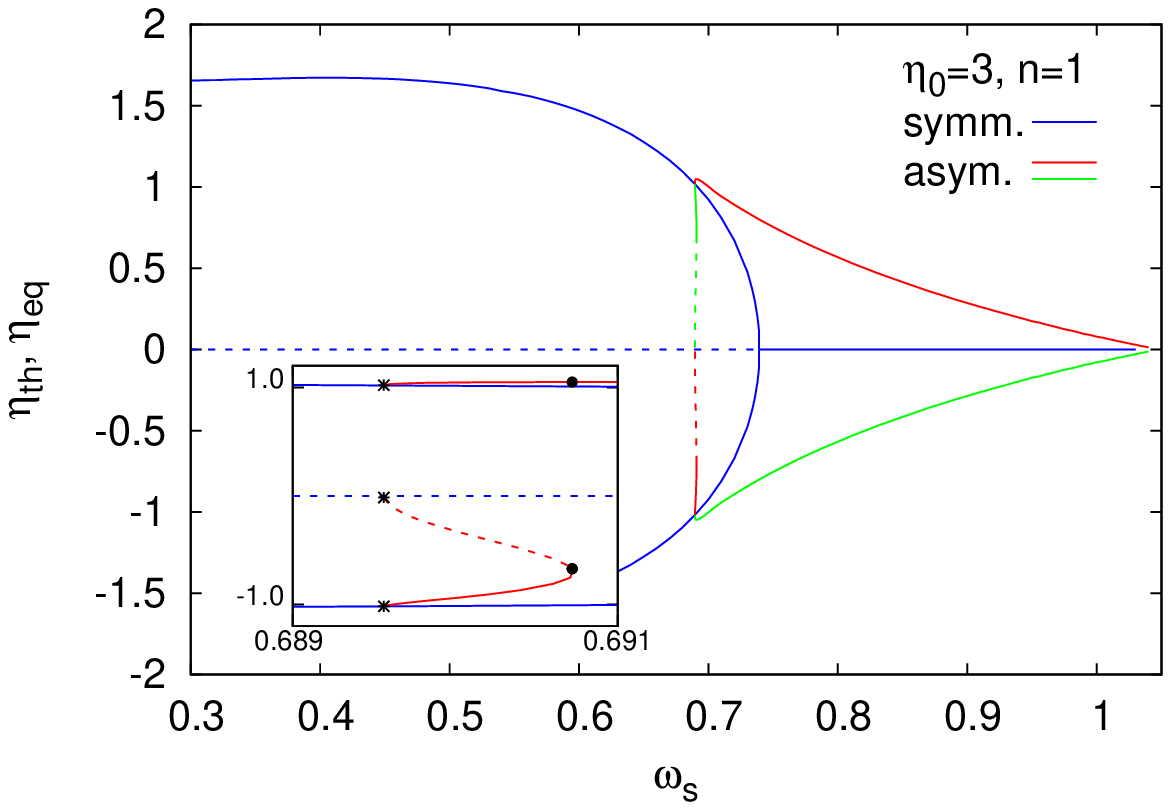}
\label{Fig5a}
}
\subfigure[][]{\hspace{-0.5cm}
\includegraphics[height=.25\textheight, angle =0]{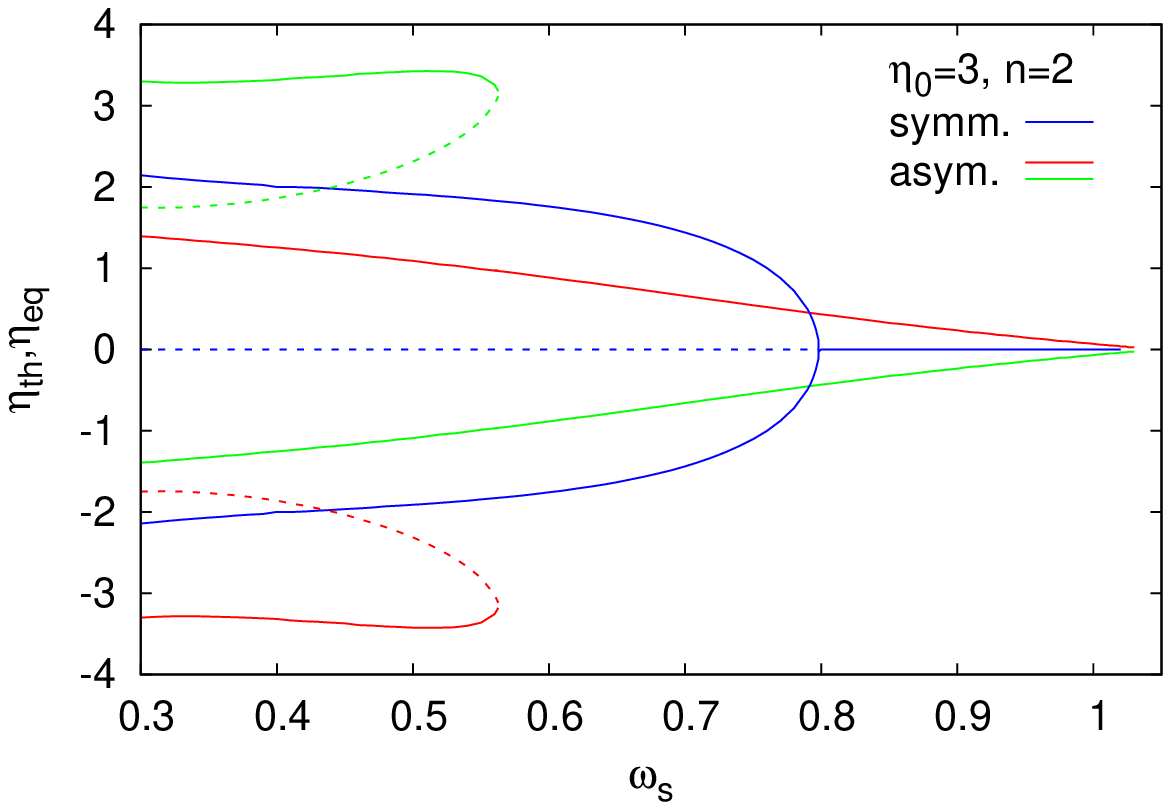}
\label{Fig5b}
}
}
\mbox{\hspace{0.2cm}
\subfigure[][]{\hspace{-1.0cm}
\includegraphics[height=.25\textheight, angle =0]{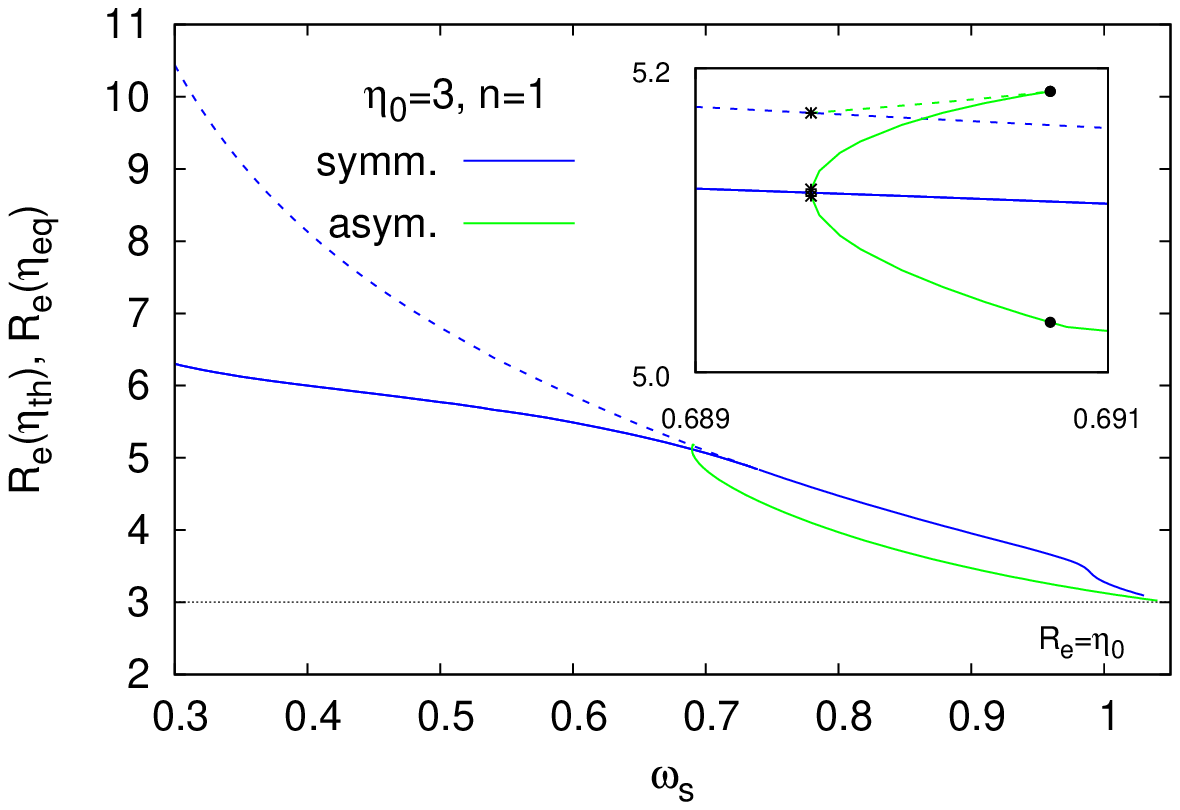}
\label{Fig5c}
}
\subfigure[][]{\hspace{-0.5cm}
\includegraphics[height=.25\textheight, angle =0]{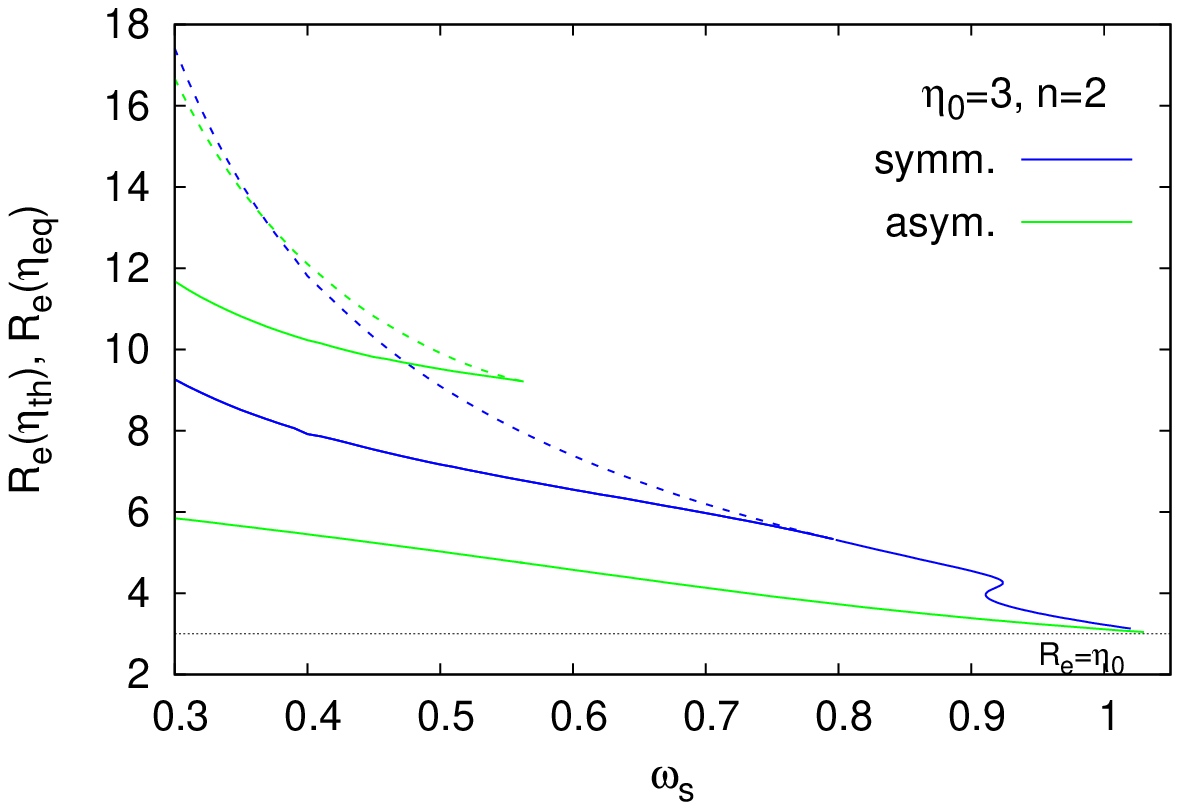}
\label{Fig5d}
}
}
\mbox{\hspace{0.2cm}
\subfigure[][]{\hspace{-1.0cm}
\includegraphics[height=.25\textheight, angle =0]{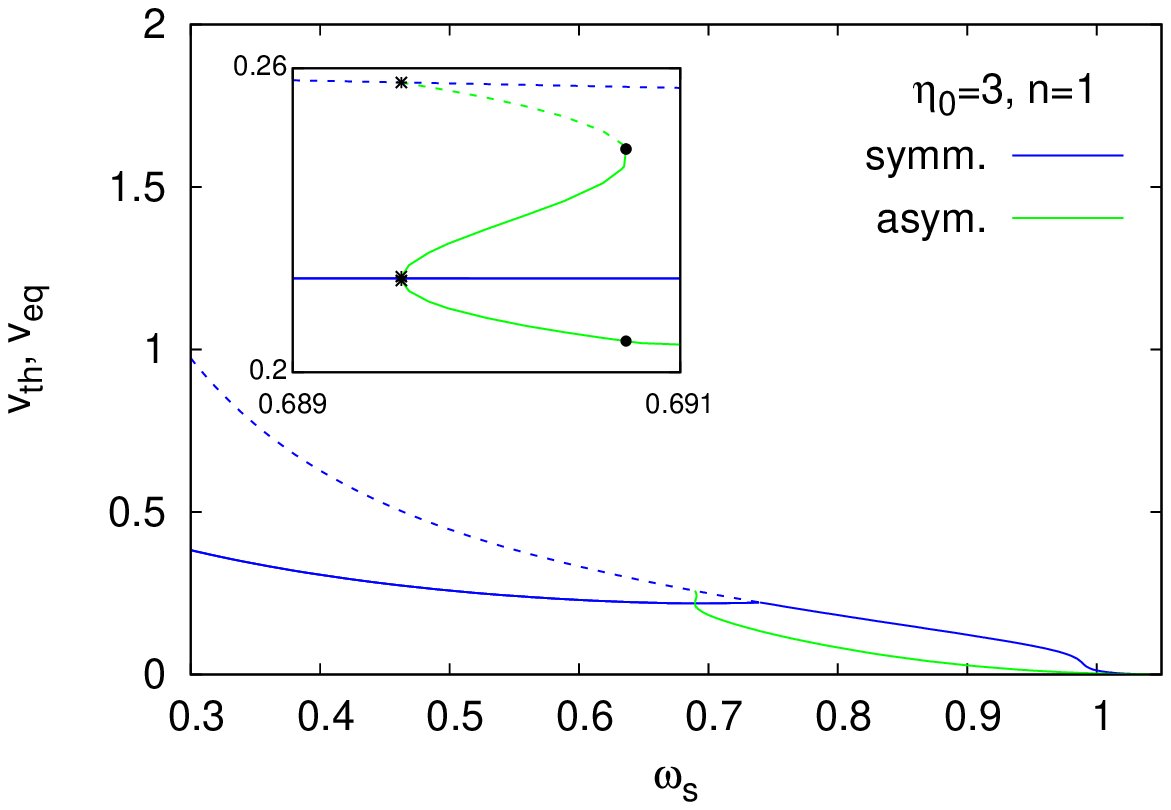}
\label{Fig5e}
}
\subfigure[][]{\hspace{-0.5cm}
\includegraphics[height=.25\textheight, angle =0]{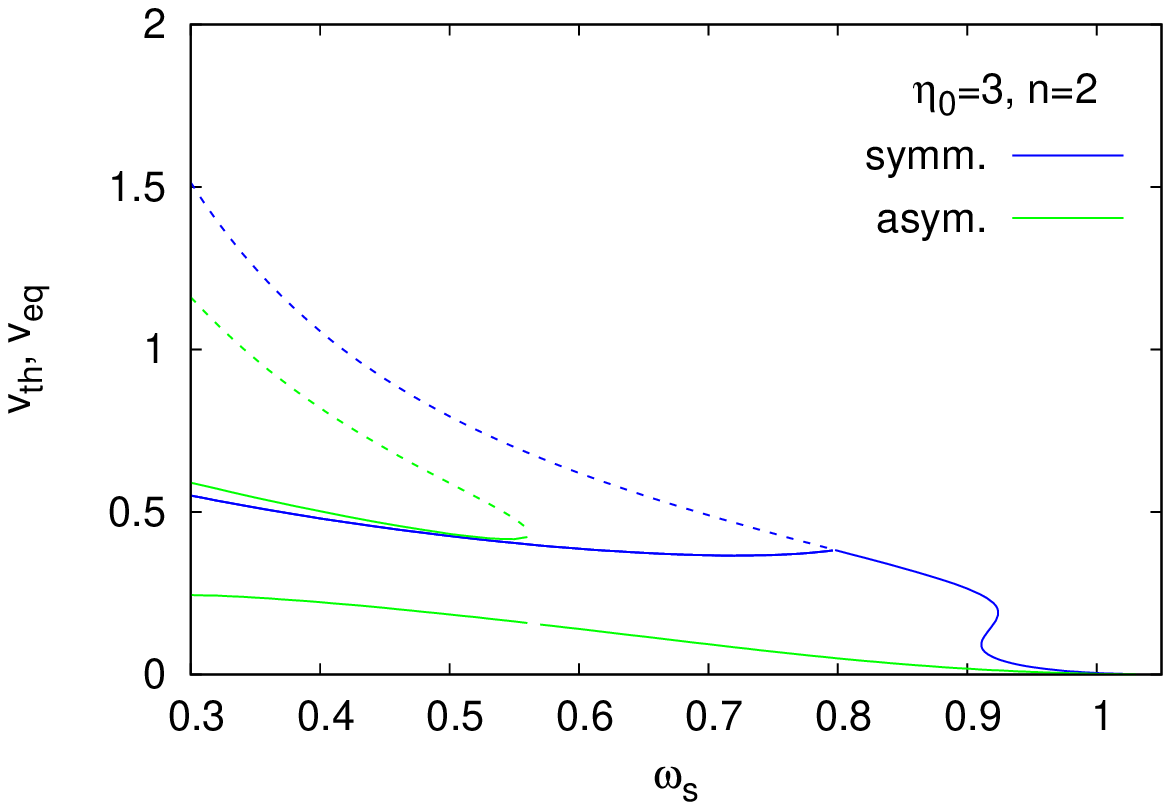}
\label{Fig5f}
}
}
\end{center}
\vspace{-0.5cm}
\caption{Geometric properties of rotating solutions:
(a) the coordinate $\eta_{\rm th}$ of the throat(s) (solid) 
and the coordinate $\eta_{\rm eq}$ of the equator (dashed) 
in the equatorial plane 
versus the boson frequency $\omega_s$ for 
rotational quantum number $n=1$ and throat parameter $\eta_0=3$;
(b) same as (a) for $n=2$ and  $\eta_0=3$;
(c) the circumferential radius $R_{\rm e}(\eta_{\rm th})$
of the throat(s) (solid) 
and the circumferential radius  $R_{\rm e}(\eta_{\rm eq})$ 
of the equator (dashed)
in the equatorial plane 
versus the boson frequency $\omega_s$ 
for rotational quantum number $n=1$ and throat parameter $\eta_0=3$;
 The dotted black line indicates the limit $\omega_s \to m_{\rm bs}$.
(d) same as (c) for $n=2$ and  $\eta_0=3$;
(e) the rotational velocity $v_{\rm th}$ of the throat(s) (solid)  
and the rotational velocity $v_{\rm eq}$ of the equator (dashed)
in the equatorial plane 
versus the boson frequency $\omega_s$ 
for rotational quantum number $n=1$ and throat parameter $\eta_0=3$;
(f) same as (e) for $n=2$ and  $\eta_0=3$.
\label{Fig5}
}
\end{figure}

\begin{figure}[t!]
\begin{center}
\mbox{\hspace{0.2cm}
\subfigure[][]{\hspace{-1.0cm}
\includegraphics[height=.25\textheight, angle =0]{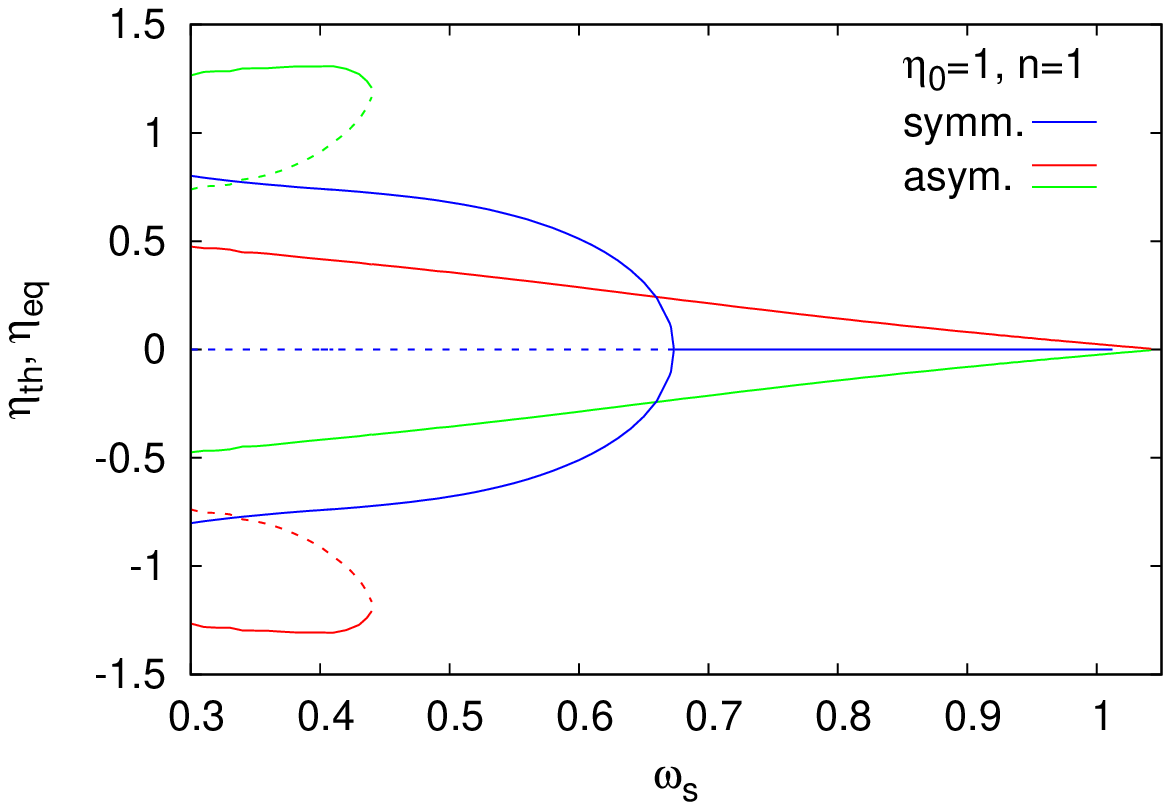}
\label{Fig6a}
}
\subfigure[][]{\hspace{-0.5cm}
\includegraphics[height=.25\textheight, angle =0]{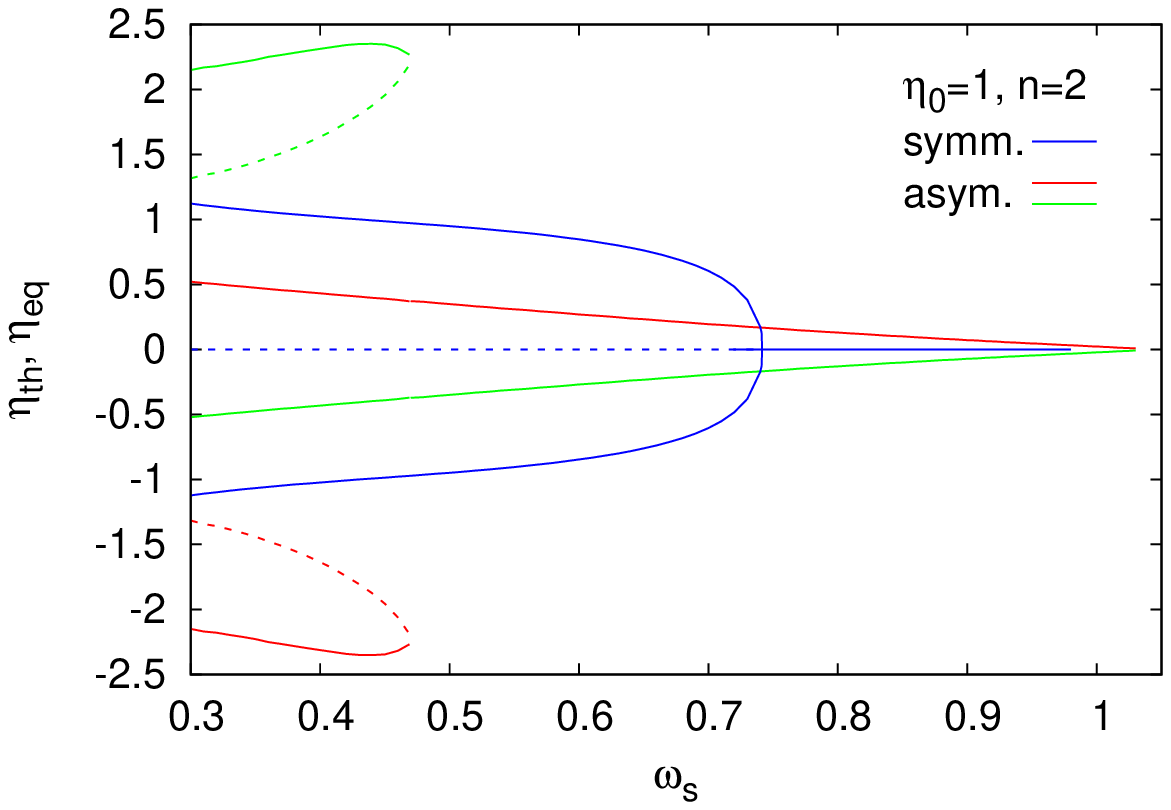}
\label{Fig6b}
}
}
\mbox{\hspace{0.2cm}
\subfigure[][]{\hspace{-1.0cm}
\includegraphics[height=.25\textheight, angle =0]{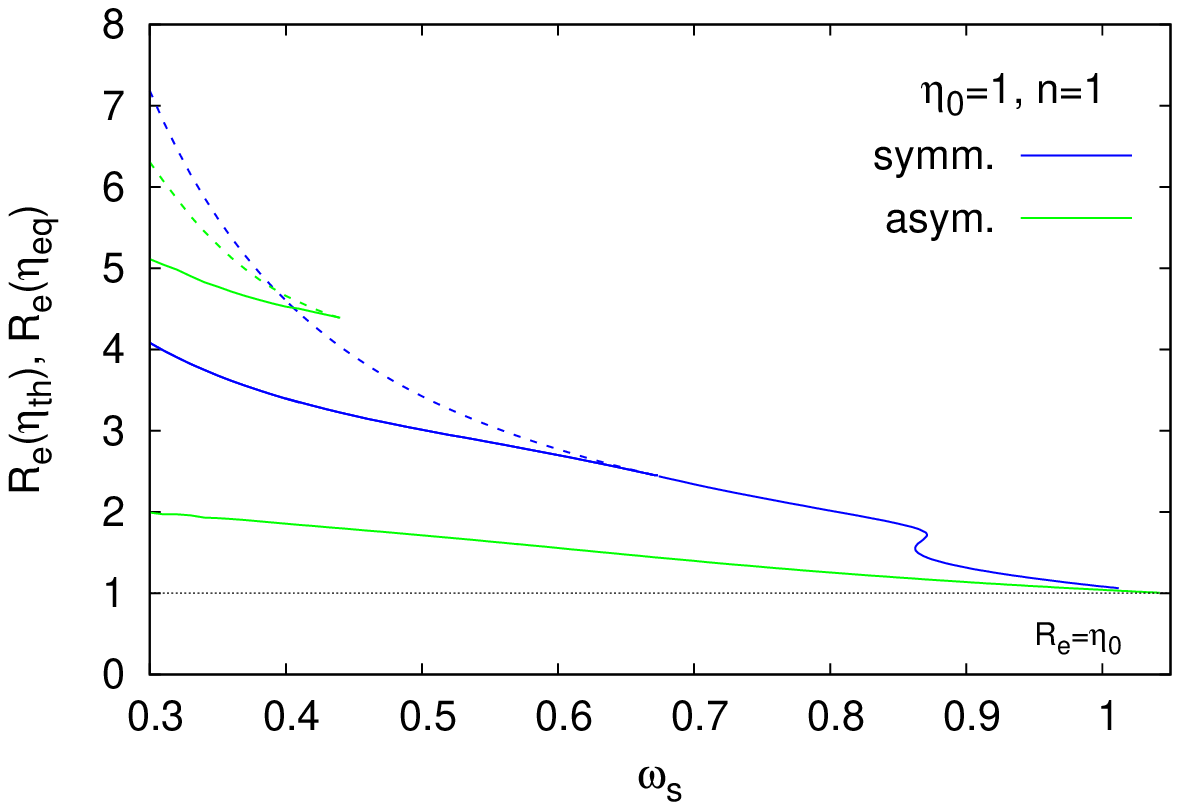}
\label{Fig6c}
}
\subfigure[][]{\hspace{-0.5cm}
\includegraphics[height=.25\textheight, angle =0]{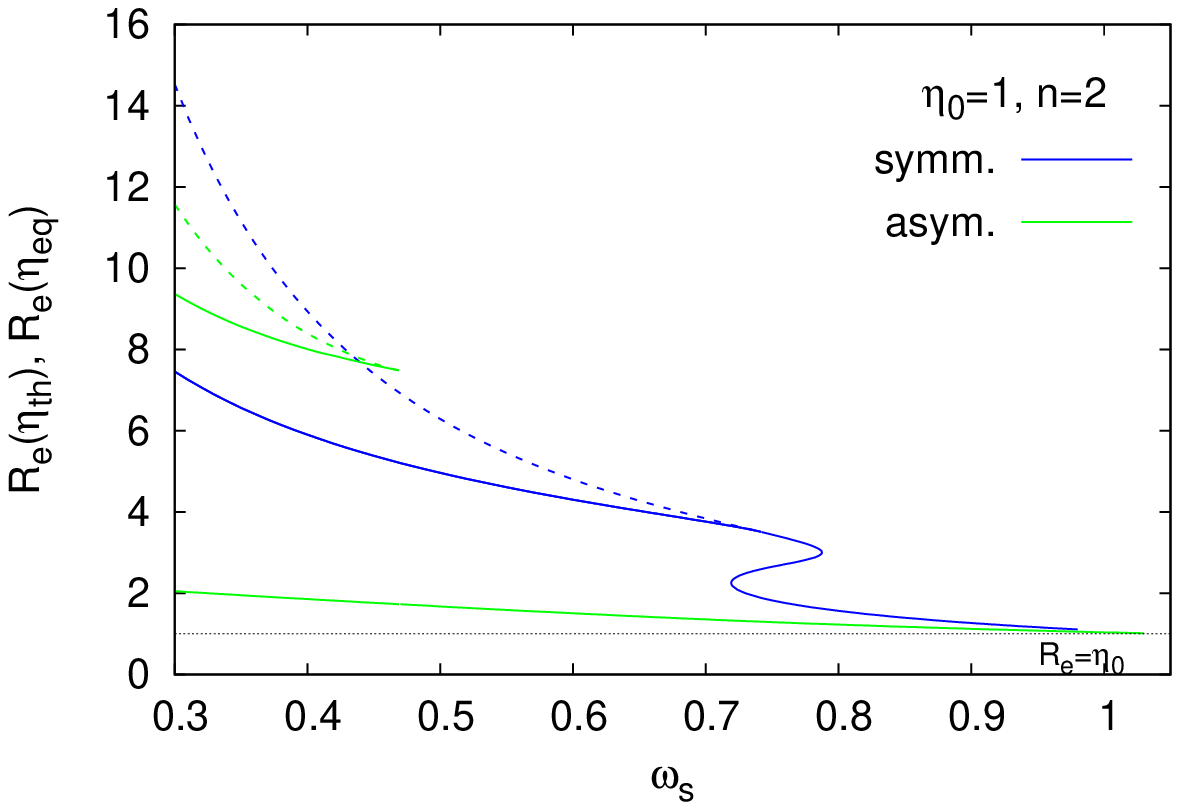}
\label{Fig6d}
}
}
\mbox{\hspace{0.2cm}
\subfigure[][]{\hspace{-1.0cm}
\includegraphics[height=.25\textheight, angle =0]{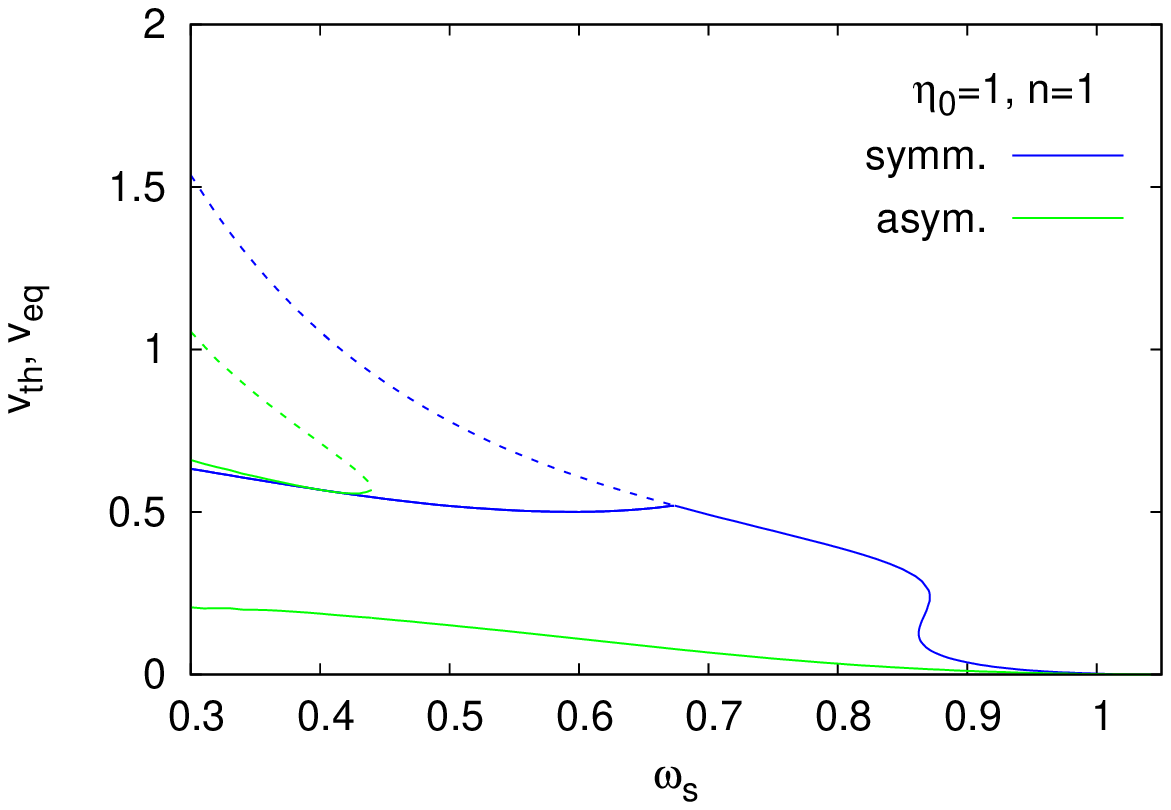}
\label{Fig6e}
}
\subfigure[][]{\hspace{-0.5cm}
\includegraphics[height=.25\textheight, angle =0]{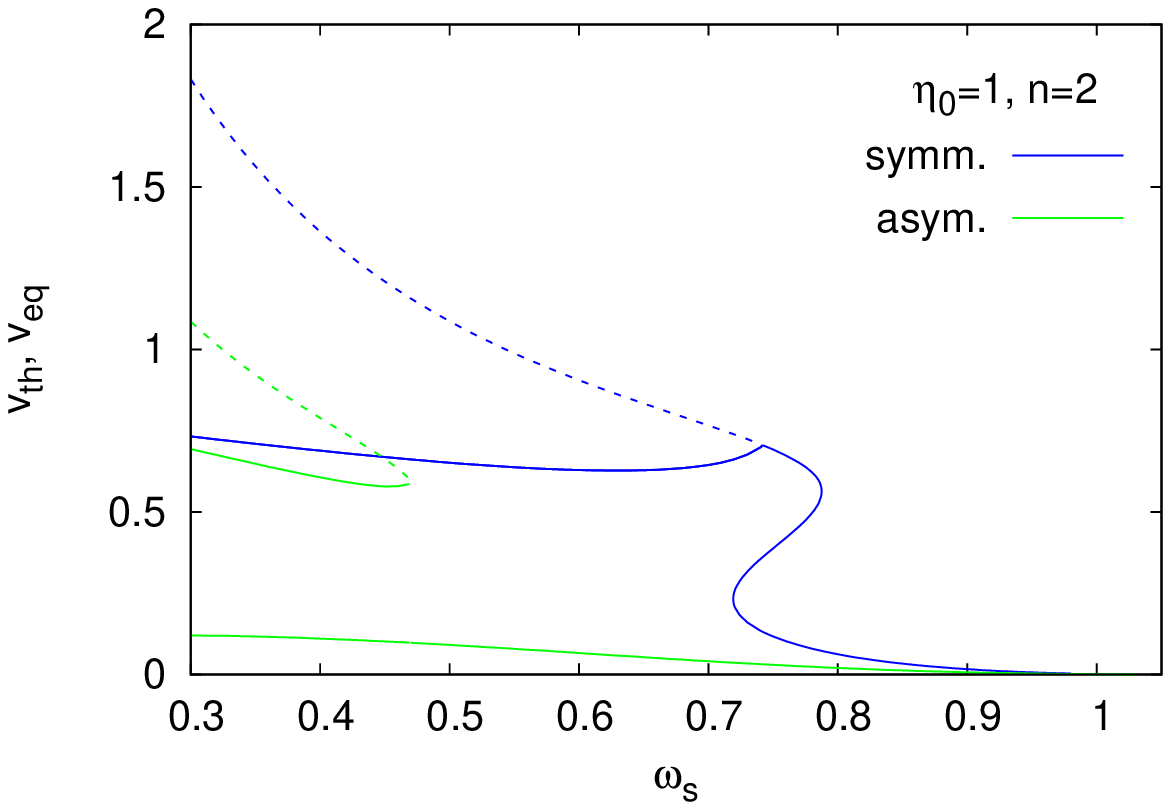}
\label{Fig6f}
}
}
\end{center}
\vspace{-0.5cm}
\caption{Geometric properties of rotating solutions:
(a) the coordinate $\eta_{\rm th}$ of the throat(s) (solid) 
and the coordinate $\eta_{\rm eq}$ of the equator (dashed) 
in the equatorial plane 
versus the boson frequency $\omega_s$ for 
rotational quantum number $n=1$ and throat parameter $\eta_0=1$;
(b) same as (a) for $n=2$ and  $\eta_0=1$;
(c) the circumferential radius $R_{\rm e}(\eta_{\rm th})$
of the throat(s) (solid) 
and the circumferential radius  $R_{\rm e}(\eta_{\rm eq})$ 
of the equator (dashed)
in the equatorial plane 
versus the boson frequency $\omega_s$ 
for rotational quantum number $n=1$ and throat parameter $\eta_0=1$;
The dotted black line indicates the limit $\omega_s \to m_{\rm bs}$.
(d) same as (c) for $n=2$ and  $\eta_0=1$;
(e) the rotational velocity $v_{\rm th}$ of the throat(s) (solid)  
and the rotational velocity $v_{\rm eq}$ of the equator (dashed)
in the equatorial plane 
versus the boson frequency $\omega_s$ 
for rotational quantum number $n=1$ and throat parameter $\eta_0=1$;
(f) same as (e) for $n=2$ and  $\eta_0=1$.
\label{Fig6}
}
\end{figure}

Let us now address the geometrical properties of the rotating solutions,
focusing on the equatorial plane.
To this end we present in Fig.\ref{Fig5} and Fig.\ref{Fig6} 
the coordinate $\eta_{\rm th}$ of the throat(s), 
the coordinate $\eta_{\rm eq}$ of the equator when present,
the circumferential radius $R_{\rm e}(\eta_{\rm th})$
of the throat(s),
the circumferential radius  $R_{\rm e}(\eta_{\rm eq})$ 
of the equator when present,
the rotational velocity $v_{\rm th}$ of the throat(s),
and the rotational velocity $v_{\rm eq}$ of the equator when present.
All quantities are shown
versus the boson frequency $\omega_s$.
Fig.\ref{Fig5} exhibits the solutions for throat parameter $\eta_0=3$,
and Fig.\ref{Fig6} those for $\eta_0=1$.

We start the discussion again with the case of rotational quantum number $n=1$
and throat parameter $\eta_0=3$,  
which features a bifurcation point of the
symmetric and asymmetric solutions at a critical boson frequency $\omega_s^{\rm cr}$.
We observe that for large values of $\omega_s$ the wormholes possess only a single throat,
until at some particular value of $\omega_s$ an equator emerges,
as shown in Fig.\ref{Fig5a}.
For the symmetric wormholes the throat residing at $\eta=0$
then degenerates into an inflection point.
For smaller boson frequency $\omega_s$ the circumferential radius $R_{\rm e}$
possesses a maximum at $\eta=0$, corresponding to an equator, while two minima are located symmetrically
with respect to  $\eta=0$,  corresponding to two throats, as demonstrated in Fig.\ref{Fig5a}.
Note, that the circumferential radius $R_{\rm e}(\eta_{\rm th})$ of the two throats is of course the same
because of symmetry. Therefore only a single solid blue curve is present 
in Fig.\ref{Fig5c}, also after the equator with circumferential radius $R_{\rm e}(\eta_{\rm eq})$
(dashed blue) has emerged.

Since the asymmetric wormhole solutions always appear in pairs, with one solution related
to the other one via reflection symmetry, $\eta \to - \eta$,
the asymmetrically located throats of these two solutions are also related via
reflection symmetry. To discern them in the figure, the throats 
of the two solutions are indicated in different colours
(red and green).
Focusing now on one of the asymmetric solutions only, there is a single throat
for large values of the boson frequency $\omega_s$. However, at a particular
value of the boson frequency, an inflection point emerges on the opposite side of the spacetime,
turning into an equator and a second throat as the boson frequency decreases.
Note, that these asymmetric wormholes with
a double throat exist only in a small region of $\omega_s$ close to the bifurcation point $\omega_s^{\rm cr}$
with the symmetric solutions.
The insets in Fig.\ref{Fig5a} and Fig.\ref{Fig5c} highlight this region with double throats.
Here the inflection point is indicated by dots, while the bifurcation with the symmetric solutions
is indicated by asterisks. 

In Fig.\ref{Fig5e} we demonstrate, that the rotation of the matter is indeed inducing a rotation
of the throat(s) and the equator, when present. We therefore exhibit the rotational velocity
$v_{\rm th}$ of the throat(s) in the equatorial plane together with the rotational velocity
$v_{\rm eq}$ of the equator. We note that for the symmetric solutions
the rotational velocity of the center $\eta=0$,
representing first a throat and then an equator,
increases monotonically with decreasing boson frequency.
When the double throat emerges, the rotational velocity of these two symmetrically located throats
remains considerably smaller that the rotational velocity of the equator.
Concerning the asymmetric solutions we note, that the rotational velocity of the single throat
is smallest. The inflection point arises with a higher rotational velocity, with the velocity of the
equator increasing and the velocity of the throat decreasing towards the bifurcation point.

In Fig.\ref{Fig5b} and \ref{Fig5d} we present the geometrical properties for 
throat parameter $\eta_0=3$ and rotational quantum number $n=2$.
The basic difference is the absence of the bifurcation of the symmetric and 
asymmetric solutions (at least for $\omega_s > 0.3$).
Thus once the asymmetric solutions develop an equator
and a second throat, these features are retained as the boson frequency decreases.
Also the geometrical properties of the solutions with the smaller throat parameter
$\eta_0=1$, shown in Fig.\ref{Fig6} 
are essentially the same as for this case.

\begin{figure}[t!]
\begin{center}
\mbox{\hspace{0.2cm}
\subfigure[][]{\hspace{-1.0cm}
\includegraphics[height=.25\textheight, angle =0]{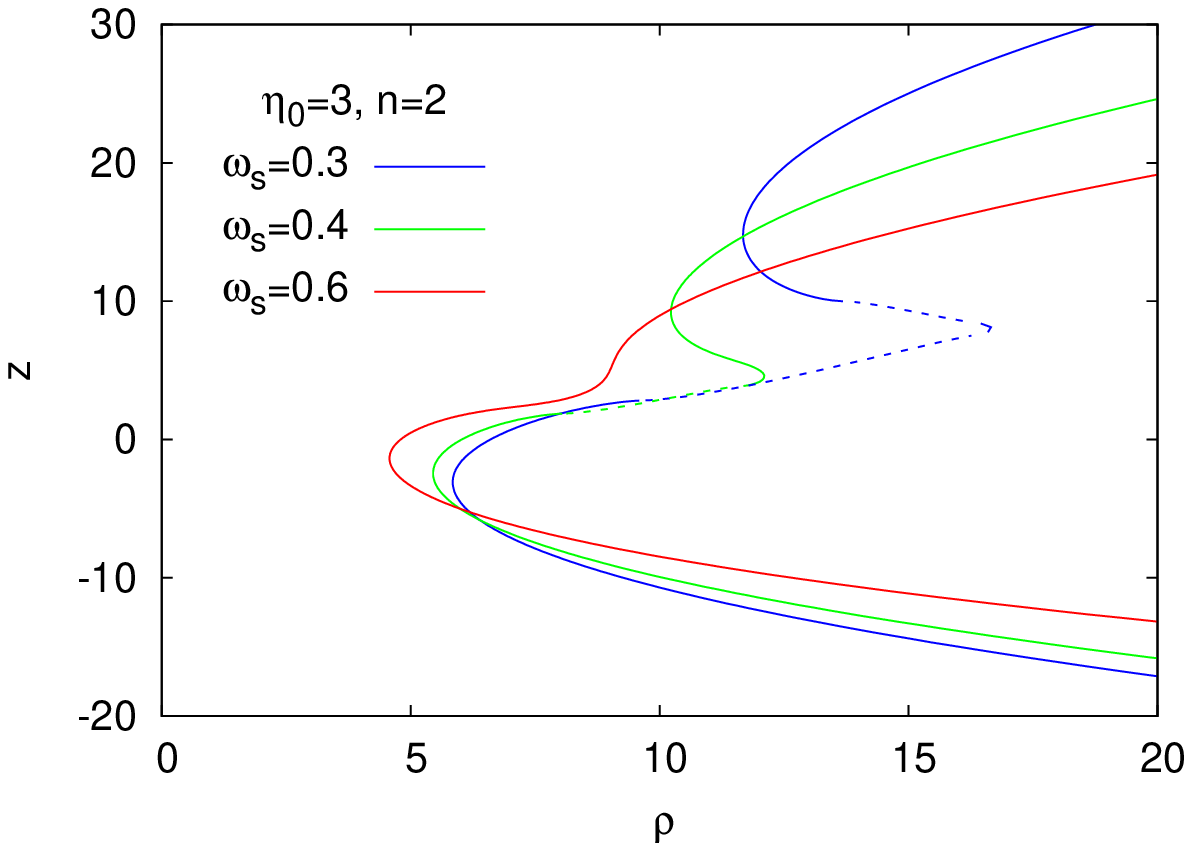}
\label{Fig7a}
}
\subfigure[][]{\hspace{-0.5cm}
\includegraphics[height=.25\textheight, angle =0]{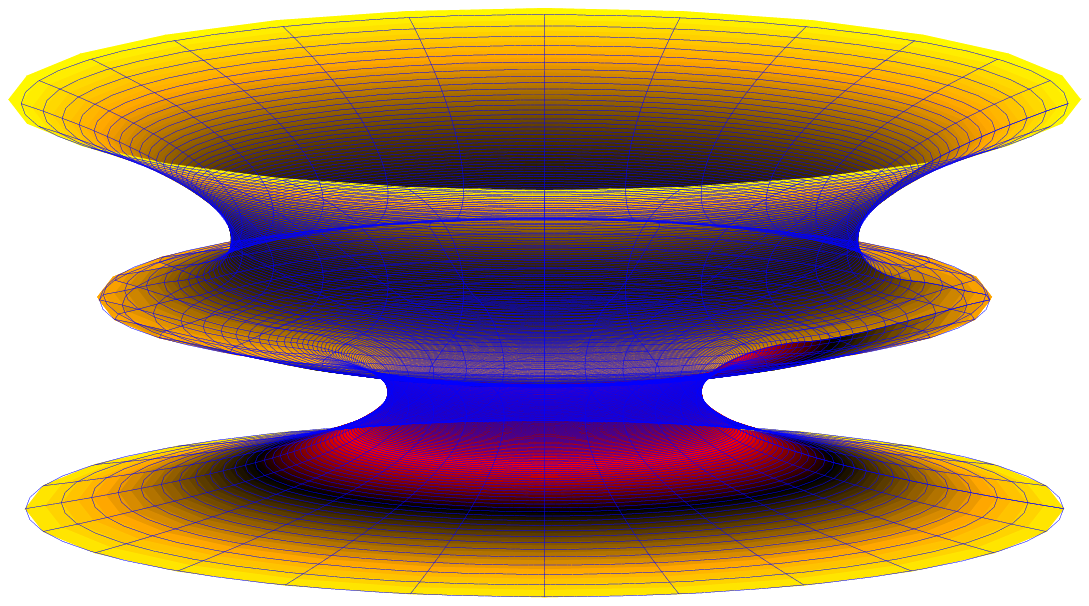}
\label{Fig7b}
}
}
\mbox{\hspace{0.2cm}
\subfigure[][]{\hspace{-1.0cm}
\includegraphics[height=.25\textheight, angle =0]{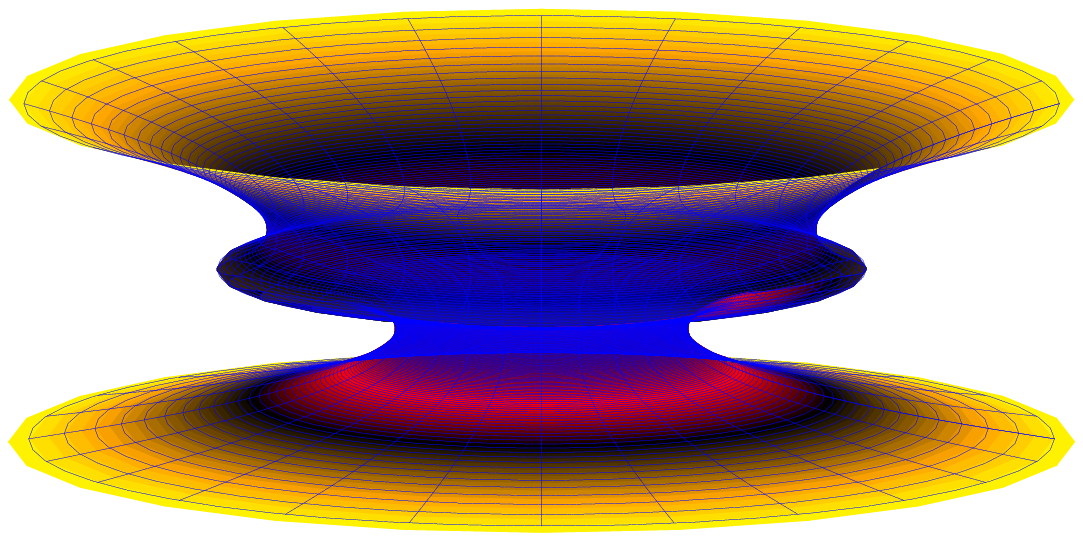}
\label{Fig7c}
}
\subfigure[][]{\hspace{-0.5cm}
\includegraphics[height=.25\textheight, angle =0]{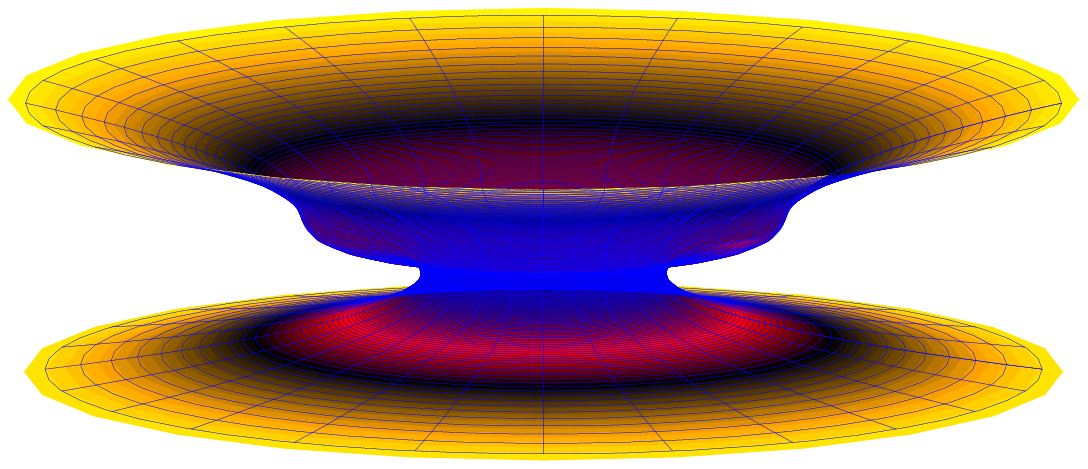}
\label{Fig7d}
}
}
\end{center}
\vspace{-0.5cm}
\caption{Geometric properties of rotating solutions:
embeddings of the equatorial plane of symmetric  and asymmetric solutions
with throat  parameter $\eta_0=3$, rotational quantum number $n=2$ and
several values of the boson frequency $\omega_s$:
(a) profile $z(\rho)$ for three embeddings with $\omega_s=0.3$, $0.4$ and $0.6$,
where a pseudo-euclidean embedding is indicated by dashed lines;
(b) 3D plot for $\omega_s=0.3$;
(c) 3D plot for $\omega_s=0.4$;
(d) 3D plot for $\omega_s=0.6$.
\label{Fig7}
}
\end{figure}

In order to get further insight into the geometry of the wormholes we show in Fig.\ref{Fig7} the 
isometric embedding of the equatorial plane for asymmetric 
wormhole solutions with throat  parameter $\eta_0=3$, rotational quantum number $n=2$ and
several values of the boson frequency $\omega_s$ as examples.
Fig.\ref{Fig7a} exhibits the profile $z(\rho)$ for all three examples, 
where a pseudo-euclidean embedding is indicated by dashed lines.
The remaining figures represent 3-dimensional embeddings,
where Fig.\ref{Fig7b} and Fig.\ref{Fig7c} correspond to asymmetric double throat solutions
($\omega_s=0.3$, resp. $\omega_s=0.4$), while Fig.\ref{Fig7d} represents an
asymmetric single throat solution with $\omega_s=0.6$.

\subsubsection{Lightrings}

\begin{figure}[t!]
\begin{center}
\mbox{\hspace{0.2cm}
\subfigure[][]{\hspace{-1.0cm}
\includegraphics[height=.25\textheight, angle =0]{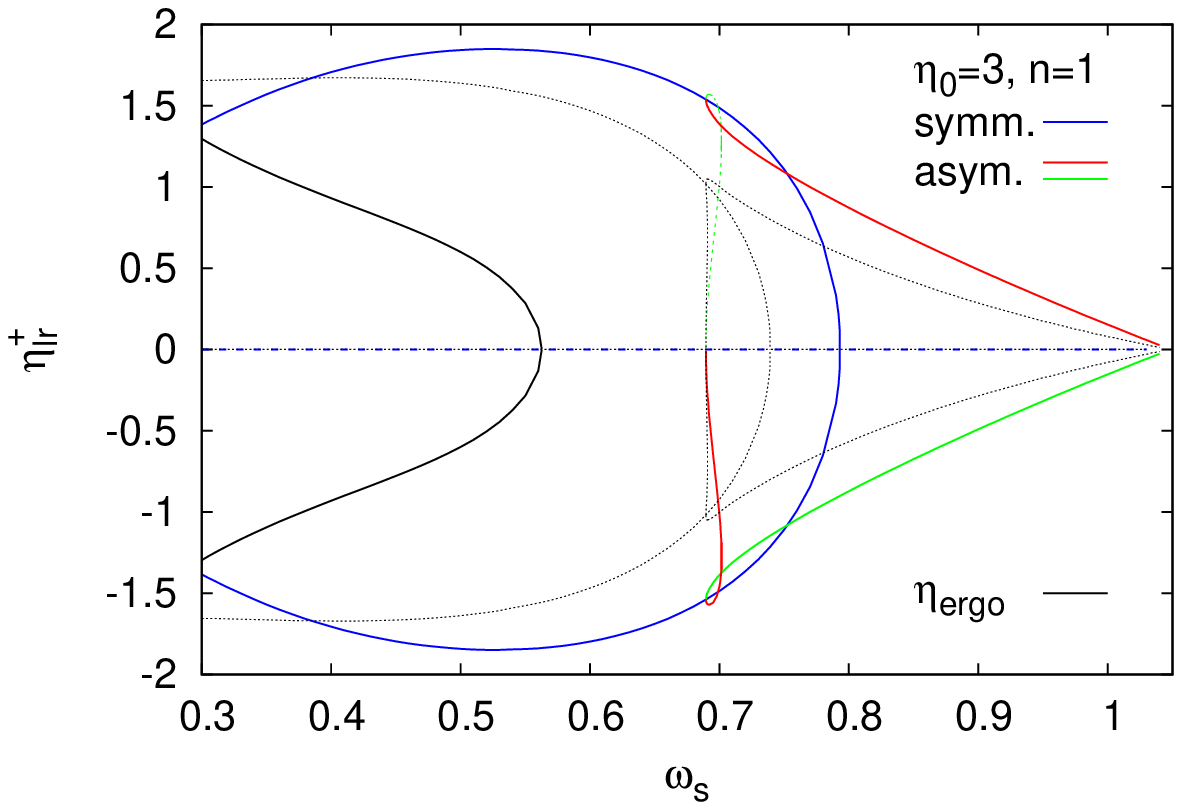}
\label{Fig8a}
}
\subfigure[][]{\hspace{-0.5cm}
\includegraphics[height=.25\textheight, angle =0]{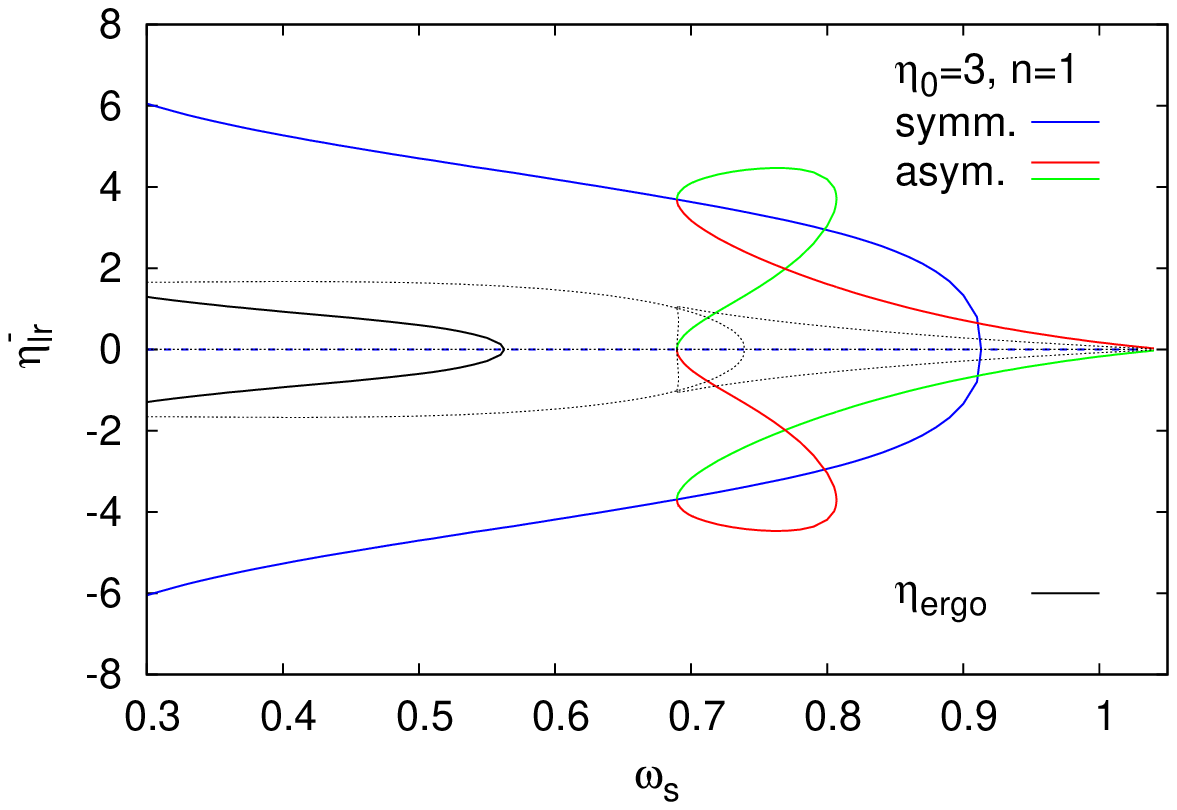}
\label{Fig8b}
}
}
\mbox{\hspace{0.2cm}
\subfigure[][]{\hspace{-1.0cm}
\includegraphics[height=.25\textheight, angle =0]{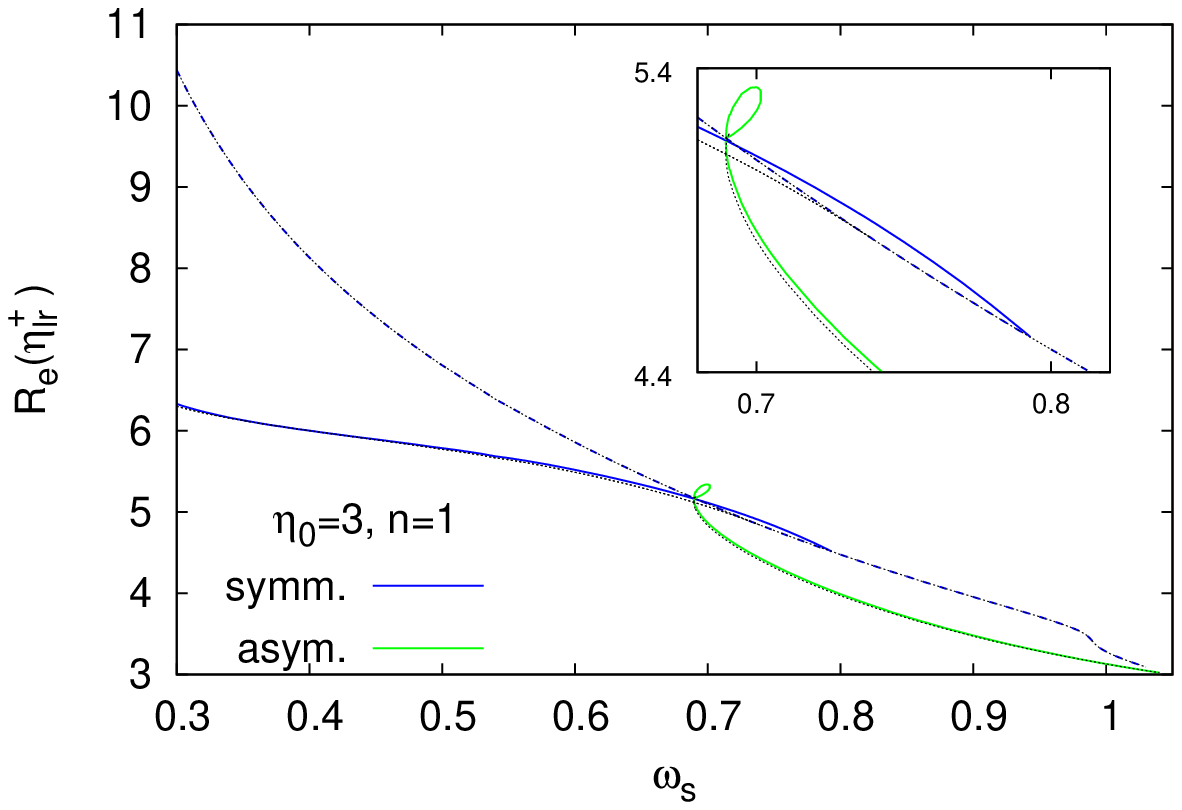}
\label{Fig8c}
}
\subfigure[][]{\hspace{-0.5cm}
\includegraphics[height=.25\textheight, angle =0]{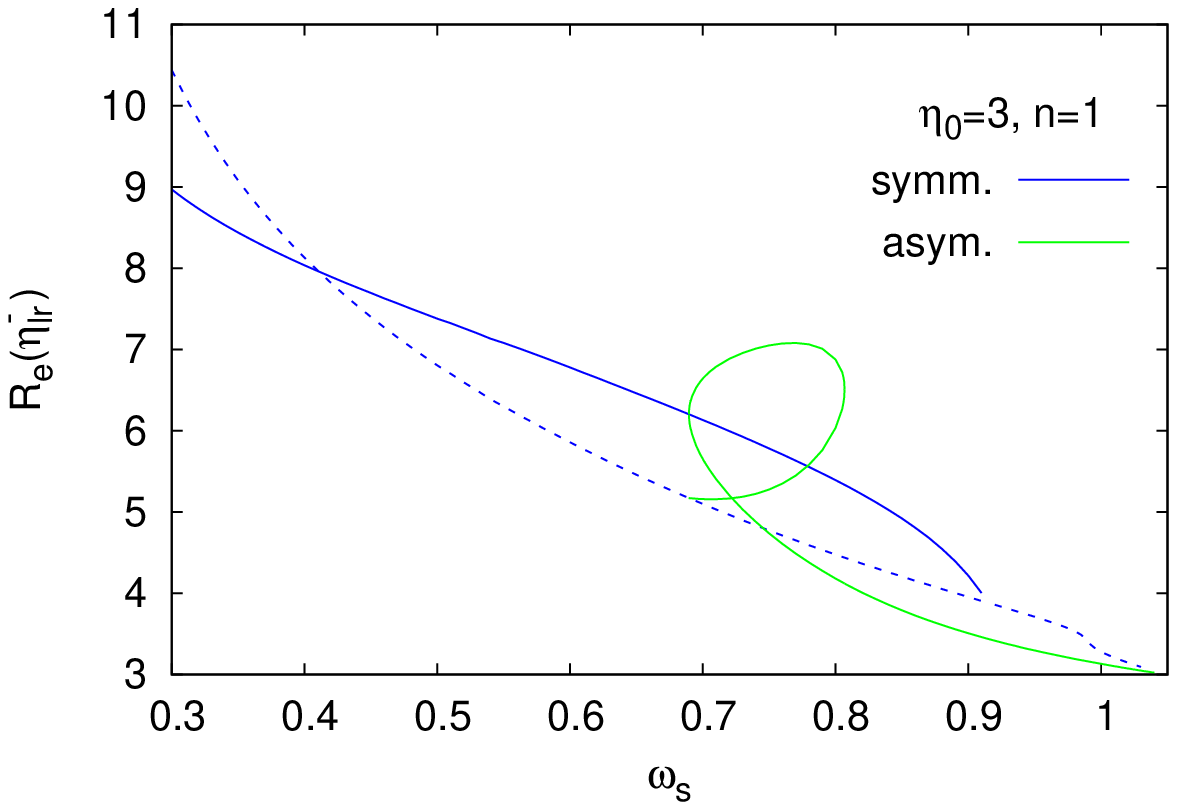}
\label{Fig8d}
}
}
\mbox{\hspace{0.2cm}
\subfigure[][]{\hspace{-1.0cm}
\includegraphics[height=.25\textheight, angle =0]{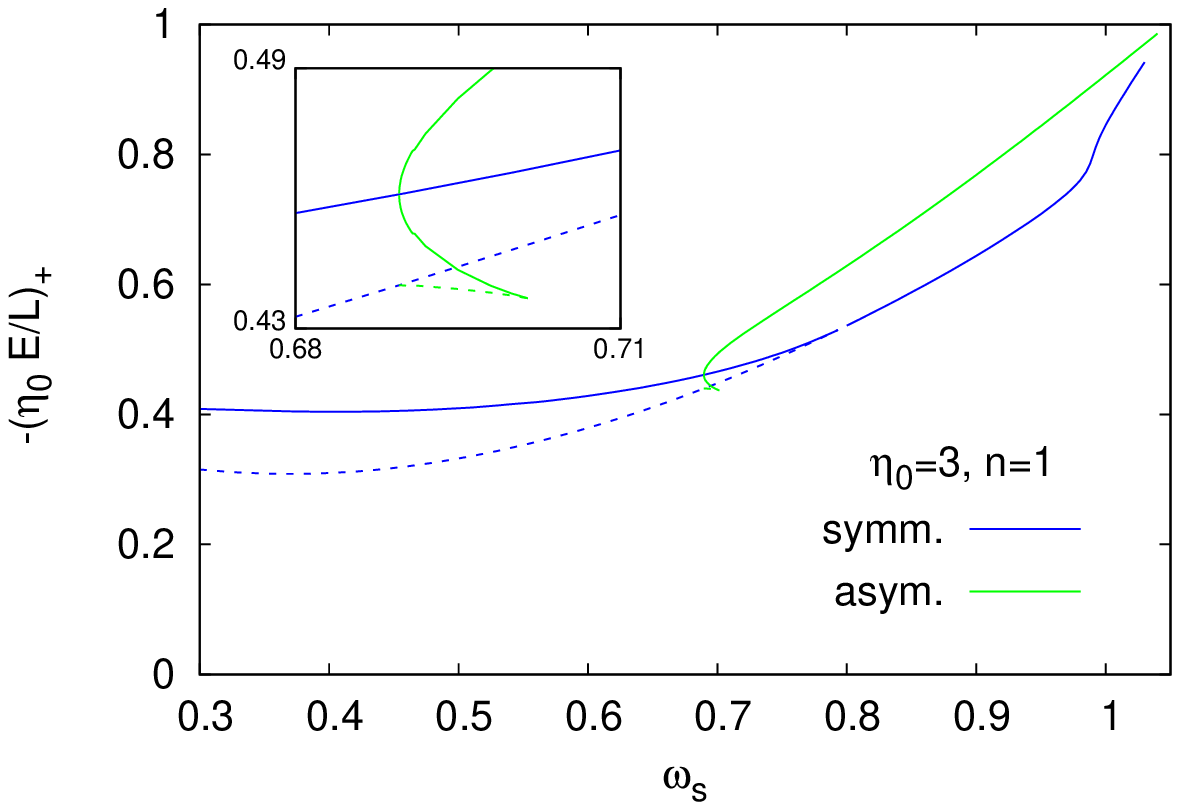}
\label{Fig8e}
}
\subfigure[][]{\hspace{-0.5cm}
\includegraphics[height=.25\textheight, angle =0]{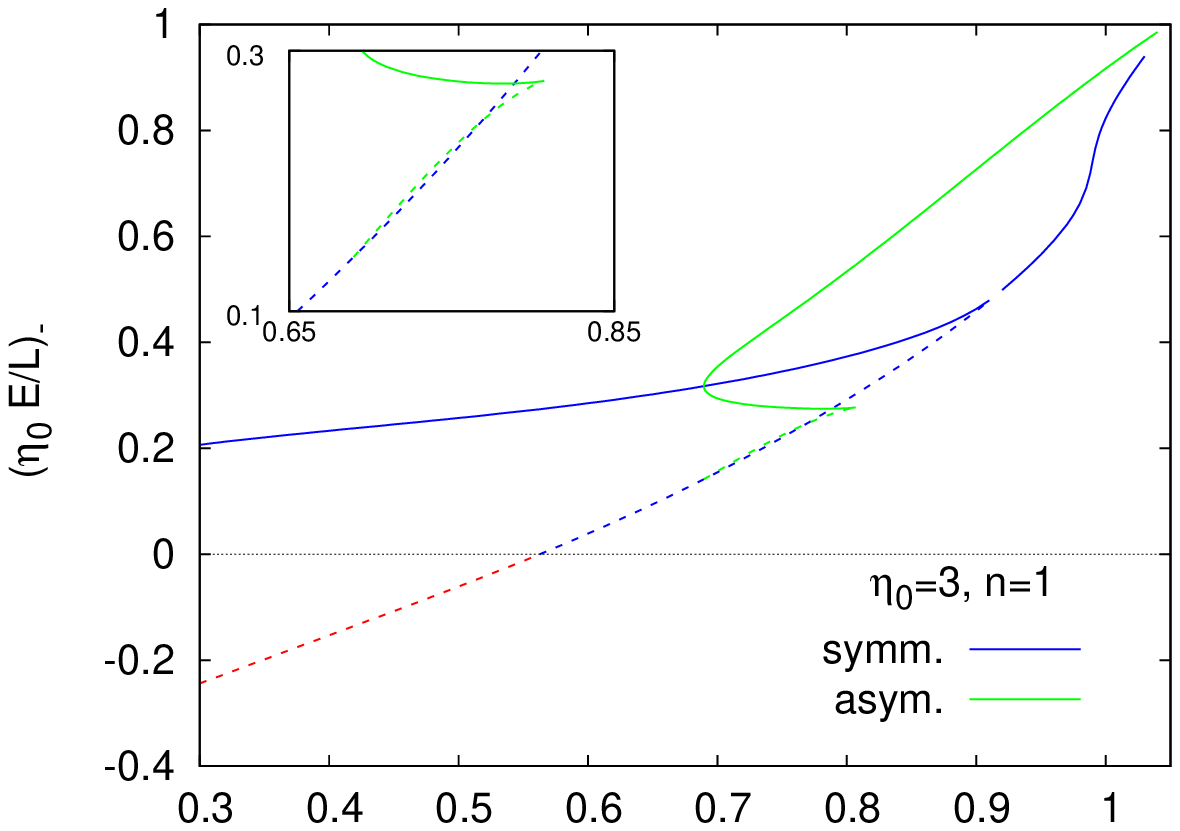}
\label{Fig8f}
}
}
\end{center}
\vspace{-0.5cm}
\caption{
Lightrings of rotating solutions
with throat parameter $\eta_0=3$
and rotational quantum number $n=1$ 
versus the boson frequency $\omega_s$:
(a) the coordinate $\eta^+_{\rm lr}$ of co-rotating
massless particles in the equatorial plane;
(b) the coordinate $\eta^-_{\rm lr}$ of counter-rotating
massless particles in the equatorial plane;
(c) the circumferential radius $R_{\rm e}(\eta^+_{\rm lr})$;
(d) the circumferential radius $R_{\rm e}(\eta^-_{\rm lr})$;
(e) the ratio of energy $E$ and angular momentum $L$ of the particle
scaled with the throat parameter for co-rotating massless particles;
(f) $\eta_0 E/L$ for counter-rotating massless particles.
Also shown are the coordinates and the circumferential radii
of the ergosurfaces (black lines).
\label{Fig8}
}
\end{figure}

\begin{figure}[t!]
\begin{center}
\mbox{\hspace{0.2cm}
\subfigure[][]{\hspace{-1.0cm}
\includegraphics[height=.25\textheight, angle =0]{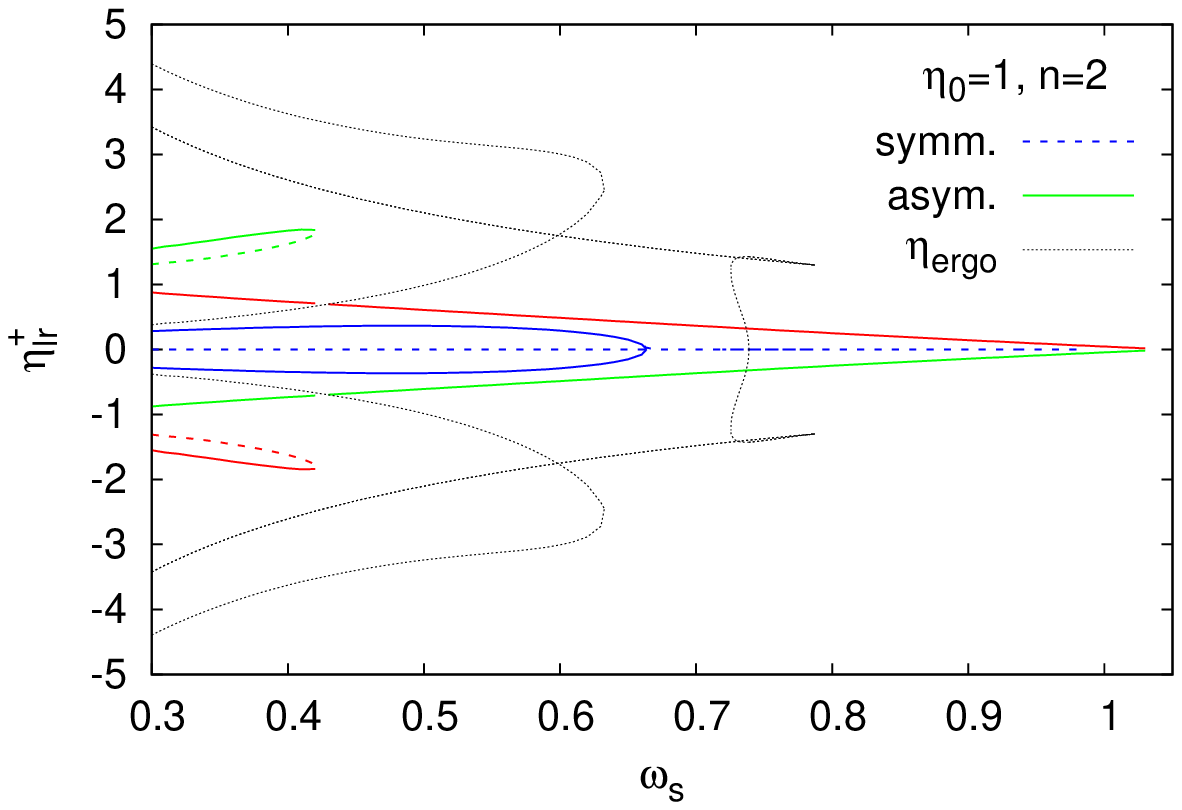}
\label{Fig9a}
}
\subfigure[][]{\hspace{-0.5cm}
\includegraphics[height=.25\textheight, angle =0]{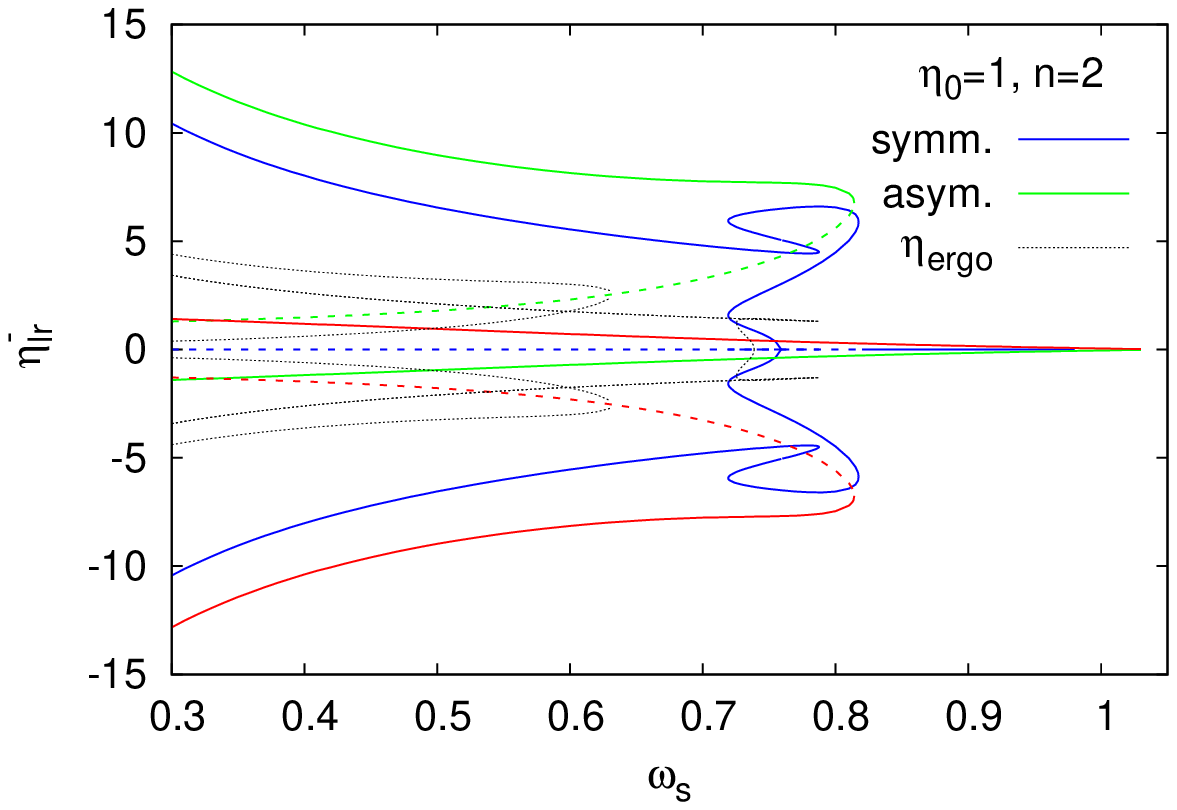}
\label{Fig9b}
}
}
\mbox{\hspace{0.2cm}
\subfigure[][]{\hspace{-1.0cm}
\includegraphics[height=.25\textheight, angle =0]{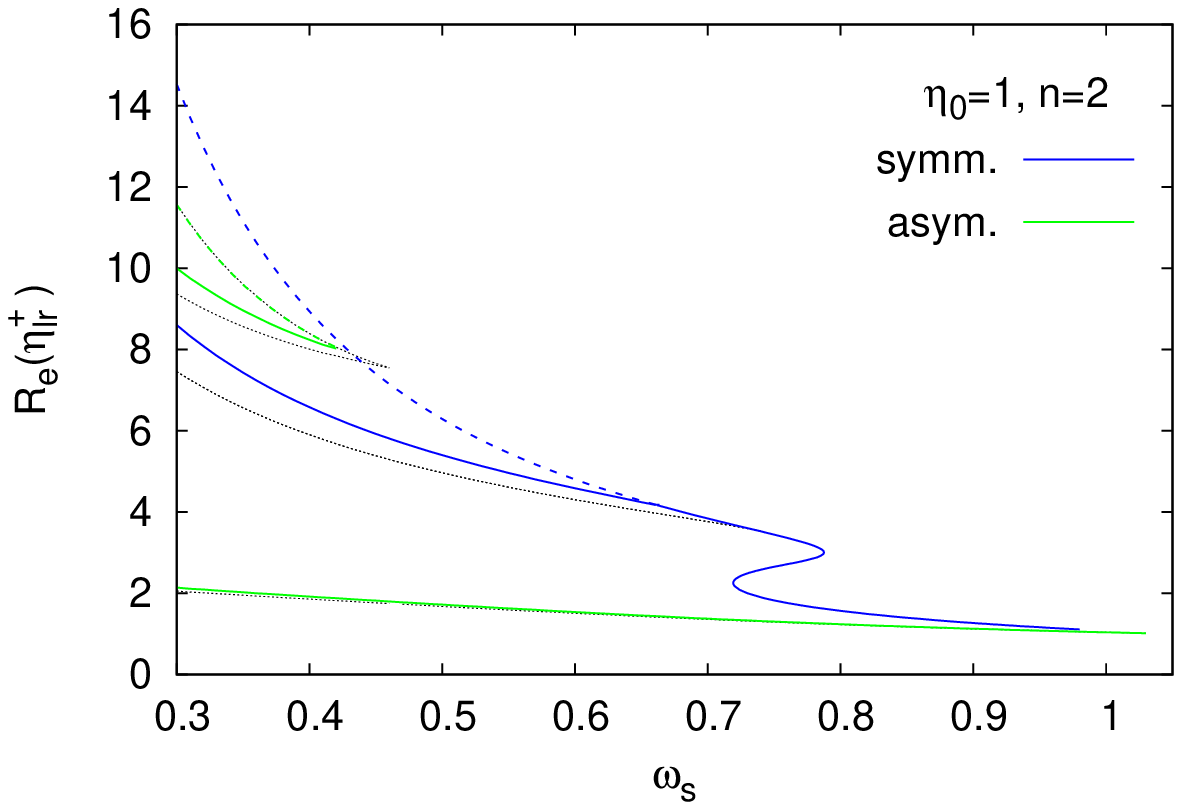}
\label{Fig9c}
}
\subfigure[][]{\hspace{-0.5cm}
\includegraphics[height=.25\textheight, angle =0]{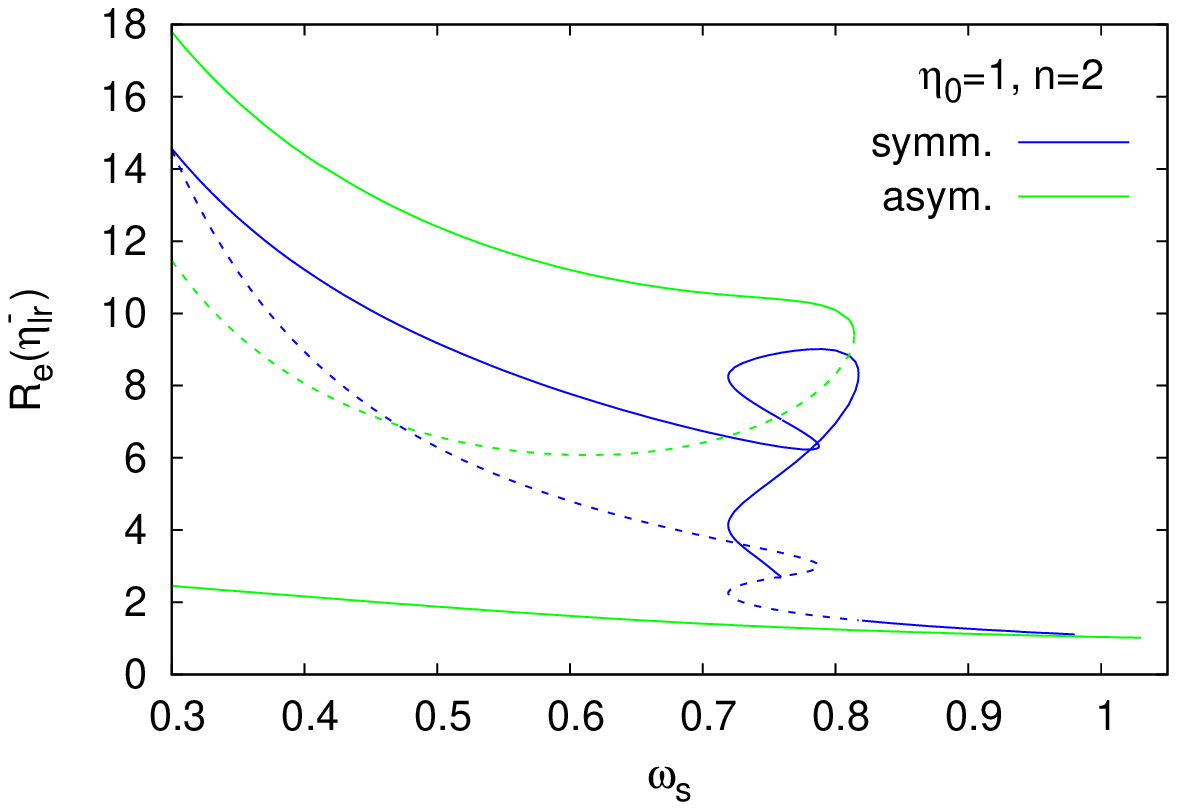}
\label{Fig9d}
}
}
\mbox{\hspace{0.2cm}
\subfigure[][]{\hspace{-1.0cm}
\includegraphics[height=.25\textheight, angle =0]{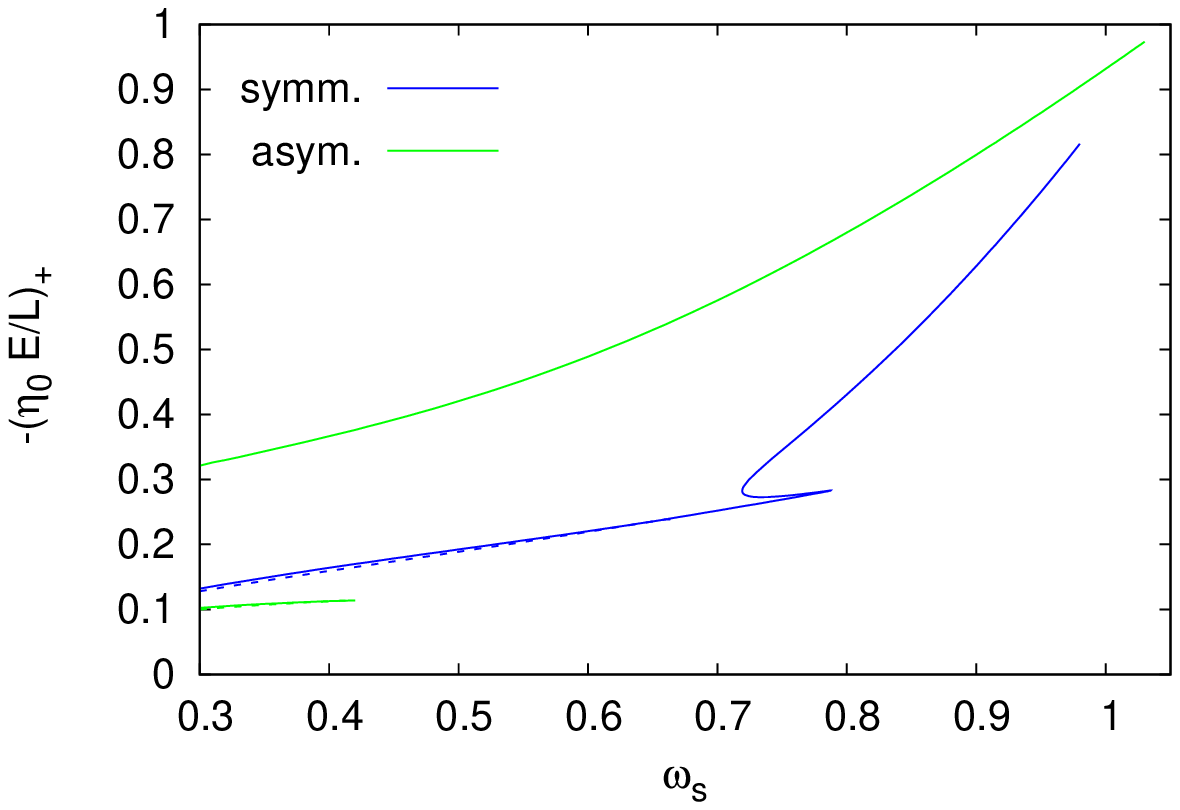}
\label{Fig9e}
}
\subfigure[][]{\hspace{-0.5cm}
\includegraphics[height=.25\textheight, angle =0]{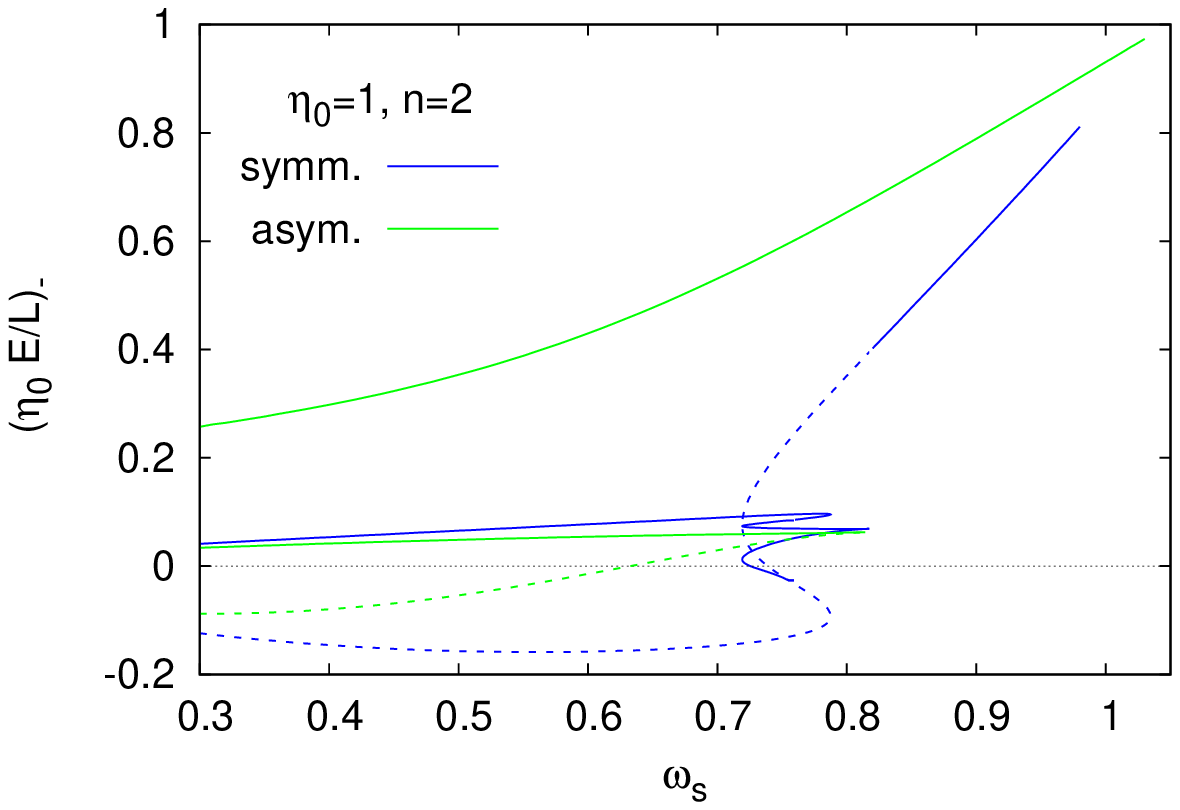}
\label{Fig9f}
}
}
\end{center}
\vspace{-0.5cm}
\caption{
Lightrings of rotating solutions
with throat parameter $\eta_0=1$
and rotational quantum number $n=2$ 
versus the boson frequency $\omega_s$:
(a) the coordinate $\eta^+_{\rm lr}$ of co-rotating
massless particles in the equatorial plane;
(b) the coordinate $\eta^-_{\rm lr}$ of counter-rotating
massless particles in the equatorial plane;
(c) the circumferential radius $R_{\rm e}(\eta^+_{\rm lr})$;
(d) the circumferential radius $R_{\rm e}(\eta^-_{\rm lr})$;
(e) the ratio of energy $E$ and angular momentum $L$ of the particle
scaled with the throat parameter for co-rotating massless particles;
(f) $\eta_0 E/L$ for counter-rotating massless particles.
Also shown are the coordinates and the circumferential radii
of the ergosurfaces (black lines).
\label{Fig9}
}
\end{figure}

At last we address the lightrings of these sets of wormhole solutions immersed in rotating bosonic matter.
In Fig.\ref{Fig8} and Fig.\ref{Fig9} we show the properties of the lightrings 
corresponding to the circular orbits of massless particles in the equatorial plane. 
Again we begin the discussion with wormhole solutions with throat parameter $\eta_0=3$ 
and rotational quantum number $n=1$, exhibited in Fig.\ref{Fig8}.

Fig.\ref{Fig8a} and \ref{Fig8b} show the coordinates $\eta^\pm_{\rm lr}$,
where the lightrings for co-rotating and
counter-rotating particles, respectively, are located.
For large values of the boson frequency $\omega_s$ a single lightring exists
for symmetric or asymmetric solutions and both co-rotating or counter-rotating particles.
This changes, when the boson frequency reaches a particular value, 
that depends on the symmetry of the wormhole solutions and the orientation of the orbits. 
When the boson frequency $\omega_s$ becomes smaller than this value, two further lightrings appear.
One of the lightrings of the symmetric wormholes is always located at the center $\eta=0$, 
i.e., either at the throat or at the equator,
whereas the other two lightrings are located symmetrically with respect to $\eta=0$ (when present).

In order to discuss the lightrings of the asymmetric wormholes, let us focus on the solution
with throat coordinate $\eta_{\rm th}>0$.
The lightring which is present already for large values of the boson frequency $\omega_s$
is always located outside the throat(s), i.e., $\eta^\pm_{\rm lr} > \eta_{\rm th}$.
This lightring merges with the lightring of the symmetric wormholes, when the 
asymmetric and symmetric solutions bifurcate.
When $\omega_s$ assumes a particular value,
a pair of lightrings emerges at some negative coordinate value $\eta^\pm_{\rm lc}$ . 
The inner one of these later merges with the lightring of the symmetric wormhole located at the equator, 
when the solutions bifurcate. The outer one merges with the corresponding lightring 
of the symmetric solutions for negative coordinate $\eta$.
In the case of counter-rotating massless particles the values of $\omega_s$, where multiple
lightrings emerge, are larger than in the co-rotating case.
Also shown in the figures are the locations of the throat(s) and the equator (dotted black) 
as well as the ergosurface (solid black).

We show in  Fig.\ref{Fig8c} and \ref{Fig8d} the circumferential radii  $R_{\rm e}(\eta^\pm_{\rm lr})$
of the lightrings
for co-rotating and counter-rotating massless particles, respectively. 
Also shown are the circumferential radii of the throat(s) $R_{\rm e}(\eta^\pm_{\rm th})$ 
and the equator $R_{\rm e}(\eta^\pm_{\rm eq})$ (dotted black) for comparison.
We note, that the circumferential radii of the lightrings of co-rotating
massless particles and the circumferential radii of the throats, respectively the equator, almost coincide, 
except for those of the pair of lightrings that emerges at the particular value of $\omega_s$.  

In  Fig.\ref{Fig8e} and \ref{Fig8f} we exhibit the ratio of the energy and angular momentum $E/L$,
scaled with the throat parameter $\eta_0$, for 
the co-rotating and counter-rotating massless particles, respectively. 
Since the wormhole spacetime carries negative angular momentum $J$ (by construction), 
the ratio $E/L$ is negative for co-rotating particles, if we assume 
positive energy and negative angular momentum $L$.
On the other hand, for counter-rotating particles $L$ is then positive.
However, as we observe from Fig.\ref{Fig8f}, the ratio $E/L$ of the lightring residing at
the equator of the symmetric wormholes becomes negative,
when $\omega_s$ is smaller than a critical value. 
Comparison with Fig.\ref{Fig8b} shows that the change of sign occurs exactly,
when the lightring enters the ergosphere.

As a second example we demonstrate the properties of the lightrings for the set of solutions
with throat parameter $\eta_0=1$ and rotational quantum number $n=2$ in  Fig.\ref{Fig9}.
This set is representative for all three sets of solutions, where we do not observe bifurcations
of symmetric and asymmetric solutions.
Consequently, the new pairs of lightrings persists to smaller boson frequencies, once they have 
appeared, unless further pairs arise, as in the case of symmetric solutions, 
where up to five lightrings of counter-rotating massless
particles can exist in a certain interval of the boson frequency.
Note, that here the branch structure of the solutions complicates the analysis again.

\section{Conclusions}

We have studied the domain of existence and the physical properties of
wormhole solutions immersed in rotating bosonic matter.
These solutions arise in General Relativity, when a complex boson field
and a phantom field are coupled minimally to gravity.
For the complex boson field we have only allowed for a mass term,
but not for self-interactions here.

The set of coupled field equations is symmetric with respect to 
reflection of the radial coordinate, $\eta \to -\eta$.
In the presence of rotating matter, the boundary conditions of the metric
and the boson field can also be chosen in a symmetric way in both
asymptotically flat spacetime regions. Clearly, symmetric
solutions then result, where the functions possess reflection symmetry.
Interestingly, however, also asymmetric solutions arise in the presence
of the non-trivial wormhole topology. However, the reflection symmetry
of the system then results in the emergence of pairs of asymmetric solutions,
where the two solutions of a pair are related via a reflection transformation,
$\eta \to -\eta$. 

The complex boson field is characterized by a harmonic time dependence
with the boson frequency $\omega_s$ and by a rotational quantum number $n$,
analogously to the case of $Q$-balls or boson stars.
The non-trivial topology of the wormhole spacetime leads, however, to a very different
dependence of the solutions on the boson frequency than in the case of
boson stars. In particular, the prominent spirals of boson stars have basically unwound,
with some backbending of the solutions remaining the only reminder of those spirals.

The absence of self-interactions of the boson field 
has pronounced consequences. In particular, there are no non-trivial solutions
of the boson field in the probe limit. Moreover, the phenomenon of spontaneous
symmetry breaking is not present, which is known to occur for non-rotating wormhole
solutions, when a sextic self-interaction is included. In that case the masses of both
asymmetric solutions are smaller than the mass of the symmetric solution
for a given particle number, when the asymmetric solutions bifurcate from the
symmetric ones. 

Here we have shown that both symmetric and asymmetric solutions can possess
an intriguing throat structure. For the symmetric solutions the center is always
either a throat or an equator. In the case of an equator the solutions represent double throat solutions,
where the throats are
located symmetrically on either side of the equator. For the asymmetric solutions
the single throat is located asymmetrically with respect to the center,
and in the case of double throat solutions,
an equator with the second throat emerge on the opposite side of the spacetime.
The wormhole solutions can also possess ergoregions, when the rotation is sufficiently fast.

Of interest is also the lightring structure of these spacetimes.
Here we have only made a rough first analysis of it,
which has revealed that in addition to the single lightring always present
further pairs of lightrings can appear. For instance, in the case of
counter-rotating particles we have observed the presence of up to five lightrings.
A more detailed study of these lightrings and, in particular, of the shadows
of these rotating wormhole solutions will be given elsewhere.

Let us briefly address the stability
of these wormhole spacetimes immersed in rotating matter.
The static Ellis wormholes are known to possess an unstable radial mode
\cite{Shinkai:2002gv,Gonzalez:2008wd,Gonzalez:2008xk,Torii:2013xba}.
For rotating wormholes such an analysis is much more involved.
However, it has been shown for five-dimensional rotating wormholes,
whose two angular momenta have the same magnitude,
that this radial instability disappears, 
when the wormhole throat rotates sufficiently fast
\cite{Dzhunushaliev:2013jja}.
This shows that rotation can have a stabilizing effect on the wormhole.
Clearly, a stability analysis of these solutions is called for.

Alternatively, one could also consider wormholes in generalized theories of gravity
by replacing General Relativity by certain well-motivated 
gravitational theories, where the phantom field would no longer be needed
\cite{Hochberg:1990is,Fukutaka:1989zb,Ghoroku:1992tz,Furey:2004rq,Lobo:2009ip,Bronnikov:2009az,Kanti:2011jz,Kanti:2011yv,Harko:2013yb},
hoping that the instability would disappear along with the phantom field.
A study of rotating wormholes in such generalized theories of gravity
should be one of our next goals to consider.

\section*{Acknowledgments}

We would like to acknowledge support by the DFG Research Training Group 1620
{\sl Models of Gravity} as well as by FP7, Marie Curie Actions, People,
International Research Staff Exchange Scheme (IRSES-606096),
COST Action CA16104 {\sl GWverse}.
BK gratefully acknowledges support
from Fundamental Research in Natural Sciences
by the Ministry of Education and Science of Kazakhstan.

%

\section*{Appendix A: Particle number}

Here we address the particle number for the asymmetric solutions.
We consider the coupling of the complex boson field to an electromagnetic gauge field in the
probe limit. Thus we neglect any backreaction of the gauge field to the boson field or the
gravitational field.
The electromagnetic field is determined from
\begin{equation}
\partial_\mu \left(\sqrt{-g}g^{\mu\sigma} g^{\nu\rho}F_{\sigma\rho}\right) = j^\nu \sqrt{-g} \ ,
\label{maxwell}
\end{equation}
where $F_{\sigma\rho} = \partial_\sigma A_\rho-\partial_\rho A_\sigma$ is the field strength of the
gauge potenial $A_\mu$, and $j^\nu$ is the current, i.e.
\begin{equation}
j^\nu = -i g^{\nu\mu}\left(\Phi^*\partial_\mu\Phi-\Phi\partial_\mu\Phi^*\right) \ .
\label{current}
\end{equation}
As ansatz for the gauge potential we choose
\begin{equation}
A_\mu dx^\mu = a_0(\eta,\theta) dt + a_\varphi(\eta,\theta) d\varphi \ .
\label{gaugepot}
\end{equation}
Substitution in the Maxwell equations (\ref{maxwell}) yields two PDEs for the 
functions $a_0(\eta,\theta)$ and $a_\varphi(\eta,\theta)$ to be solved numerically. As boundary conditions
we take 
\begin{equation}
a_0(\pm\infty,\theta)=a_\varphi(\pm\infty,\theta)=0 \ , 
\partial_\theta a_0(\eta,0)=a_\varphi(\eta,0)=0 \ , 
\partial_\theta a_0(\eta,\pi/2)=\partial_\theta a_\varphi(\eta,\pi/2)=0 \ . 
\label{gaugebdcond}
\end{equation}
The charges $Q_\pm$ are now determined from the asymptotic behaviour of $a_0$,
\begin{equation}
a_0 \to  \pm\frac{Q_\pm}{\eta}\ {\rm  as} \ \eta \to \pm\infty \ .
\label{echarge}
\end{equation}

\section*{Appendix B: 
\boldmath Limit $\omega_s \to 0$ ($n=0$) \unboldmath}

We here address the limit $\omega_s \to 0$ of the non-rotating solutions.
In Fig.\ref{Fig1} we have seen, that both the mass and the particle number
seem to diverge as the boson frequency approaches its minimal value,
$\omega_s \to 0$.
At the same time the boson field itself seems to vanish,
a fact that appears hard to reconcile  with the divergence of mass and
particle number.
On the other hand, however, we have seem in the figure,
that the quantity $\omega_s Q/M \to 1$ for $\omega_s \to 0$.

\begin{figure}[t!]
\begin{center}
\mbox{\hspace{0.2cm}
\subfigure[][]{\hspace{-1.0cm}
\includegraphics[height=.25\textheight, angle =0]{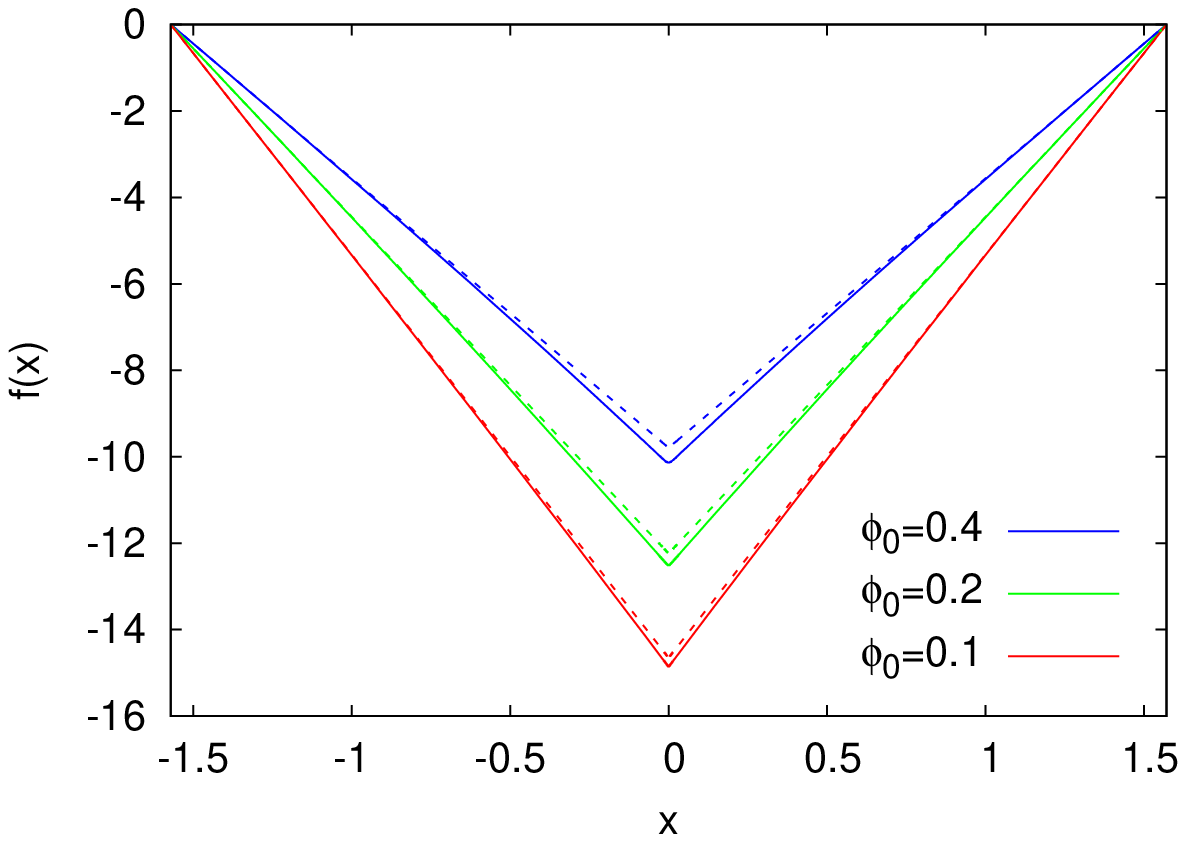}
\label{Fig10a}
}
\subfigure[][]{\hspace{-0.5cm}
\includegraphics[height=.25\textheight, angle =0]{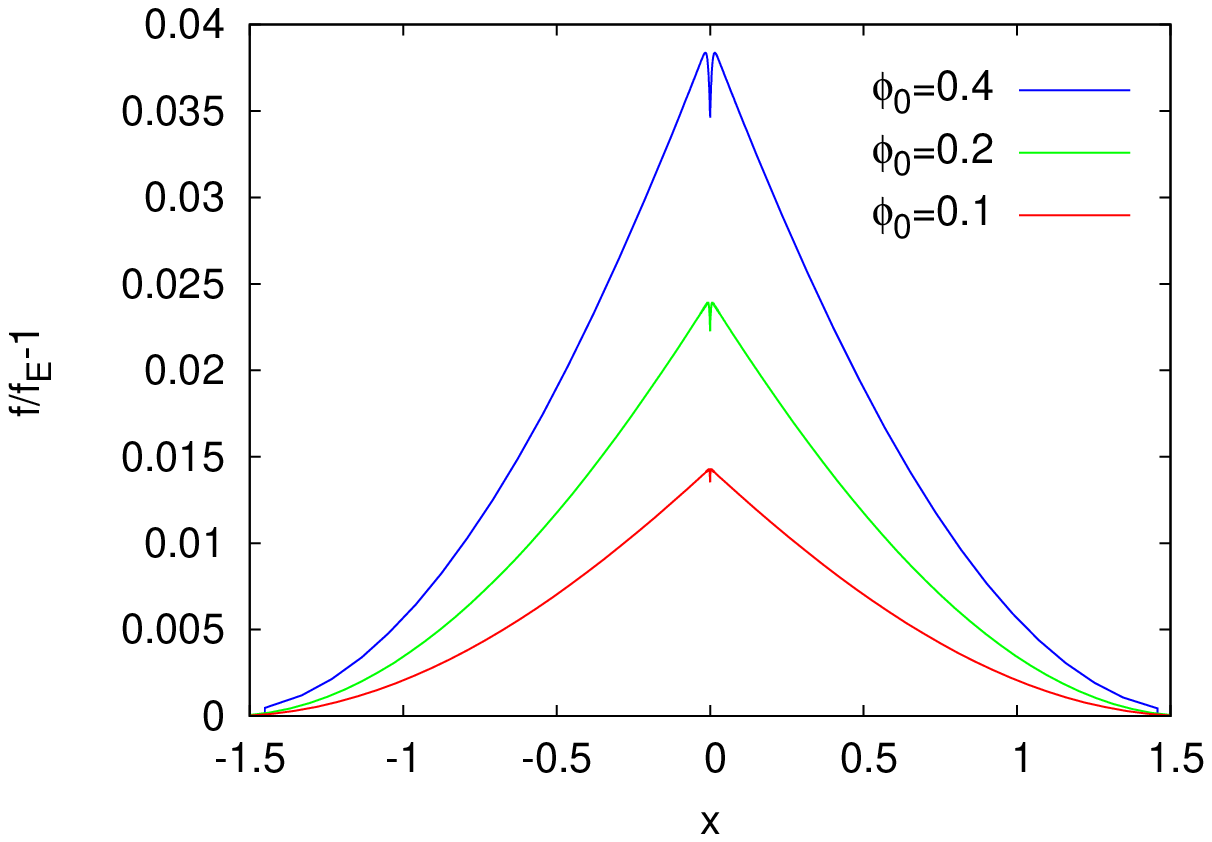}
\label{Fig10b}
}
}
\mbox{\hspace{0.2cm}
\subfigure[][]{\hspace{-1.0cm}
\includegraphics[height=.25\textheight, angle =0]{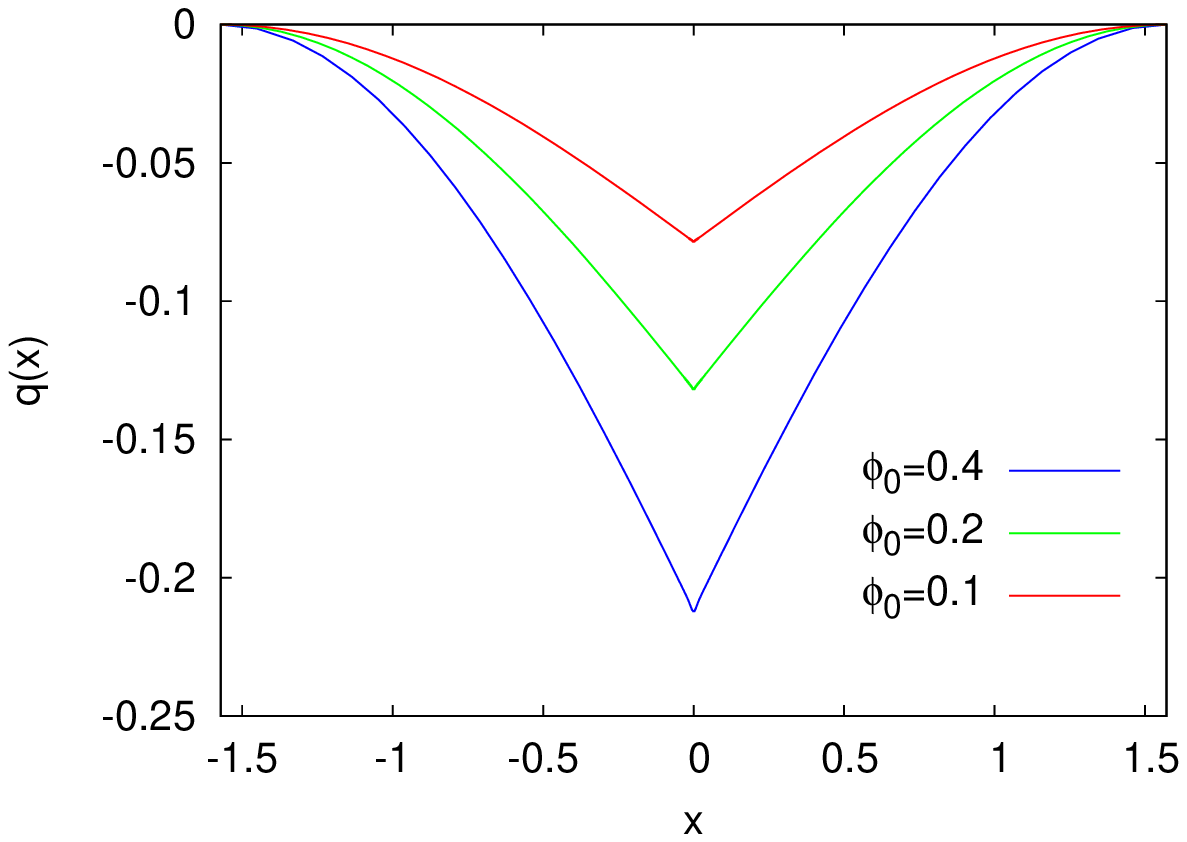}
\label{Fig10c}
}
\subfigure[][]{\hspace{-0.5cm}
\includegraphics[height=.25\textheight, angle =0]{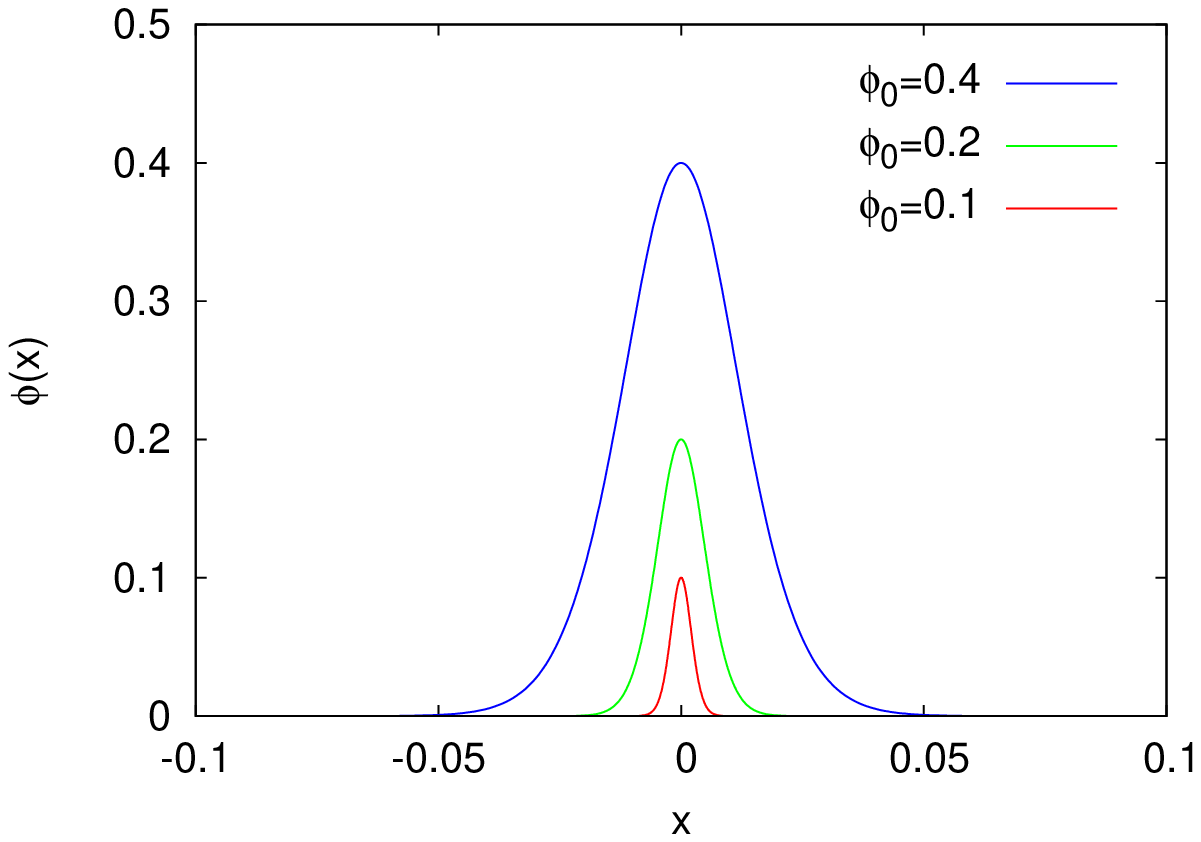}
\label{Fig10d}
}
}
\end{center}
\vspace{-0.5cm}
\caption{Non-rotating solutions ($n=0$)
for throat parameter  $\eta_0=3$ and 
central value of the boson field $\phi_0=0.4$, $0.2$ and $0.1$:
(a) the metric function $f(x)$ versus the compactified radial coordinate
$x=\arctan(\eta/\eta_0)$,
the dotted curves show the function $f_{\rm E}(x)$
of the symmetrized Ellis wormholes with the same values of the mass;
(b) the relative difference between $f(x)$ and $f_{\rm E}(x)$;
(c) the metric function $q(x)$;
(d) the boson function $\phi(x)$ close to the center $x=0$.
\label{Fig10}
}
\end{figure}

In order to get some insight into this intriguing situation, let us inspect the
behaviour of the solutions in this limit.
To this end we exhibit in Fig.\ref{Fig10} the metric functions $f$ and $q$ for a
sequence of solutions with decreasing central values of the boson field $\phi_0$.
We observe from Fig.\ref{Fig10a} that the function $f$ assumes a wedge form 
whose minimum decreases as $\phi_0$ decreases. The dotted lines show 
the symmetrized massive Ellis wormholes whose metric function reads
\begin{equation}
f_{\rm E}(x) = \frac{2 M}{\eta_0} \left(|x|-\frac{\pi}{2}\right) \ , \ \ \ \ \
x=\arctan\left( \frac{\eta}{\eta_0} \right) \ ,
\end{equation}
and which possess the same mass $M$ as the wormholes immersed in bosonic matter.

Fig.\ref{Fig10b} shows the relative difference $f(x)/f_{\rm E}(x)-1$. We observe that
this difference becomes smaller with decreasing $\phi_0$ and seems to vanish
as $\phi_0 \to 0$.
On the other hand, we note from Fig.\ref{Fig10c} that the metric function $q(x)$ vanishes as 
$\phi_0$ tends to zero, which is also consistent with the case of the massive Ellis wormhole.
Fig.\ref{Fig10d} shows the boson field on a small interval around the center $x=0$. We note that the
function $\phi(x)$ decreases along with its maximum $\phi_0$. At the same time the interval where $\phi(x)$ is
noticeably different from zero diminishes with decreasing  $\phi_0$.

Some insight may come from the field equation of the boson field. For the non-rotating
case it reads
\begin{equation}
\left(e^{\frac{q}{2}} h \phi'\right)' 
=e^{\frac{3q}{2}-f}\left[m_{\rm bs}^2-\omega_s^2 e^{-f}\right] \phi h \ .
\label{nreqphi}
\end{equation}
Integration from $-\infty$ to $\infty$ yields
\begin{equation}
\left. e^{\frac{q}{2}} h \phi'\right|_{-\infty}^{\infty} 
=\int_{-\infty}^{\infty}{e^{\frac{3q}{2} -f}\left[m_{\rm bs}^2-\omega_s^2 e^{-f}\right] \phi h } d\eta \ .
\label{intnreqphi}
\end{equation}
Since the boson field decays exponentially the left hand side vanishes. Consequently, the right hand 
side also has to vanish. Taking into account that $\phi$ is non-negative, we conclude that the term
in the brackets has to change sign. We observe that $f$ assumes its (only) minimum at $\eta=0$, which implies
that the term in the brackets also assumes its minimum at $\eta=0$ and is negative there. Thus we arrive 
at the condition ($f_0=f(0)$)
\begin{equation}
m_{\rm bs}^2-\omega_s^2 e^{-f_0} \leq 0 \ \ \ \Longleftrightarrow \ \ \ 
e^{f_0} \leq \frac{\omega_s^2}{m_{\rm bs}^2} \ .
\label{condf0}
\end{equation}
Consequently, we conclude that $f_0 \to -\infty$ as $\omega_s \to 0$.

\begin{figure}[t!]
\begin{center}
\mbox{\hspace{0.2cm}
\subfigure[][]{\hspace{-1.0cm}
\includegraphics[height=.25\textheight, angle =0]{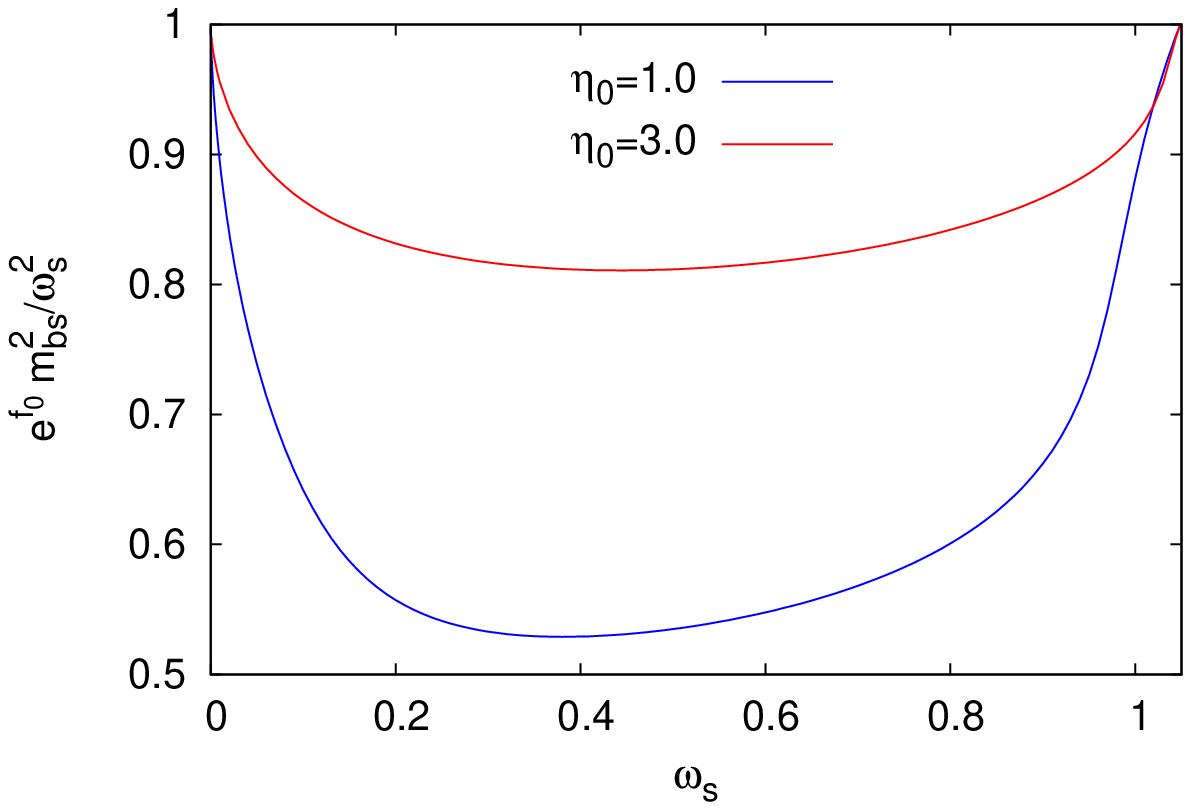}
\label{Fig11a}
}
\subfigure[][]{\hspace{-0.5cm}
\includegraphics[height=.25\textheight, angle =0]{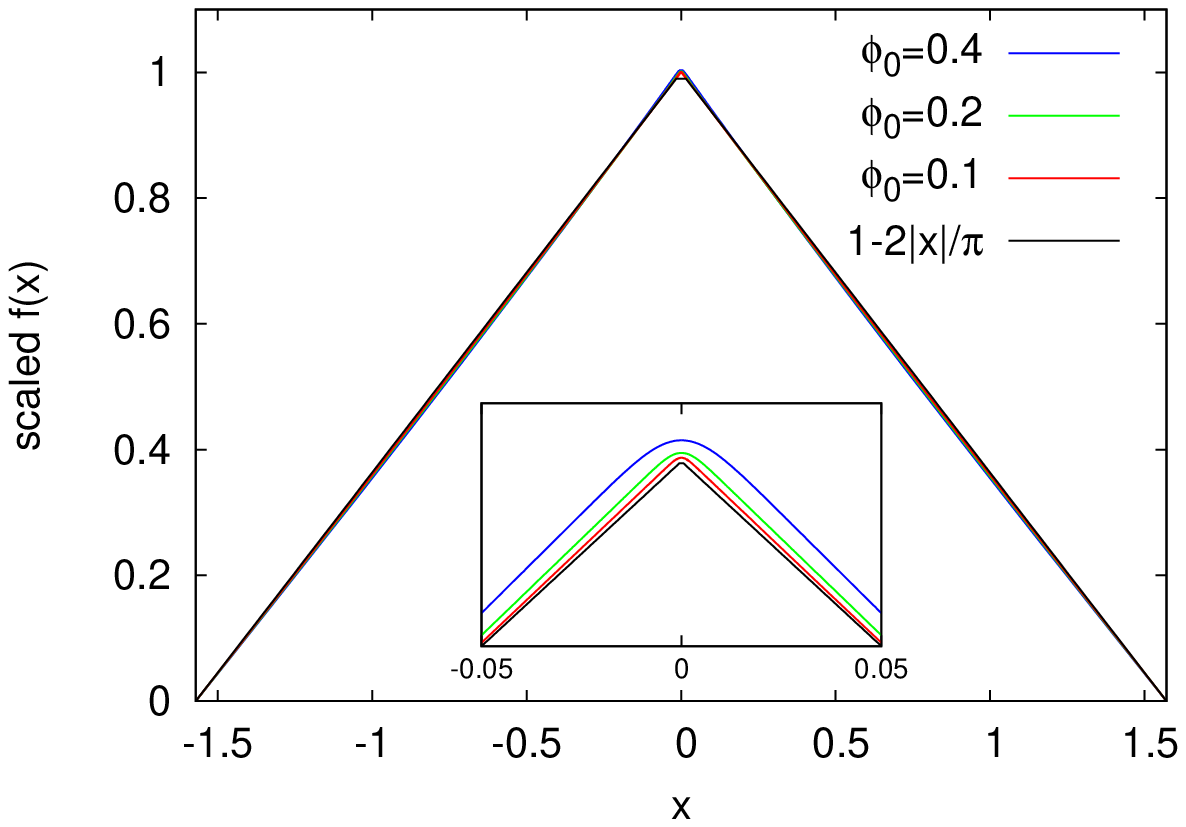}
\label{Fig11b}
}
}
\mbox{\hspace{0.2cm}
\subfigure[][]{\hspace{-1.0cm}
\includegraphics[height=.25\textheight, angle =0]{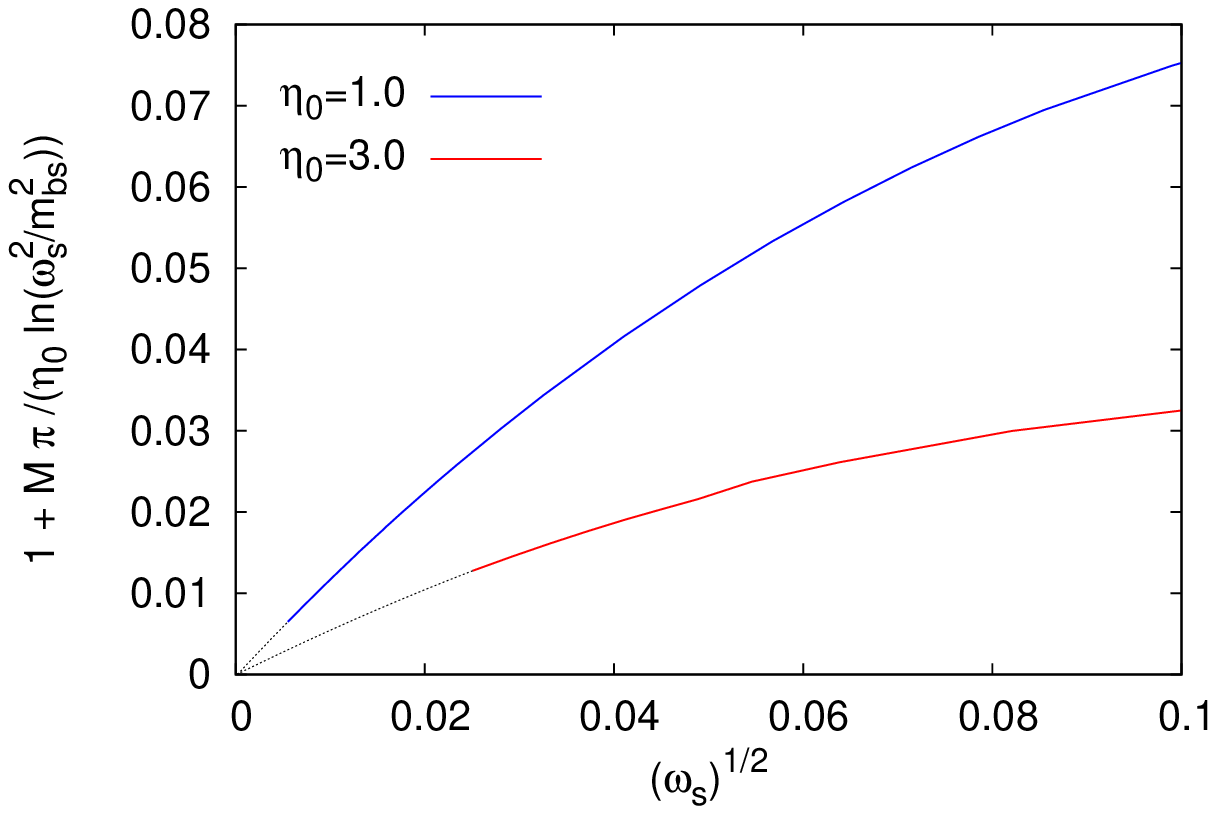}
\label{Fig11c}
}
}
\end{center}
\vspace{-0.5cm}
\caption{Non-rotating solutions:
(a) the ratio $e^{f_0}m_{\rm bs}^2/\omega_s^2$ versus the boson frequency $\omega_s$
for throat parameter $\eta_0=1$ and $3$;
(b) the scaled functions $f(x)/\ln(\omega_s^2/m_{\rm bs}^2)$ 
versus the compactified coordinate $x=\arctan(\eta/\eta_0)$ for 
throat parameter $\eta_0=3$ and central values of the boson field
$\phi_0 = 0.4$, $0.2$, $0.1$, also shown is the function $1-2|x|/\pi$;
(c) the quantity
 $1 + M \pi/\eta_0\ln(\omega_s^2/m_{\rm bs}^2)$ versus the square root of
the boson frequency $\sqrt{\omega_s}$ for throat parameter
 $\eta_0=1$ and $3$ (with thin dotted extrapolation to $\omega_s=0$).
\label{Fig11}
}
\end{figure}

In Fig.\ref{Fig11a} we show the ratio 
$e^{f_0}m_{\rm bs}^2/\omega_s^2$ versus the boson frequency $\omega_s$ for 
throat parameters $\eta_0=1$ and $3$. 
We note that this 
ratio tends to the value one as $\omega_s \to 0$. However, in this limit the integrand in Eq.~(\ref{intnreqphi})
becomes non-negative and solutions cease to exist, except for the case that the
boson field $\phi$ vanishes identically.

In  Fig.\ref{Fig11b} we now exhibit the scaled metric function $f(x)/\ln(\omega_s^2/m_{\rm bs}^2)$ 
for throat parameter $\eta_0=3$ and decreasing central values of the boson field
$\phi_0 = 0.4$, $0.2$, $0.1$ together with the function $1-2|x|/\pi$, which is proportional
to the function $f_{\rm E}$ of the symmetrized Ellis wormholes. We note that there is hardly
any difference between the functions discernible. In order to illustrate the difference
we have to zoom in considerably, as illustrated in the inset of the figure.

Assuming that in the limit $\omega_s \to 0$ the metric function behaves like 
$f(x)= \ln(\omega_s^2/m_{\rm bs}^2)(1-2|x|/\pi)$, we conclude that the mass 
behaves like $M/\eta_0 = -\ln(\omega_s^2/m_{\rm bs}^2)/\pi$. 
This is demonstrated in Fig.\ref{Fig11c} where
we exhibit the quantity 
$ 1 + M \pi/\left(\eta_0\ln(\omega_s^2/m_{\rm bs}^2)\right)$ versus the square root
of the boson frequency $\sqrt{\omega_s}$ for small values of $\omega_s$.
For $\omega_s \to 0$ this quantity indeed goes to zero.


\clearpage

\begin{thebibliography}{99}
  


\bibitem{Ellis:1973yv}
  H.~G.~Ellis,
  J.\ Math.\ Phys.\  {\bf 14}, 104 (1973).

\bibitem{Bronnikov:1973fh}
  K.~A.~Bronnikov,
  Acta Phys.\ Polon.\  {\bf B4}, 251 (1973).

\bibitem{Kodama:1978dw}
  T.~Kodama,
  Phys.\ Rev.\  {\bf D18}, 3529 (1978).

\bibitem{Ellis:1979bh}
  H.~G.~Ellis,
  Gen.\ Rel.\ Grav.\  {\bf 10}, 105 (1979).

\bibitem{Morris:1988cz}
  M.~S.~Morris, K.~S.~Thorne,
  Am.\ J.\ Phys.\  {\bf 56}, 395 (1988).

\bibitem{Morris:1988tu}
  M.~S.~Morris, K.~S.~Thorne and U.~Yurtsever,
  Phys.\ Rev.\ Lett.\  {\bf 61}, 1446 (1988).

\bibitem{Lobo:2005us}
  F.~S.~N.~Lobo,
  Phys.\ Rev.\ D {\bf 71}, 084011 (2005).

\bibitem{Lobo:2017oab} 
  F.~S.~N.~Lobo,
  Fundam.\ Theor.\ Phys.\  {\bf 189}, pp. (2017).


\bibitem{Dzhunushaliev:2011xx} 
  V.~Dzhunushaliev, V.~Folomeev, B.~Kleihaus and J.~Kunz,
  JCAP {\bf 1104}, 031 (2011).

\bibitem{Dzhunushaliev:2012ke} 
  V.~Dzhunushaliev, V.~Folomeev, B.~Kleihaus and J.~Kunz,
  Phys.\ Rev.\ D {\bf 85}, 124028 (2012).

\bibitem{Dzhunushaliev:2013lna} 
  V.~Dzhunushaliev, V.~Folomeev, B.~Kleihaus and J.~Kunz,
  Phys.\ Rev.\ D {\bf 87}, 104036 (2013).

\bibitem{Dzhunushaliev:2014mza} 
  V.~Dzhunushaliev, V.~Folomeev, B.~Kleihaus and J.~Kunz,
  Phys.\ Rev.\ D {\bf 89}, 084018 (2014)

\bibitem{Aringazin:2014rva} 
  A.~Aringazin, V.~Dzhunushaliev, V.~Folomeev, B.~Kleihaus and J.~Kunz,
  JCAP {\bf 1504}, 005 (2015).

\bibitem{Dzhunushaliev:2016ylj} 
  V.~Dzhunushaliev, V.~Folomeev, B.~Kleihaus and J.~Kunz,
  JCAP {\bf 1608}, 030 (2016).

\bibitem{Charalampidis:2013ixa} 
  E.~Charalampidis, T.~Ioannidou, B.~Kleihaus and J.~Kunz,
  Phys.\ Rev.\ D {\bf 87}, 084069 (2013).


\bibitem{Dzhunushaliev:2014bya} 
  V.~Dzhunushaliev, V.~Folomeev, C.~Hoffmann, B.~Kleihaus and J.~Kunz,
  Phys.\ Rev.\ D {\bf 90}, 124038 (2014).

\bibitem{Hoffmann:2017jfs} 
  C.~Hoffmann, T.~Ioannidou, S.~Kahlen, B.~Kleihaus and J.~Kunz,
  Phys.\ Rev.\ D {\bf 95}, 084010 (2017).

\bibitem{Hoffmann:2017vkf} 
  C.~Hoffmann, T.~Ioannidou, S.~Kahlen, B.~Kleihaus and J.~Kunz,
  Phys.\ Lett.\ B {\bf 778}, 161 (2018).


\bibitem{Jetzer:1991jr}
  P.~Jetzer,
  Phys.\ Rept.\  {\bf 220}, 163 (1992).

\bibitem{Lee:1991ax}
  T.~D.~Lee, Y.~Pang,
  Phys.\ Rept.\  {\bf 221}, 251 (1992).

\bibitem{Schunck:2003kk}
  F.~E.~Schunck, E.~W.~Mielke,
  Class.\ Quant.\ Grav.\  {\bf 20}, R301 (2003).

\bibitem{Liebling:2012fv}
  S.~L.~Liebling and C.~Palenzuela,
  Living Rev.\ Rel.\  {\bf 15}, 6 (2012).

\bibitem{Kashargin:2007mm}
  P.~E.~Kashargin and S.~V.~Sushkov,
  Grav.\ Cosmol.\  {\bf 14}, 80 (2008).

\bibitem{Kashargin:2008pk}
  P.~E.~Kashargin and S.~V.~Sushkov,
  Phys.\ Rev.\ D {\bf 78}, 064071 (2008).

\bibitem{Kleihaus:2014dla}
  B.~Kleihaus and J.~Kunz,
  Phys.\ Rev.\ D {\bf 90}, 121503 (2014).

\bibitem{Chew:2016epf}
  X.~Y.~Chew, B.~Kleihaus and J.~Kunz,
  Phys.\ Rev.\ D {\bf 94}, 104031 (2016).

\bibitem{Kleihaus:2017kai}
  B.~Kleihaus and J.~Kunz,
  Fundam.\ Theor.\ Phys.\  {\bf 189}, 35 (2017).


\bibitem{Abe:2010ap}
  F.~Abe,
  Astrophys.\ J.\  {\bf 725}, 787 (2010).

\bibitem{Toki:2011zu}
  Y.~Toki, T.~Kitamura, H.~Asada and F.~Abe,
  Astrophys.\ J.\  {\bf 740}, 121 (2011).

\bibitem{Takahashi:2013jqa}
  R.~Takahashi and H.~Asada,
  Astrophys.\ J.\  {\bf 768}, L16 (2013).

\bibitem{Cramer:1994qj}
  J.~G.~Cramer, R.~L.~Forward, M.~S.~Morris, M.~Visser, G.~Benford and G.~A.~Landis,
  Phys.\ Rev.\ D {\bf 51}, 3117 (1995).

\bibitem{Safonova:2001vz}
  M.~Safonova, D.~F.~Torres and G.~E.~Romero,
  Phys.\ Rev.\ D {\bf 65}, 023001 (2002).

\bibitem{Perlick:2003vg}
  V.~Perlick,
  Phys.\ Rev.\ D {\bf 69}, 064017 (2004).

\bibitem{Nandi:2006ds}
  K.~K.~Nandi, Y.~Z.~Zhang and A.~V.~Zakharov,
  Phys.\ Rev.\ D {\bf 74}, 024020 (2006).

\bibitem{Nakajima:2012pu}
  K.~Nakajima and H.~Asada,
  Phys.\ Rev.\ D {\bf 85}, 107501 (2012).

\bibitem{Tsukamoto:2012xs}
  N.~Tsukamoto, T.~Harada and K.~Yajima,
  Phys.\ Rev.\ D {\bf 86}, 104062 (2012).

\bibitem{Kuhfittig:2013hva}
  P.~K.~F.~Kuhfittig,
  Eur.\ Phys.\ J.\ C {\bf 74}, 2818 (2014).

\bibitem{Tsukamoto:2016zdu}
  N.~Tsukamoto and T.~Harada,
  Phys.\ Rev.\ D {\bf 95}, 024030 (2017).


\bibitem{Bambi:2013nla}
  C.~Bambi,
  Phys.\ Rev.\ D {\bf 87}, 107501 (2013).

\bibitem{Nedkova:2013msa}
  P.~G.~Nedkova, V.~K.~Tinchev and S.~S.~Yazadjiev,
  Phys.\ Rev.\ D {\bf 88}, 124019 (2013).

\bibitem{Zhou:2016koy}
  M.~Zhou, A.~Cardenas-Avendano, C.~Bambi, B.~Kleihaus and J.~Kunz,
  Phys.\ Rev.\ D {\bf 94}, 024036 (2016).

\bibitem{Lamy:2018zvj} 
  F.~Lamy, E.~Gourgoulhon, T.~Paumard and F.~H.~Vincent,
  arXiv:1802.01635 [gr-qc].




\bibitem{Strocchi:2008gsa} 
  F.~Strocchi,
  Lect.\ Notes Phys.\  {\bf 732}, 1 (2008).


\bibitem{Schunck:1996}
 F. E. Schunck and E. W. Mielke,
 in {\sl Relativity and Scientific Computing: Computer Algebra, Numerics, Visualization},
 eds.~F.~W. Hehl, R.~A. Puntigam, and H. Ruder,
 (Springer Berlin Heidelberg, 1996) 

\bibitem{Schunck:1996he} 
  F.~E.~Schunck and E.~W.~Mielke,
  Phys.\ Lett.\ A {\bf 249}, 389 (1998).

\bibitem{Ryan:1996nk} 
  F.~D.~Ryan,
  Phys.\ Rev.\ D {\bf 55}, 6081 (1997).

\bibitem{Yoshida:1997qf} 
  S.~Yoshida and Y.~Eriguchi,
  Phys.\ Rev.\ D {\bf 56}, 762 (1997).

\bibitem{Schunck:1999pm} 
  F.~E.~Schunck and E.~W.~Mielke,
  Gen.\ Rel.\ Grav.\  {\bf 31}, 787 (1999).

\bibitem{Kleihaus:2005me}
  B.~Kleihaus, J.~Kunz and M.~List,
  Phys.\ Rev.\  D {\bf 72}, 064002 (2005).

\bibitem{Kleihaus:2007vk}
  B.~Kleihaus, J.~Kunz, M.~List and I.~Schaffer,
  Phys.\ Rev.\  D {\bf 77}, 064025 (2008).


\bibitem{Hoffmann}
work in progress.

\bibitem{schonauer:1989}
W. Sch\"onauer and R. Wei\ss ,
 J. Comput. Appl. Math. 27, 279 (1989) 279;
 \\
 M. Schauder, R. Wei\ss\ and W. Sch\"onauer,
 {\it The CADSOL Program Package},
 Universit\"at Karlsruhe, Interner Bericht Nr. 46/92 (1992).

\bibitem{Hartmann:2013tca}
  B.~Hartmann, J.~Riedel and R.~Suciu,
  Phys.\ Lett.\ B {\bf 726}, 906 (2013).




\bibitem{Shinkai:2002gv}
  H.~-a.~Shinkai and S.~A.~Hayward,
  Phys.\ Rev.\ D {\bf 66}, 044005 (2002).

\bibitem{Gonzalez:2008xk}
  J.~A.~Gonzalez, F.~S.~Guzman, and O.~Sarbach,
  Class.\ Quant.\ Grav.\  {\bf 26}, 015011 (2009).

\bibitem{Gonzalez:2008wd}
  J.~A.~Gonzalez, F.~S.~Guzman, and O.~Sarbach,
  Class.\ Quant.\ Grav.\  {\bf 26}, 015010 (2009).




\bibitem{Torii:2013xba} 
  T.~Torii and H.~a.~Shinkai,
  Phys.\ Rev.\ D {\bf 88}, 064027 (2013).

\bibitem{Dzhunushaliev:2013jja}
  V.~Dzhunushaliev, V.~Folomeev, B.~Kleihaus, J.~Kunz and E.~Radu,
  Phys.\ Rev.\ D {\bf 88}, 124028 (2013).








\bibitem{Hochberg:1990is}
  D.~Hochberg,
  Phys.\ Lett.\  {\bf B251}, 349 (1990).

\bibitem{Fukutaka:1989zb}
  H.~Fukutaka, K.~Tanaka, K.~Ghoroku,
  Phys.\ Lett.\  {\bf B222}, 191 (1989).

\bibitem{Ghoroku:1992tz}
  K.~Ghoroku, T.~Soma,
  Phys.\ Rev.\  {\bf D46}, 1507 (1992).

\bibitem{Furey:2004rq}
  N.~Furey, A.~DeBenedictis,
  Class.\ Quant.\ Grav.\  {\bf 22}, 313 (2005).

\bibitem{Lobo:2009ip} 
  F.~S.~N.~Lobo and M.~A.~Oliveira,
  Phys.\ Rev.\ D {\bf 80}, 104012 (2009).

\bibitem{Bronnikov:2009az}
  K.~A.~Bronnikov and E.~Elizalde,
  Phys.\ Rev.\  D {\bf 81}, 044032 (2010).

\bibitem{Kanti:2011jz} 
  P.~Kanti, B.~Kleihaus and J.~Kunz,
  Phys.\ Rev.\ Lett.\  {\bf 107}, 271101 (2011).

\bibitem{Kanti:2011yv}
  P.~Kanti, B.~Kleihaus and J.~Kunz,
  Phys.\ Rev.\ D {\bf 85}, 044007 (2012).
%

\bibitem{Harko:2013yb} 
  T.~Harko, F.~S.~N.~Lobo, M.~K.~Mak and S.~V.~Sushkov,
  Phys.\ Rev.\ D {\bf 87}, 067504 (2013).








\end{thebibliography}
\end{document}